\documentclass{article} 
\usepackage{iclr2025_conference,times}

\usepackage{amsmath,amsfonts,bm}

\def\eqref#1{equation~\ref{#1}}

\def\1{\bm{1}}

\def\vh{{\bm{h}}}

\def\vx{{\bm{x}}}
\def\vy{{\bm{y}}}

\def\mH{{\bm{H}}}

\def\mY{{\bm{Y}}}

\DeclareMathAlphabet{\mathsfit}{\encodingdefault}{\sfdefault}{m}{sl}
\SetMathAlphabet{\mathsfit}{bold}{\encodingdefault}{\sfdefault}{bx}{n}

\def\gB{{\mathcal{B}}}

\def\gD{{\mathcal{D}}}

\def\gP{{\mathcal{P}}}

\def\gR{{\mathcal{R}}}

\def\gT{{\mathcal{T}}}

\def\gX{{\mathcal{X}}}

\DeclareMathOperator*{\argmax}{arg\,max}
\DeclareMathOperator*{\argmin}{arg\,min}

\usepackage[english]{babel}

\definecolor{hypercolor}{HTML}{1f72d1}
\usepackage[colorlinks=true,linkcolor=hypercolor,urlcolor=hypercolor,citecolor=hypercolor,filecolor=hypercolor]{hyperref}
\usepackage{url}

\usepackage[utf8]{inputenc} 
\usepackage[T1]{fontenc}    
\usepackage{booktabs}       
\usepackage{bbm}
\usepackage{amsfonts}       
\usepackage{dsfont}
\usepackage{amsmath}
\usepackage{amssymb}
\usepackage{mathtools}

\usepackage{caption}
\usepackage{subcaption}
\usepackage{graphicx}
\usepackage{nicefrac}       
\usepackage{microtype}      
\usepackage[normalem]{ulem}
\usepackage{multirow}
\useunder{\uline}{\ul}{}

\usepackage{algorithm}
\usepackage[noend]{algpseudocode}
\usepackage{algorithmicx}

\usepackage{enumitem}
\setlist[enumerate]{leftmargin=22pt}
\setlist[itemize]{leftmargin=22pt}
\setlist[enumerate,1]{label=\arabic*.}
\setlist[enumerate,2]{label=\arabic*.}

\newcommand{\inlinesubsubsection}[1]{%
  \par\noindent
  \refstepcounter{subsubsection}
  \textbf{\thesubsubsection\quad #1}%
  \quad
}

\definecolor{highlight}{RGB}{0,0,255}  

\newcommand{\recognizer}{\tau}

\title{Procedural Synthesis of \\
Synthesizable Molecules}

\author{
Michael Sun$^{1}$, Alston Lo$^{1}$, Minghao Guo$^1$, Jie Chen$^2$, Connor Coley$^3$, Wojciech Matusik$^1$ \\
$^1$MIT CSAIL \quad $^2$MIT-IBM Watson AI Lab, IBM Research \quad $^3$MIT Chemical Engineering \\
\texttt{\{msun415,alston,wojciech\}@csail.mit.edu},
\texttt{chenjie@us.ibm.com}, \\
\texttt{ccoley@mit.edu}
}

\iclrfinalcopy 
\begin{document}

\maketitle

\begin{abstract}
Designing synthetically accessible molecules and recommending analogs to unsynthesizable molecules are important problems for accelerating molecular discovery. We reconceptualize both problems using ideas from program synthesis. Drawing inspiration from syntax-guided synthesis approaches, we decouple the syntactic skeleton from the semantics of a synthetic tree to create a bi-level framework for reasoning about the combinatorial space of synthesis pathways. Given a molecule we aim to generate analogs for, we iteratively refine its skeletal characteristics via Markov Chain Monte Carlo simulations over the space of syntactic skeletons. Given a black-box oracle to optimize, we formulate a joint design space over syntactic templates and molecular descriptors and introduce evolutionary algorithms that optimize both syntactic and semantic dimensions synergistically. Our key insight is that once the syntactic skeleton is set, we can amortize over the search complexity of deriving the program's semantics by training policies to fully utilize the fixed horizon Markov decision process imposed by the syntactic template. We demonstrate performance advantages of our bi-level framework for synthesizable analog generation and synthesizable molecule design. Notably, our approach offers the user explicit control over the resources required to perform synthesis and biases the design space towards simpler solutions, which is particularly promising for autonomous synthesis platforms. Code is at \url{https://github.com/shiningsunnyday/SynthesisNet}.
\end{abstract}

\section{Introduction}
\indent The discovery of new molecular entities is central to advancements in fields such as pharmaceuticals \citep{zhavoronkov2019,lyu2019}, materials science \citep{hachmann2011,janet2020}, and environmental engineering \citep{zimmerman2020,yao2021}. Traditional make-design-test workflows for molecular design typically rely on labor-intensive methods that involve a high degree of trial and error \citep{lengeling2018}. Systematic and data-efficient approaches that minimize costly experimental trials are the key to accelerating these processes \citep{coley2020a, coley2020b, gao2022}. In recent years, a large number of molecular generative models has been proposed \citep{cao2018, ma2018, simonovsky2018, you2018, li2018, samanta2020, liu2018, jin2018, jin2020, guo2022, sun2024}. However, few of their outputs are feasible to make and proceed to experimental testing due to their lack of consideration for synthesizability \citep{gao2020}.
This has motivated recent methods that integrate design and synthesis into a single workflow, aiming to optimize both processes simultaneously \citep{button2019,bradshaw2019,bradshaw2020,gao2021,swanson2024} which significantly closes the gap between the design and make steps, reducing cycle time significantly \citep{koscher2023,volkamer2023,mccorkindale2023}. This development is spurred by the curation of a small but robust collection of expert reaction templates that are inspired by real-world reactions and defined in close collaboration with chemists \cite{hartenfeller2011}. This workflow facilitates de novo applications by imposing synthetic accessibility by design, recasting molecular design as navigating the space of possible synthetic procedures over a set of building blocks and forward reaction steps, as defined in \cite{vinkers2003}.
However, these methods still face computational challenges, particularly in navigating the combinatorial explosion of potential synthetic procedures \citep{smith1997}.

Inspired by techniques in program synthesis, particularly syntax-guided synthesis \citep{alur2013}, our method decouples the syntactical template of a synthetic procedure (the \emph{skeleton}) from their chemical semantics (the \emph{substance}). This bifurcation allows for a more granular optimization process, wherein the syntactical and semantic aspects of reaction pathways can be optimized independently yet synergistically.
Our methodology employs a bi-level optimization strategy. The upper level optimizes the syntactic template of the synthetic pathway, and the lower level fine-tunes the molecular descriptors within that given structural framework. This dual-layered approach is facilitated by a surrogate policy, implemented by a graph neural network, that propagates embeddings top-down following the topological order of the syntactical skeleton. This ensures that each step in the synthetic pathway is optimized in context, respecting the overarching structural strategy while refining the molecular details. We address the combinatorial explosion in the number of programs using tailored strategies for fixed horizon Markov decision process (MDP) environments. This algorithm amortizes the complexity of the search space through predictive modeling and simulation of Markov Chain Monte Carlo (MCMC) processes \citep{metropolis1953, hastings1970, gilks1995}, focusing on the generation and evaluation of syntactical skeletons. By leveraging the inductive biases from retrosynthetic analysis without resorting to retrosynthesis search, our approach combines accuracy and efficiency in ``synthesizing" synthetic pathways. In summary, the contributions of this work are:
\begin{enumerate}
    \item  We reconceptualize molecule design and synthesis as a conditional program synthesis problem, establishing common terminology for bridging the two fields.
    \item We propose a bi-level framework that decouples the syntactical skeleton of a synthetic tree from the program semantics. Then, we introduce amortized algorithms within our framework for the tasks of synthesizable analog recommendation and synthesizable molecule design.
    \item  We demonstrate improvements across multiple dimensions of performance for both tasks, and include in-depth visualizations and analyses for understanding the source of our method's efficacy as well as its limitations.
\end{enumerate}


\section{Related Works}\label{sec:related}

\subsection{Synthesizable Analog Generation}
\noindent The problem of synthesizable analog generation aims to find molecules close to the target molecule that are \textit{synthesizable}. Closely related but distinct from this problem is computer-assisted retrosynthetic analysis, which has developed through the decades \citep{corey1985} in tandem with computers, and is now known as retrosynthetic planning due to its resemblance to more classical tests of AI based around planning. 
As retrosynthetic planning is done by working backwards (top-down), partial success is not straightforward to define. In other domains, procedural modeling is a bottom-up generation process that generates analogs by design \citep{merrell2010, merrell2011, muller2006, merrell2023}. Thus, synthesizable analog generation warrants more specialized methods. Prior works such as \citet{dolfus2022, levin2023} address this by performing alterations of existing retrosynthesis routes, but this constrained approach severely limits the diversity of analogs. Instead, we neither start from a search route nor constrain the search route, but instead, extract analogs via the iterative refinement of the program's syntactical skeleton with inner loop decoding of the program semantics in a bi-level setup. 
We implement the iterative refinement phase using an MCMC sampler with a stationary distribution governed by similarity to the target being conditioned on. This is a common technique used to search over procedural models of buildings, shapes, and furniture arrangements \citep{merrell2011, talton2011, yu2011}, and we showcase its efficacy for the new application domain of molecules.

\subsection{Synthesizable Molecule Design}\label{sec:2-3}
\noindent The problem of synthesizable molecule design is to design the synthetic pathway, or \textit{program}, whose output molecule optimizes some property oracle function. Note that unlike generic molecular optimization approaches, the design space is reformulated to guarantee synthesizability by construction. The early works to follow this formulation \citep{vinkers2003, hartenfeller2011, button2019} use machine learning to assemble molecules by iteratively selecting building blocks and virtual reaction templates to enumerate a library, with recent works such as \citet{swanson2024} obtaining experimental validation. The key computational challenge these methods must address is how to best navigate the combinatorial search space of synthetic pathways. Prior efforts using bottom-up generation \citep{bradshaw2019,bradshaw2020} probabilistically model synthetic trees as sequences of actions. These adopt an encoder-decoder approach to map between a continuous latent space and a complex combinatorial space. This results in low reconstruction accuracy and hinders the method on the task of conditional generation. Instead, SynNet \citep{gao2021} admits a unified framework for solving analog generation and molecule design by formulating the problem as an infinite-horizon MDP and doing amortized tree generation conditioned on Morgan fingerprints. We show improvements to both tasks through our novel formulation.

\subsection{Program Synthesis}\label{sec:2-2}
Program synthesis is the problem of synthesizing a function $f$ from a set of primitives and operators to meet some correctness specification. 
A program synthesis problem entails: (1) a background theory $\mathsf{T}$ that is the vocabulary for constructing formulas, (2) a correctness specification, i.e., a logical formula involving the output of $f$ and $\mathsf{T}$, and (3) a set of expressions $L$ that $f$ can take on, described by a context-free grammar $G_L$.
In molecular synthesis, we can formulate $\mathsf{T}$ as containing operators for chemical reactions, constants for reagents, molecular graph isomorphism checking, and so forth. The correctness specification for finding a synthesis route for a molecule $M$ is simply $f(\gB)=M$, where $\gB$ is a set of building blocks, and we seek to find an implementation $f$ from $L$ to meet the specification. A coarser specification is to match the molecule's fingerprint, $\text{FP}(f(\gB)) = \text{FP}(M)$, and as shown by \citet{gao2021}, this relaxed formulation enables both analog generation and fingerprint-based molecule optimization. Our key innovation takes inspiration from the line of work surrounding syntax-guided synthesis \citep{alur2013, schkufza2013}. Syntax guidance explicitly constrains $G_L$, which reduces the set of implementations $f$ can take on \citep{alur2013} and facilitates more accurate amortized inference. Further discussion on the connections between program synthesis and molecular synthesis and the similarities to the literature of retrosynthesis are given in App. \ref{app:program}.

\begin{figure}[h!]
 \centering
 \includegraphics[width=\textwidth]{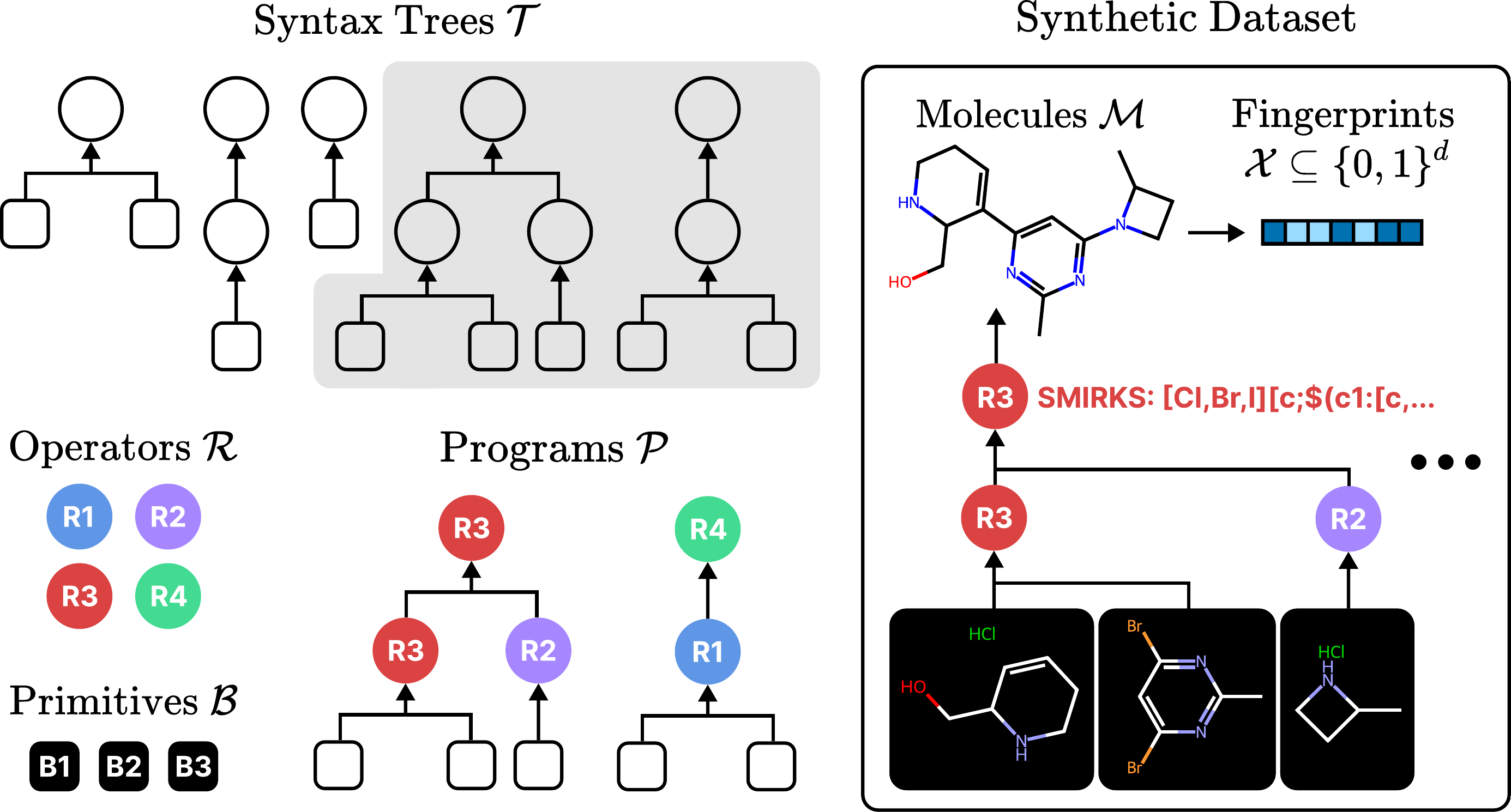}
 \caption{Program synthesis terminology for modeling synthesis pathways.}
 \label{fig:fig1}
\end{figure}

\section{Methodology}
\setcounter{secnumdepth}{2}

\subsection{Problem Definition}

For clarity, we recapitulate the problems of interest discussed in Section \ref{sec:related}: 

\textbf{Synthesizable analog generation} is the inverse task of inferring the program and inputs that best reconstructs a target molecule. {Denoting the set of reactions as $\mathcal{R}$ and the set of building blocks as $\gB$, we obtain a compact yet expressive design space $\gP$: all non-trivial, attributed binary trees where each internal node corresponds to a reaction. Drawing parallels to the program synthesis literature, we call $\gP$ the \textit{program} space. The problem can now be formalized: given a space of molecules $\mathcal{M}$, learn a mapping $F \colon \mathcal{M} \rightarrow \gP \times \gB^*, M \mapsto (P, B)$ such that $B$ can be assigned to the leaf nodes of $P$ and running the reaction(s) in $P$ in a bottom-up manner (by recursively feeding the product of a node’s reaction to its parent) produces a molecule $M_*$ with minimal ``distance" to $M$. Using program execution notation, the objective is stated as: $\argmin_{(P,B)\in \gP \times \gB^*} \text{dist}(P(B),M)$.}

\textbf{Synthesizable molecule design} is the forward task of finding the program and inputs whose output optimizes a property oracle function. {The oracle can represent property predictors, simulation, experimental validation, etc. but are black-box in nature. The objective is $\argmin_{(P,B)} \text{Oracle}(P(B))$.

Moving forwards, we identify $(P, B)$ with its output $M_*$ when it simplifies notation. Stripping the semantics from a program leaves behind a syntactic skeleton, which lies in the space $\gT$ of non-trivial binary trees. Figure \ref{fig:fig1} illustrates the discussed terminologies\footnote{Please refer to Figure 3 of \cite{gao2021} for example chemical illustrations of the two tasks.}.}



\setcounter{secnumdepth}{3}
\subsection{Solution Overview}
Herein, we use expert-defined reaction templates, a popular abstraction codifying deterministic graph transformation patterns in the SMIRKS formal language. SMIRKS assumes the reactants are ordered for defining how atoms and bonds are transformed from reactants to products. Since templates are designed to encompass the most common and robust chemical transformations, ours are restricted to uni-molecular (e.g., isomerization, ring closure, or fragmentation reactions) or bi-molecular (e.g., additions, substitutions, or coupling reactions) reactions. In practice, we featurize molecules using Morgan fingerprints with radius 2 and $d = 2048$ bits, which is a common molecular representation in both predictive and design tasks. This means that $F$ is now technically a map over fingerprint space $\gX \subseteq \{0, 1\}^d$. It is then natural to use the Tanimoto distance between fingerprints as our notion of molecular distance. 


\begin{figure}[h!]
    \centering
    \makebox[1\textwidth][c]{%
        \scalebox{1}[1.05]{\includegraphics[width=1.05\textwidth]{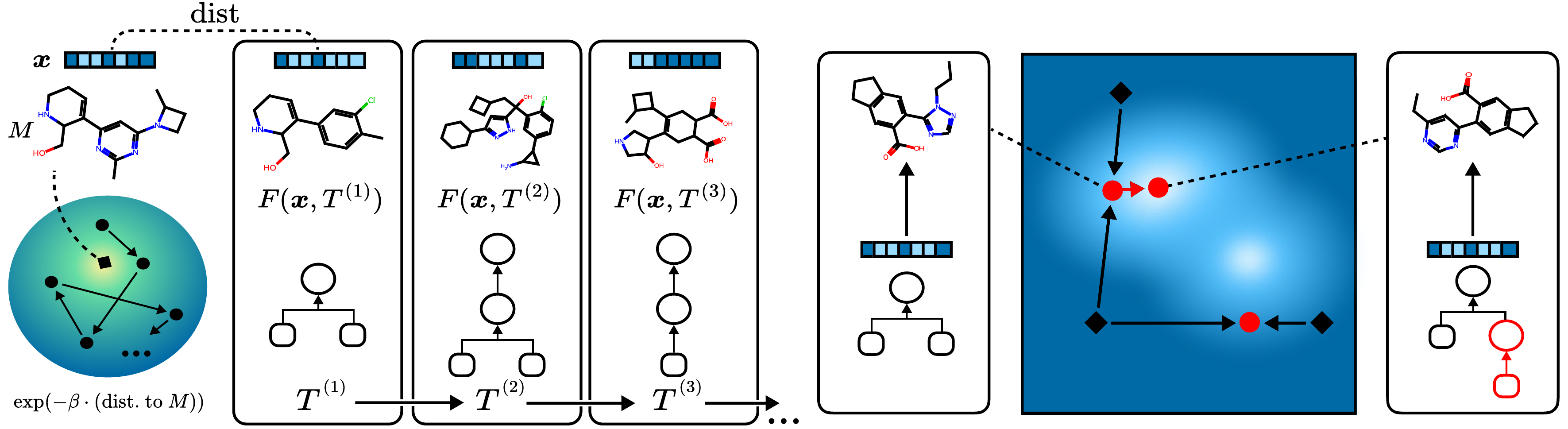}}        
      }
      \vspace{0.3em}
     \caption{(Left) 
     Our Metropolis-Hastings algorithm in Section \ref{sec:3-3} iteratively refines the syntax tree skeleton towards the stationary distribution which is proportional to the inverse distance to our target molecule $M$. (Right) Our genetic algorithm over the joint design space $\gX\times\gT$ in Section \ref{sec:3-4} combines the strategies of semantic crossover ($\rightarrow$) and syntactical mutation ($\textcolor{red}{\rightarrow}$) to encourage both global improvement and local exploration.}
     \label{fig:mcmc}

\end{figure}

\subsection{Bilevel Syntax-Semantic Synthesizable Analog Generation}\label{sec:3-3}



Given a molecule $M \in \mathcal{M}$, we aim to find a program and inputs $(P, B)$ whose output $M_*$ is most similar to $M$. Suppose first that the syntax $T$ of $P$ was given. Then, we would use two learned policies $\pi_{\mathcal{R}}$ and $\pi_\mathcal{{B}}$ that iteratively attribute reactions and building blocks to the nodes of $T$ in topological order (Section \ref{sec:inner}). Recalling Section \ref{sec:2-2}, this can be seen as parameterizing the derivation process of a context-free grammar $G_\gP$. We give further details in App. \ref{app:grammar} and discuss how these derivations are constrained through syntax. 

The application of our policies essentially specifies a syntax-conditioned program synthesis map $F \colon \mathcal{M} \times \gT \to \gP \times \gB^*$. However, $T$ is not known during inference, so to remove this dependence, we consider two further strategies. One is to learn a classifier $\recognizer{} \colon \mathcal{M} \rightarrow \gT$ to predict the most likely syntax tree of the program that produces $M$. We implement $\recognizer{}$ with a multi-layer perceptron (MLP) trained under a standard classification task.\footnote{In practice, we restrict to a finite subset of $\gT$ by imposing a maximum number of reaction nodes.} However, a single prediction may be inadequate, since there can be multiple skeletons corresponding to the same molecule. Instead, our second strategy uses a bi-level setup, where the outer loop explores syntax space $\gT$ through invocations of the inner loop $F$, which explores the program's semantics. This approach is further made efficient by amortizing the training of the inner loop.

\subsubsection{Outer loop: Syntax Tree Structure Optimization}\label{sec:3-3-2}

We simulate a Markov process on $\gT$ for discovering skeletons whose decoding will maximize the similarity between $M$ and the program’s output (Figure \ref{fig:mcmc} (Left)). The details for how we bootstrap and apply $\gT$ is in App. \ref{app:eda}. We adopt the Metropolis-Hastings algorithm with proposal distribution $q(T\,|\, T_0) \propto \exp(-\lambda\, d_{\gT}(T,T_0))$ and scoring function $ f(T) \propto \exp(-\beta\, d_{\mathcal{M}}(M, F(M,T)))$,
where $\lambda$ and $\beta$ are parameters that trade-off exploration with exploitation, $d_{\gT}$ is the tree edit distance, and $d_{\mathcal{M}}$ is the Tanimoto distance. In other words, we use the inner loop to score candidates in $\gT$. 


\subsubsection{Inner loop: Inference of Tree Semantics}\label{sec:inner}

We now formulate syntax-conditioned derivations under $G_{\gP}$ as a finite-horizon MDP. 

\textbf{State space:} To bridge $\gT$ and $\gP \times \gB^*$, we introduce an intermediate state space of partial programs $\partial \gP$ consisting of all possible partial programs arising from the following modifications to any syntax tree $T \in \gT$: (1) prepend a new root node of $T$, and attribute it with a fingerprint from $\gX$, or (2) attribute some internal (resp.~leaf) nodes of $T$ with elements of $\mathcal{R}$ (resp.~$\gB$). We further require that if a node in $T$ is filled, then so is its parent. Intuitively, $\partial \gP$ comprises all partially filled-in trees in $\gT$ obeying topological order (with an added root node attached to a molecular fingerprint). 

\textbf{Action space:} At a state $S \in \partial \gP$, the actions are to attribute any frontier node, i.e., unfilled nodes whose parents are filled, with an item from $\mathcal{R}$ (resp.~$\gB$) if the node is internal (resp.~a leaf).

\textbf{Policy network:} We parameterize the policies $\pi_{\mathcal{R}} \colon \partial \gP \rightarrow \mathcal{R^*}$ and $\pi_{\gB} \colon \partial \gP \rightarrow \mathcal{B^*}$ with separate graph neural networks. Given a partial program $S$ as input, the former predicts the reactions that should be attributed to internal frontier nodes as an $|\mathcal{R}|$-label classification problem, while the latter does so for building blocks and leaf nodes. However, $|\gB| \gg |\mathcal{R}|$ is very large, so $\pi_{\gB}$ instead predicts a 256-dimensional embedding at each nodes which implicitly specifies the building block whose 256-bit Morgan fingerprint is closest. 

\textbf{Training:} We train both policies using supervised policy learning. Key to this approach is the dataset used for training constructed using Algorithm \ref{alg:pretrain}. See App. \ref{app:policy-network} for further details.

\begin{algorithm}[H]
\caption{Construction of training dataset.}
\label{alg:pretrain}
\begin{algorithmic}[1]
\Require{A synthetic dataset $\gD_0 \subseteq \gP \times \gB^*$ of programs (Section \ref{sec:4-1-1}).}
\State $\gD \gets \varnothing$
\For{each $(P, B) \in \gD_0$}
    \State Turn $(P, B)$ into a fully-filled program $T \in \partial \gP$ whose root is attributed with $\text{FP}(P, B)$.
    \For{each $\Lambda \in 2^T$ containing the root and closed under $\text{parent}(\cdot)$ }
        \State $\text{Frontier}(\Lambda) \gets \{i \in T \mid i \notin \Lambda \text{ and } \text{parent}(i) \in \Lambda \}$
        \State Populate node features $\mH$ and labels $\mY$ based on $P$ and $B$ (App. \ref{app:policy-network})
        \For{$i \in T - \Lambda$}
            \State Mask the feature in $\mH$ corresponding to node $i$
        \EndFor 
        \For{$i \in T - \text{Frontier}(\Lambda)$}
            \State Mask the label in $\mY$ corresponding to node $i$
        \EndFor 
        \State $\gD \gets \gD \cup \{(T, \mH, \mY)\}$
    \EndFor
\EndFor 
\Return $\gD$
\end{algorithmic}
\end{algorithm}

\begin{figure}[h!]

 \centering
     \centering
     \includegraphics[width=0.9\textwidth]{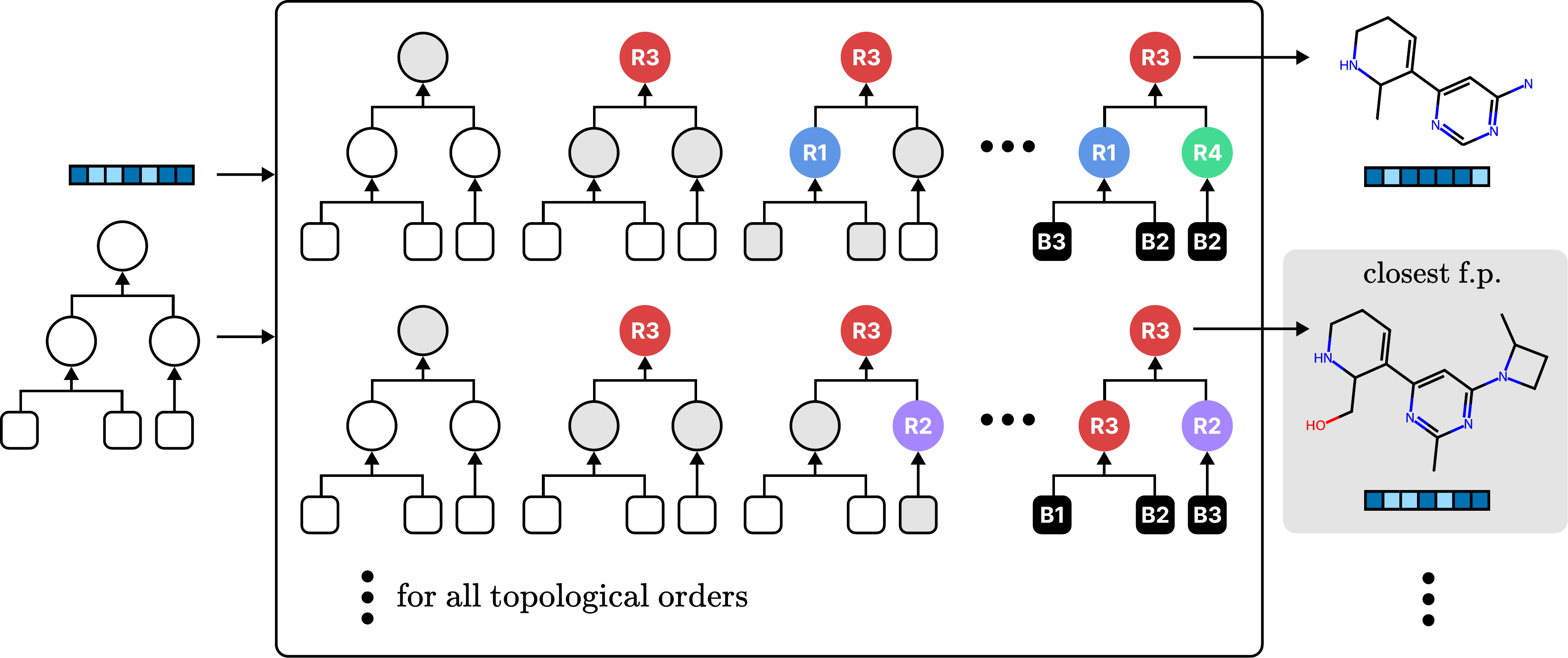}
     \caption{Illustration of our decoding scheme $F$: (Left) The input is a Morgan fingerprint $\vx$ and syntax skeleton $T$; (Middle) Decode once for every topological ordering of the tree, tracking all partial programs with a stack; (Right) Execute all decoded programs, then returning the closest analog which minimizes distance to $\vx$.}
     \label{fig:fig3}
 \vspace{-1em}
\end{figure}

\subsection{Bi-level Syntax-Semantic Synthesizable Molecular Design}\label{sec:3-4}
The task of synthesizable molecule design is to find a program $P$ and building blocks $B$ whose $M_*$ maximizes a property of interest. Given the learned policies from synthetic planning, we apply the inner loop procedure $F \colon \gX \times \gT \rightarrow \gP \times \gB^*$ as a surrogate, casting the optimization problem as one over the joint design space $\gX \times \gT$.
We approach this problem with a genetic algorithm (GA) over fingerprint space $\gX$ that leverages this extra dimension of syntax $\gT$ (Figure \ref{fig:mcmc} (Right)). The seed population is obtained by sampling random bit vectors. To generate a child $\vx$ from two parents $\vx_1$ and $\vx_2$, we combine \textit{semantic} crossover with \textit{syntactical} mutation, reminiscent of our bi-level approach for analog generation:

\textbf{Semantic crossover:} We first generate $\vx_*$ by combining bits from both $\vx_1$ and $\vx_2$ and possibly mutating a small number of bits of the  result.

\textbf{Syntactic mutation:} We set $T = \recognizer(\vx_*)$ and apply random edit(s) to obtain a perturbed tree $T'$. Together, $(\vx_*, T)$ and $(\vx_*, T')$ form a sibling pool. Applying the surrogate $F$ to each child 
gives two sibling SMILES that are turned into two sibling  fingerprints. We fit a Gaussian process on past individuals
and select the sibling with the highest expected improvement as the favoured child $\vx$.


Intuitively, semantic crossover optimizes for chemical semantics by combining existing ones from the mating pool, while syntactic mutation explores syntactic analogs of individuals of interest. Each child is then given a fitness score under the property oracle, and the top scoring unique individuals are retained into the next generation (i.e., elitist selection). Further details and hyperparameters are given in App. \ref{app:ga}. 

\section{Experiments}



\subsection{Experiment Setup}
\subsubsection{Data Generation}\label{sec:4-1-1}

We use 91 reaction templates from \citet{hartenfeller2011, button2019} representative of common synthetic reactions. They consist primarily of ring formation and fusing reactions but also peripheral modifications. Starting from 147,505 purchaseable compounds from Enamine Building Blocks, we follow the same steps as \citet{gao2021} to generate 600,000 synthetic trees. Filtering by $\text{QED} >0.5$ of the product molecules leaves 227,808 synthetic trees (136,684 for training, 45,563 for validation, and 45,561 for testing), which are then preprocessed into programs to construct our final datasets. We bootstrap our set of syntactic templates based on those observed in the training set, resulting in 1117 syntactic skeleton classes. Additional statistics on these syntactic templates and insights on their coverage are given in App. \ref{app:eda}. Further analyses of the structure-property relationship and a detailed case study are given in App. \ref{app:recognizer}.

\subsubsection{Baselines} 

We evaluate against all 25 molecular optimization methods benchmarked in the large-scale study by \citet{gao2022}. Most of these are divided into three categories based on the molecular representation used: string, graph, or synthesis trees. Synthesis methods restrict the design space to only products of robust template-based reactions, so for fair comparison, we also consider intra-category rankings. We report the synthetic accessibility (SA) score \citep{ertl2009} of the optimized molecules to cross-verify the synthesizability of synthesis-based methods as well a investigate the performance trade-off imposed by constraining for template-compatible synthesizability.

\subsubsection{Computational Efficiency}\label{sec:4-1-3}
\textbf{Constructing $\mathcal{D}$}. Alg. \ref{alg:pretrain} suggests $|\mathcal{D}|=O(2^{\max_{\mathcal{T}}|T|} |\mathcal{D}_0| )$ because we compute all the (closed under $\text{parent}(\cdot)$) masks per $P \in \mathcal{D}_0$. However, we don't need to explicitly store $\mathcal{D}$. We only need to precompute the (closed under $\text{parent}(\cdot)$) masks for each $T\in \hat{\mathcal{T}_k}$ ($k$ fixed and small), so the running time is $O(\sum_{T\in \hat{\mathcal{T}_k}} 2^{|T|})$, independent of $\mathcal{D}_0$. Besides, the actual number of masks is far smaller than $2^{|T|}$, and high $|T|$ is less represented in $\mathcal{D}_0$, since large programs are less likely to be sampled, so in practice $|\mathcal{D}|$ is much smaller than $2^{\max_{\mathcal{T}}|T|} |\mathcal{D}_0|$. Detailed statistics are given in App \ref{app:eda}. \\\\
\textbf{Training with $\mathcal{D}$}. During training, we flat-index into $\mathcal{D}_0$ and their precomputed masks to perform a pass over $\mathcal{D}$. Despite the larger dataset, we find the total training steps to actually be comparable with \cite{gao2021}. To scale to larger $k$, we propose a stratified sampling strategy that does only a $O(\mathcal{D}_0)$ pass per epoch, has positive support over $\mathcal{D}$, and represents each $(P,B) \in \mathcal{D}_0$ equally. The idea is to sample a constant number of masks $C$ per $P,B\in \mathcal{D}_0$, and re-sample each epoch. We show this has complexity linear to the \textit{size} of $\mathcal{D}_0$ per epoch but empirically converges in much fewer training steps than SynNet for $C=1$. Details are in App \ref{app:convergence}. We further include an ablation study on the downstream task performance of this simplified training strategy in App \ref{app:ablate-simple}.

\subsubsection{Evaluation Metrics} 

\textbf{Synthesizable analog generation.} We evaluate the ability to generate a diverse set of structural analogs to a given input molecule using the Recovery Rate (RR, whether the most similar analog is exactly the target), Average Similarity (as measured by Tanimoto distance to the input), SA score, and Internal Diversity (average pairwise Tanimoto distance).

\textbf{Synthesizable molecule design.} We evaluate our method's ability to optimize 15 oracle functions \citep{huang2022} relevant to real-world drug discovery:
\begin{enumerate}
    \item \textbf{Bioactivity predictors} (GSK3$\beta$, JNK3, DRD2) that estimate responses against targets related to the pathogenesis of various diseases such as Alzheimer’s disease \citep{koch2015inhibitors} based on experimental data \citep{sun2017excape}, and whose inhibitors are the basis for many antipsychotics and have shown promise for treating diseases like Parkinson's schizophrenia and bipolar disorder \citep{madras2013history}.
    \item \textbf{Structural profiles} (Median1, Median2, Rediscovery) that primarily focus on maximizing structural similarity to multiple molecules, which is useful for designing molecules fitting a more multifaceted structural profile \citep{brown2004graph}. The rediscovery oracle focuses on hit expansion around a specific drug.
    \item \textbf{Multi-property objectives} (Osimertinib, 6 others) that use real drugs as a basis for optimizing additional pharmacological properties, mimicking real-world drug discovery.
    \item \textbf{Docking Simulations} (M\textsuperscript{pro}, DRD3) against M\textsuperscript{pro}, the main protease of SARS-Cov-2, and DRD3, which has its own leaderboard with a particular focus on sample efficiency.
\end{enumerate}
In addition to the average score of the top $k$ molecules, we particularly focus on sample efficiency, i.e., the top-$k$ AUC as described in \citet{gao2022}.

\begin{table}[ht]
\caption{We generate 5 unique analog molecules conditioned on an input molecule $M$ and sort them by decreasing similarity to $M$. For SynNet, we follow their beam search strategy and produce analogs using the top 5 beams. For Ours ($\recognizer$), we sample the top 5 syntactic templates from $\recognizer{}$. Ours ($\recognizer$, rev) is the same except we use a bottom-up decoding process, and it is included as an ablation for Section \ref{sec:ablation-order}. Then, we evaluate how similar, diverse, and structurally simple the first $k$ molecules are. The best method is bolded. For three-way comparisons, the second best method is underlined.}
\label{tab:1}
\centering
\small 
\resizebox{0.9\textwidth}{!}{
\begin{tabular}{llcc@{\hskip4pt}c@{\hskip4pt}cc@{\hskip4pt}c@{\hskip4pt}cc@{\hskip4pt}c}
\toprule
& & & \multicolumn{3}{c}{Avg.\ Sim. $\uparrow$} & \multicolumn{3}{c}{SA $\downarrow$} & \multicolumn{2}{c}{Diversity $\uparrow$} \\
Dataset & Method & RR\ $\uparrow$ & Top-1 & Top-3 & Top-5 & Top-1 & Top-3 & Top-5 & Top-3 & Top-5 \\ 
\midrule
\multirow{2}{*}{Train Set} 
 & Ours ($\recognizer$, rev) & 79.3 \% & 0.923 & 0.632 & 0.569 & \textbf{3.072} & \textbf{2.795} & \textbf{2.716} & \textbf{0.615} & \textbf{0.657} \\
 & Ours ($\recognizer$) & \textbf{88.1\%} & \textbf{0.958} & \textbf{0.704} & \textbf{0.626} & 3.099 & 2.928 & 2.852 & 0.532 & 0.615 \\ \midrule
\multirow{3}{*}{Test Set} 
 & SynNet & {\ul 46.3\%} & {\ul 0.766} & \textbf{0.622} & \textbf{0.566} & 3.108 & 3.057 & 3.035 & 0.525 & 0.584 \\
 & Ours ($\recognizer$, rev) & 40.8 \% & 0.749 & 0.548 & 0.487 & \textbf{2.970} & \textbf{2.743} & \textbf{2.659} & \textbf{0.640} & \textbf{0.685} \\
 & Ours ($\recognizer$) & \textbf{52.3\%} & \textbf{0.799} & {\ul 0.588} & {\ul 0.548} & {\ul 3.075} & {\ul 2.895} & {\ul 2.856} & {\ul 0.609} & {\ul 0.653} \\ 
\midrule
\multirow{3}{*}{ChEMBL} 
& SynNet & 4.9\% & 0.499, & 0.436, & 0.394 & 2.669, & 2.685, & 2.697 & 0.644, & 0.693 \\
& Ours ($\recognizer$) & {\ul 7.6\%} & {\ul 0.531,} & {\ul 0.443,} & {\ul 0.396} & {\ul 2.544,} & {\ul 2.510,} & {\ul 2.460} & {\ul 0.675,} & {\ul 0.727} \\
& Ours (MCMC) & \textbf{9.2\%} & \textbf{0.532,} & \textbf{0.486,} & \textbf{0.432} & \textbf{2.364,} & \textbf{2.310,} & \textbf{2.263} & \textbf{0.765,} & \textbf{0.759} \\
\bottomrule
\end{tabular}
}
\end{table}

\subsection{Results on Synthesizable Analog Generation}\label{sec:4-2-1}

In Table \ref{tab:1}, we see our method outperforming SynNet across both dimensions of similarity (how ``analog'' compounds are) and diversity (how different the compounds are). Additionally, our method achieves lower SA Score, which is a proxy for synthetic accessibility that rewards simpler molecules. Guided by a set of simple yet expressive syntactic templates, our model simultaneously produces more diverse and structurally \textit{simple} molecules without sacrificing one for the other. Additionally, our policy network is well-suited to navigate these simple yet horizon structures, enabling a $6\%$ higher reconstruction accuracy after training on the same dataset. Combining these three dimensions, we can conclude our method is the superior one for the task of synthesizable analog generation. To better understand which design choices are responsible for the performance, we provided a comprehensive analysis of the policy network in App. \ref{app:policy-network}. We begin in App. \ref{app:aux} by elaborating on the main distinction of our method vs existing works, highlight the novelty of our formulation, and motivate an auxiliary training task that takes inspiration from cutting-edge ideas in inductive program synthesis. We then perform several key ablations in App. \ref{app:ablate}, using concrete examples to highlight success and failure cases. Lastly, we perform a step-by-step walkthrough of our decoding algorithm in App. \ref{app:attn}, visualizing the evolution of attention weights to showcase the full-horizon awareness of our surrogate and the dynamic incorporation of new information. These analyses shed insights into why our surrogate works, and points to future extensions to make it even better.

\begin{table}[!htb]
\caption{Our model's average performance across 13 TDC oracles compared to baselines compiled in \citet{gao2022}. We limit to 1000 oracle calls each run and normalize oracle outputs to $[0, 1]$. We report to mean score, AUC, and SA scores \citep{ertl2009} of the top 10 molecules. Methods are categorized by \citet{gao2022} and, for brevity, we display only the top three algorithms within each category with respect their AUC. The best method per column is bolded and the best synthesis-based method is underlined. See App. \ref{app:full-results} for the full results and experiment details.}
\label{tab:average-results}
\centering
\small
\resizebox{0.7\textwidth}{!}{
\begin{tabular}{llc@{\hskip6pt}cc@{\hskip6pt}cc@{\hskip6pt}c}
\toprule
& & \multicolumn{2}{c}{Score $\uparrow$} & \multicolumn{2}{c}{AUC $\uparrow$} & \multicolumn{2}{c}{SA $\downarrow$} \\
Category & Method & Value & Rank & Value & Rank & Value & Rank \\
\midrule
\multirow{2}{*}{Screening} & Screening &0.426 &20 &0.377 &20 &3.097 &8 \\
& MolPAL &0.472 &16 &0.444 &15 &3.018 &4 \\
\midrule
\multirow{4}{*}{String}
& REINVENT &0.697 &2 &0.607 &2 &3.415 &9 \\
& REINVENT-SELFIES &0.682 &3 &0.578 &4 &3.791 &15 \\
& STONED &0.609 &8 &0.555 &6 &5.550 &24 \\
& 7 rows omitted $\cdots$ & & & & & & \\
\midrule
\multirow{4}{*}{Graph}
& Graph-GA & \textbf{0.701} &1 &0.601 &3 &3.982 &17 \\
& GPBO &0.642 &6 &0.570 &5 &3.954 &16 \\
& DST &0.555 &10 &0.479 &11 &4.146 &20 \\
&  7 rows omitted $\cdots$ & & & & & & \\
\midrule
\multirow{4}{*}{Synthesis} 
& SynNet &0.578 &9 &0.545 &7 &3.075 &6 \\
& DoG-Gen &0.634 &7 &0.511 &9 &2.793 &2 \\
& DoG-AE &0.460 &18 & 0.450 &14 &2.857 &3 \\
\cmidrule{2-8}
& Ours & {\ul 0.670} &4 & {\ul\textbf{0.608}} &1 & {\ul\textbf{2.739}} &1 \\
\bottomrule
\end{tabular}
}
\end{table}

\begin{table}[!htb]
\caption{AutoDock Vina scores against DRD3 and M\textsuperscript{pro}, limited to 5000 oracle calls. For ZINC (Screening), we use numbers from TDC's DRD3 Leaderboard, and for SynNet, we report both their paper's numbers (*) and our reproduced results.  We also report the top 3 binders for M\textsuperscript{pro} for the real-world case study in App. \ref{app:docking-study}.}
\label{tab:docking}
\centering
\small
\resizebox{0.7\textwidth}{!}{\begin{tabular}{llcccccccc}\toprule
& & & \multicolumn{5}{c}{Score $\uparrow$} & AUC $\uparrow$ & SA $\downarrow$ \\
Target &Method &\#Calls &1st &2nd &3rd &Top-10 &Top-100 & Top-10 & Top-10 \\\midrule
\multirow{4}{*}{DRD3} 
&ZINC & $-$ &12.8 &$-$ &$-$ &12.59 &12.08 &$-$ &$-$ \\
&SynNet* &5000 &12.3 &$-$ &$-$ &12.02 &11.13 &$-$ &$-$ \\
&SynNet &5000 &10.8 & 10.4 & 10.3 &10.30 &9.20 &9.55 &2.59 \\
&Ours &5000 & \textbf{13.7} & \textbf{13.1} & \textbf{13.1} &\textbf{13.01} &\textbf{12.13} & \textbf{11.91} & \textbf{2.13} \\
\midrule 
\multirow{3}{*}{M\textsuperscript{pro}} 
&SynNet &5000 &8.3 &8.3 &8.2 &8.09 &7.46 &7.60 & \textbf{2.27} \\
&Ours &5000 & \textbf{9.9} &\textbf{9.7} &\textbf{9.7} & \textbf{9.54} & \textbf{9.02} & \textbf{9.01} & 2.59 \\
\cmidrule{2-10}
&SynNet* &$-$ & 10.5 &9.3 &9.3 &$-$ &$-$ &$-$ &$-$ \\
&Ours &10000 & \textbf{10.8} & \textbf{10.7} & \textbf{10.6} &$-$ &$-$ &$-$ &$-$ \\
\bottomrule
\end{tabular}}
\end{table}


\begin{table}[!htb]
\caption{(Left) Ablation of sibling pool generation strategies on JNK3: (edits) mutates the syntax, ($\tau$) uses the top skeletons predicted from $\recognizer{}$, and (flips) doesn't consider skeleton and instead flips random bits in the fingerprint. (Right) Ablation of SynNet with our Bayesian optimization (BO) acquisition over a sibling pool generated by beam search. \textbf{Seeds} and \textbf{All} are the average scores of the initial and final populations.}
\label{tab:sibling}
\centering
\small
\begin{minipage}{0.4\linewidth}
\resizebox{\textwidth}{!}{
\begin{tabular}{lcccccc}
\toprule
& 1st  & 2nd  & 3rd  & Top-10  & Top-100  & Diversity  \\ \midrule
Ours (edits) & \textbf{0.88} & \textbf{0.88} & \textbf{0.87} & \textbf{0.86} & \textbf{0.8} & \textbf{0.61} \\
Ours ($\tau$) & \textbf{0.88} & \textbf{0.88} & \textbf{0.87} & 0.83  & 0.74  & 0.55  \\
Ours (flips) & 0.87  & 0.87  & 0.86  & 0.84  & 0.77  & 0.49  \\ \bottomrule
\end{tabular}
}
\end{minipage}
\hspace{0.01\linewidth}
\begin{minipage}{0.57\linewidth}
\resizebox{\textwidth}{!}{
\begin{tabular}{llccccc}
\toprule
Oracle & Method & Top-1 & Top-10 & Top-100 & Seeds & All \\ 
\midrule
\multirow{3}{*}{GSK3$\beta$} 
& SynNet & 0.94 & 0.907 & 0.815 & $0.050 \pm 0.051$ & $0.803 \pm 0.041$ \\
& SynNet + BO & {0.85} & {0.684} & {0.471} & {$0.013 \pm 0.024$} & {$0.447 \pm 0.090$} \\
& Ours & \textbf{0.98} & \textbf{0.967} & \textbf{0.944} & $0.074 \pm 0.055$ & $\textbf{0.941} \pm \textbf{0.012}$ \\ 
\midrule
\multirow{3}{*}{JNK3} 
& SynNet & 0.80 & 0.758 & 0.719 & $0.032 \pm 0.025$ & $0.715 \pm 0.017$ \\
& SynNet + BO & {0.31} & {0.241} & {0.143} & {$0.006 \pm 0.012$} & {$0.134 \pm 0.039$} \\
& Ours & \textbf{0.88} & \textbf{0.862} & \textbf{0.800} & $0.059 \pm 0.053$ & $\textbf{0.792} \pm \textbf{0.030}$ \\ 
\midrule
\multirow{3}{*}{DRD2} 
& SynNet & \textbf{1.000} & \textbf{1.000} & 0.998 & $0.007 \pm 0.018$ & $0.996 \pm 0.003$ \\
& SynNet + BO & {0.982} & {0.963} & {0.722} & {$0.005 \pm 0.018$} & {$0.672 \pm 0.147$} \\
& Ours & \textbf{1.000} & \textbf{1.000} & \textbf{1.000} & $0.024 \pm 0.056$ & $\textbf{1.000} \pm \textbf{0.000}$ \\ 
\bottomrule
\end{tabular}
}
\end{minipage}
\end{table}

\subsection{Results on Synthesizable Molecule Design}\label{sec:4-2-2}


In Table \ref{tab:average-results}, we see our method outperforming \textit{all} synthesis-based methods on average across the 13 TDC oracles for all considered metrics -- average score, AUC, and SA score. Surprisingly, our method stays competitive with the state-of-the-art string and graph methods in terms of average score (ranking 4th) but being considerably more sample-efficient at finding the top molecules (ranking 1st for top-1/10 AUC). We see evidence that a synthesizability-constrained design space does not sacrifice end performance when reaping benefits of enhanced synthetic accessibility and sample efficiency.

The AutoDock Vina scores reflect our method's strength in real-world ligand design tasks. Our best binders against M\textsuperscript{pro} in Table \ref{tab:docking} are significantly better than nearly all known inhibitors from virtual screening or literature \citep{ghahremanpour2020, zhang2021}. For example, \citet{zhang2021} report a best score of $-8.5$. Our best binders against DRD3 also rank us 3rd on the \href{https://tdcommons.ai/benchmark/docking_group/drd3/}{TDC Leaderboard} (as of Aug.~2024). We present additional analysis of the best binders for our two docking targets in App. \ref{app:docking-study}.

\subsection{Ablation Studies}\label{sec:4-3}

In Tables \ref{tab:1} and \ref{tab:sibling}, we analyze findings from four carefully designed ablation studies (1 in \ref{sec:ablation-order}, 2 \& 3 in \ref{sec:ablation-analog}, 4 in \ref{sec:ablation-template}) to justify the key design decisions that differentiate our method from the predecessor SynNet as well as other synthesis-based methods that have similar modules. 

\inlinesubsubsection{Top-down vs Bottom-up Decoding.}\label{sec:ablation-order}

Should we decode top-down or bottom-up? This separation exists between retrosynthesis vs molecular design. Baselines that serialize the construction of synthesis trees \citep{bradshaw2019, bradshaw2020, gao2021} adopt the latter. Given only knowledge of the target molecule, the model has to predict the first building block, the first reaction, and so on. Our empirical findings confirm observations by \citet{gao2021} that the first few steps are difficult, due to inherent ambiguity -- multiple valid choices exist. Our method resolves this bottleneck by reformulating the (Markov) state as partial syntax trees, where holes are reactions and building blocks left to predict. This state captures the horizon structure, so we can learn tailored policies for the fixed horizon. We reintroduce the inductive bias of retrosynthetic analysis to procedural synthesis. Conditioned on the syntax (skeleton), we show a top-down filling order outperforms bottom-up, with two intuitive explanations: (1) there are orders of magnitude less reactions than building blocks (91 vs.\ 147,505) and (2) it is easier to reason backwards from the specification (target molecule) which reactions lead to the product. To demonstrate these factors compensate for any gains from bottom-up pruning of applicable reactions, we perform an additional ablation in Table \ref{tab:1}. Instead of the MDP enforcing we fill in a skeleton top-down, we fill the skeleton from the bottom-up. We retrain the model by pretraining on inverted masks, and decode by following every possible topological order of the skeleton with edges reversed. The results show this cannot reconstruct the training data as well and struggles on generalization. 


\inlinesubsubsection{{How Analog Generation Capabilities Translate to Design Capabilities.}}\label{sec:ablation-analog}

Why does our superior analog generation capability translate to better performance when used as an offspring generator within molecule optimization? An ideal surrogate takes as input a fingerprint and outputs multiple synthesizable analogs, creating diverse offspring(s) that balance local neighborhood exploration with global exploitation. SynNet does the former by mutating the fingerprint directly, whereas the key insight of our syntax-guided method is to mutate the syntactic skeleton instead, doing so via editing mutations. Table \ref{tab:sibling} (Left) shows edit-based mutations are superior to the top recognition strategy used for analog generation ((Skeleton) in Table \ref{tab:1}) and the trivial strategy of ignoring the skeleton and flipping individual bits to obtain siblings. This suggests edits to the skeleton better preserves the locality bias within the GA. Higher population diversity and average scores for $k\in \{10,100\}$ suggest the same symbiotic relationship between diversity and similarity in analog generation is also the key enabler to better GA optimization. Our GA benefits from an inner loop sibling acquisition within the crossover operation, acquiring the highest expected improvement sibling to expend an oracle call on. It can be argued this extra mechanism is why our method gets better results and makes for an unfair comparison with SynNet, or that this is a method-agnostic hack to improve GA performance. In Table \ref{tab:sibling} (Right), we show SynNet endowed with a similar mechanism in its crossover operation (generate an offspring pool using the top beams then apply a BO acquisition step on top) didn't improve, but actually downgraded the performance. We hypothesize that SynNet's optimization trajectory is \textit{derailed} by the additional variation to its sibling pool, reducing local movements within the output space that a syntactic editing approach naturally preserves. Thus, we believe the performance gains of this mechanism is \textit{unlocked} by our syntax-driven approach.

\inlinesubsubsection{Extrapolation to Unseen Templates.}\label{sec:ablation-template}

How dependent is our framework on the initial set of syntactic templates? Can SynthesisNet generalize to more templates? Our ablation study in App. \ref{app:extrapolate} evaluate whether the model can extrapolate to new template classes. We investigate: 1) how well SynthesisNet extrapolates to programs whose structural template was unseen during training, 2) how well the framework can incorporate new templates at test time, and 3) how robust overall task performance is to them. To answer these questions, we hold out $\approx 25\%$ of $\hat{\mathcal{T}}$ to be the test set, remove all programs belonging to those templates, retrain, and analyze changes in task performances, both holistic results and results specific to the held-out templates. Our results reveal minor performance drop, and in some instances, \textit{improved} results along some metrics. Our analysis reveals the surrogate model is robust and unsensitive to missing templates, and the framework can also incorporate unseen templates in the online phases of the downstream tasks. We encourage future works to study the exact scaling laws between templates and performance, and we hope our initial findings unlock new directions for scaling up our solutions to achieve greater coverage and impact.

\section{Discussion \& Conclusion}

We reconceptualize synthesis pathway design using a program synthesis approach, introducing a bi-level framework that separates the syntactical skeleton of a synthetic tree from its chemical semantics. Our learning algorithms leverage the tree horizon structure, improving performance on key metrics of analog generation and de novo design. By decoupling syntax and semantics, we effectively navigate a rich design space, integrating design and synthesis into a single workflow that reduces discovery cycle time. Our framework offers control over synthesis resources and biases towards simpler solutions, with the exciting prospect of integration with autonomous synthesis platforms \citep{coley2019}.

\bibliography{iclr2025_conference}

\begin{thebibliography}{77}
\providecommand{\natexlab}[1]{#1}
\providecommand{\url}[1]{\texttt{#1}}
\expandafter\ifx\csname urlstyle\endcsname\relax
  \providecommand{\doi}[1]{doi: #1}\else
  \providecommand{\doi}{doi: \begingroup \urlstyle{rm}\Url}\fi

\bibitem[Alur et~al.(2013)Alur, Bodik, Juniwal, Martin, Raghothaman, Seshia, Singh, Solar-Lezama, Torlak, and Udupa]{alur2013}
Rajeev Alur, Rastislav Bodik, Garvit Juniwal, Milo M.~K. Martin, Mukund Raghothaman, Sanjit~A. Seshia, Rishabh Singh, Armando Solar-Lezama, Emina Torlak, and Abhishek Udupa.
\newblock Syntax-guided synthesis.
\newblock In \emph{2013 Formal Methods in Computer-Aided Design}, pp.\  1--8, 2013.
\newblock \doi{10.1109/FMCAD.2013.6679385}.

\bibitem[Bengio et~al.(2021)Bengio, Jain, Korablyov, Precup, and Bengio]{bengio2021flow}
Emmanuel Bengio, Moksh Jain, Maksym Korablyov, Doina Precup, and Yoshua Bengio.
\newblock Flow network based generative models for non-iterative diverse candidate generation.
\newblock \emph{Advances in Neural Information Processing Systems}, 34:\penalty0 27381--27394, 2021.

\bibitem[Bengio et~al.(2023)Bengio, Lahlou, Deleu, Hu, Tiwari, and Bengio]{bengio2023gflownet}
Yoshua Bengio, Salem Lahlou, Tristan Deleu, Edward~J Hu, Mo~Tiwari, and Emmanuel Bengio.
\newblock Gflownet foundations.
\newblock \emph{The Journal of Machine Learning Research}, 24\penalty0 (1):\penalty0 10006--10060, 2023.

\bibitem[Bradshaw et~al.(2019)Bradshaw, Paige, Kusner, Segler, and Hern{\'a}ndez-Lobato]{bradshaw2019}
John Bradshaw, Brooks Paige, Matt~J Kusner, Marwin Segler, and Jos{\'e}~Miguel Hern{\'a}ndez-Lobato.
\newblock A model to search for synthesizable molecules.
\newblock \emph{Advances in Neural Information Processing Systems}, 32, 2019.

\bibitem[Bradshaw et~al.(2020)Bradshaw, Paige, Kusner, Segler, and Hern{\'a}ndez-Lobato]{bradshaw2020}
John Bradshaw, Brooks Paige, Matt~J Kusner, Marwin Segler, and Jos{\'e}~Miguel Hern{\'a}ndez-Lobato.
\newblock Barking up the right tree: an approach to search over molecule synthesis dags.
\newblock \emph{Advances in neural information processing systems}, 33:\penalty0 6852--6866, 2020.

\bibitem[Brown et~al.(2004)Brown, McKay, Gilardoni, and Gasteiger]{brown2004graph}
Nathan Brown, Ben McKay, Fran{\c{c}}ois Gilardoni, and Johann Gasteiger.
\newblock A graph-based genetic algorithm and its application to the multiobjective evolution of median molecules.
\newblock \emph{Journal of chemical information and computer sciences}, 44\penalty0 (3):\penalty0 1079--1087, 2004.

\bibitem[Bunel et~al.(2018)Bunel, Hausknecht, Devlin, Singh, and Kohli]{bunel2018}
Rudy Bunel, Matthew Hausknecht, Jacob Devlin, Rishabh Singh, and Pushmeet Kohli.
\newblock Leveraging grammar and reinforcement learning for neural program synthesis.
\newblock \emph{arXiv preprint arXiv:1805.04276}, 2018.

\bibitem[Button et~al.(2019)Button, Merk, Hiss, and Schneider]{button2019}
Alexander Button, Daniel Merk, Jan~A Hiss, and Gisbert Schneider.
\newblock Automated de novo molecular design by hybrid machine intelligence and rule-driven chemical synthesis.
\newblock \emph{Nature machine intelligence}, 1\penalty0 (7):\penalty0 307--315, 2019.

\bibitem[Chen et~al.(2020)Chen, Li, Dai, and Song]{chen2020}
Binghong Chen, Chengtao Li, Hanjun Dai, and Le~Song.
\newblock Retro*: learning retrosynthetic planning with neural guided a* search.
\newblock In \emph{Proceedings of the 37th International Conference on Machine Learning}, ICML'20. JMLR.org, 2020.

\bibitem[Chen et~al.(2021)Chen, Lu, Rajeswaran, Lee, Grover, Laskin, Abbeel, Srinivas, and Mordatch]{chen2021decision}
Lili Chen, Kevin Lu, Aravind Rajeswaran, Kimin Lee, Aditya Grover, Misha Laskin, Pieter Abbeel, Aravind Srinivas, and Igor Mordatch.
\newblock Decision transformer: Reinforcement learning via sequence modeling.
\newblock \emph{Advances in neural information processing systems}, 34:\penalty0 15084--15097, 2021.

\bibitem[Chen et~al.(2018)Chen, Liu, and Song]{chen2018}
Xinyun Chen, Chang Liu, and Dawn Song.
\newblock Execution-guided neural program synthesis.
\newblock In \emph{International Conference on Learning Representations}, 2018.

\bibitem[Coley et~al.(2019)Coley, Thomas, Lummiss, Jaworski, Breen, Schultz, Hart, Fishman, Rogers, Gao, Hicklin, Plehiers, Byington, Piotti, Green, Hart, Jamison, and Jensen]{coley2019}
Connor~W. Coley, Dale~A. Thomas, Justin A.~M. Lummiss, Jonathan~N. Jaworski, Christopher~P. Breen, Victor Schultz, Travis Hart, Joshua~S. Fishman, Luke Rogers, Hanyu Gao, Robert~W. Hicklin, Pieter~P. Plehiers, Joshua Byington, John~S. Piotti, William~H. Green, A.~John Hart, Timothy~F. Jamison, and Klavs~F. Jensen.
\newblock A robotic platform for flow synthesis of organic compounds informed by ai planning.
\newblock \emph{Science}, 365\penalty0 (6453):\penalty0 eaax1566, 2019.
\newblock \doi{10.1126/science.aax1566}.
\newblock URL \url{https://www.science.org/doi/abs/10.1126/science.aax1566}.

\bibitem[Coley et~al.(2020{\natexlab{a}})Coley, Eyke, and Jensen]{coley2020a}
Connor~W Coley, Natalie~S Eyke, and Klavs~F Jensen.
\newblock Autonomous discovery in the chemical sciences part i: Progress.
\newblock \emph{Angewandte Chemie International Edition}, 59\penalty0 (51):\penalty0 22858--22893, 2020{\natexlab{a}}.

\bibitem[Coley et~al.(2020{\natexlab{b}})Coley, Eyke, and Jensen]{coley2020b}
Connor~W Coley, Natalie~S Eyke, and Klavs~F Jensen.
\newblock Autonomous discovery in the chemical sciences part ii: outlook.
\newblock \emph{Angewandte Chemie International Edition}, 59\penalty0 (52):\penalty0 23414--23436, 2020{\natexlab{b}}.

\bibitem[Corey et~al.(1985)Corey, Long, and Rubenstein]{corey1985}
Elias~James Corey, Alan~K Long, and Stewart~D Rubenstein.
\newblock Computer-assisted analysis in organic synthesis.
\newblock \emph{Science}, 228\penalty0 (4698):\penalty0 408--418, 1985.

\bibitem[De~Cao \& Kipf(2018)De~Cao and Kipf]{cao2018}
Nicola De~Cao and Thomas Kipf.
\newblock Molgan: An implicit generative model for small molecular graphs.
\newblock \emph{arXiv preprint arXiv:1805.11973}, 2018.

\bibitem[Dolfus et~al.(2022)Dolfus, Briem, and Rarey]{dolfus2022}
Uschi Dolfus, Hans Briem, and Matthias Rarey.
\newblock Synthesis-aware generation of structural analogues.
\newblock \emph{Journal of Chemical Information and Modeling}, 62\penalty0 (15):\penalty0 3565--3576, 2022.

\bibitem[Ellis et~al.(2019)Ellis, Nye, Pu, Sosa, Tenenbaum, and Solar-Lezama]{ellis2019}
Kevin Ellis, Maxwell Nye, Yewen Pu, Felix Sosa, Josh Tenenbaum, and Armando Solar-Lezama.
\newblock Write, execute, assess: Program synthesis with a repl.
\newblock \emph{Advances in Neural Information Processing Systems}, 32, 2019.

\bibitem[Ertl \& Schuffenhauer(2009)Ertl and Schuffenhauer]{ertl2009}
Peter Ertl and Ansgar Schuffenhauer.
\newblock Estimation of synthetic accessibility score of drug-like molecules based on molecular complexity and fragment contributions.
\newblock \emph{Journal of cheminformatics}, 1:\penalty0 1--11, 2009.

\bibitem[Flynn(2014)]{flynn2014}
Alison~B Flynn.
\newblock How do students work through organic synthesis learning activities?
\newblock \emph{Chemistry Education Research and Practice}, 15\penalty0 (4):\penalty0 747--762, 2014.

\bibitem[Gao \& Coley(2020)Gao and Coley]{gao2020}
Wenhao Gao and Connor~W Coley.
\newblock The synthesizability of molecules proposed by generative models.
\newblock \emph{Journal of chemical information and modeling}, 60\penalty0 (12):\penalty0 5714--5723, 2020.

\bibitem[Gao et~al.(2021)Gao, Mercado, and Coley]{gao2021}
Wenhao Gao, Roc{\'\i}o Mercado, and Connor~W Coley.
\newblock Amortized tree generation for bottom-up synthesis planning and synthesizable molecular design.
\newblock \emph{arXiv preprint arXiv:2110.06389}, 2021.

\bibitem[Gao et~al.(2022)Gao, Fu, Sun, and Coley]{gao2022}
Wenhao Gao, Tianfan Fu, Jimeng Sun, and Connor Coley.
\newblock Sample efficiency matters: a benchmark for practical molecular optimization.
\newblock \emph{Advances in neural information processing systems}, 35:\penalty0 21342--21357, 2022.

\bibitem[Gao et~al.(2024)Gao, Luo, and Coley]{gao2024generative}
Wenhao Gao, Shitong Luo, and Connor~W Coley.
\newblock Generative artificial intelligence for navigating synthesizable chemical space.
\newblock \emph{arXiv preprint arXiv:2410.03494}, 2024.

\bibitem[Ghahremanpour et~al.(2020)Ghahremanpour, Tirado-Rives, Deshmukh, Ippolito, Zhang, Cabeza~de Vaca, Liosi, Anderson, and Jorgensen]{ghahremanpour2020}
Mohammad~M Ghahremanpour, Julian Tirado-Rives, Maya Deshmukh, Joseph~A Ippolito, Chun-Hui Zhang, Israel Cabeza~de Vaca, Maria-Elena Liosi, Karen~S Anderson, and William~L Jorgensen.
\newblock Identification of 14 known drugs as inhibitors of the main protease of sars-cov-2.
\newblock \emph{ACS medicinal chemistry letters}, 11\penalty0 (12):\penalty0 2526--2533, 2020.

\bibitem[Gilks et~al.(1995)Gilks, Best, and Tan]{gilks1995}
Wally~R Gilks, Nicky~G Best, and Keith~KC Tan.
\newblock Adaptive rejection metropolis sampling within gibbs sampling.
\newblock \emph{Journal of the Royal Statistical Society Series C: Applied Statistics}, 44\penalty0 (4):\penalty0 455--472, 1995.

\bibitem[Guo et~al.(2024)Guo, Yu, Li, Zhang, Wang, Li, and Dong]{guo2024retrosynthesis}
Jiasheng Guo, Chenning Yu, Kenan Li, Yijian Zhang, Guoqiang Wang, Shuhua Li, and Hao Dong.
\newblock Retrosynthesis zero: Self-improving global synthesis planning using reinforcement learning.
\newblock \emph{Journal of Chemical Theory and Computation}, 2024.

\bibitem[Guo et~al.(2022)Guo, Thost, Li, Das, Chen, and Matusik]{guo2022}
Minghao Guo, Veronika Thost, Beichen Li, Payel Das, Jie Chen, and Wojciech Matusik.
\newblock Data-efficient graph grammar learning for molecular generation.
\newblock \emph{arXiv preprint arXiv:2203.08031}, 2022.

\bibitem[Hachmann et~al.(2011)Hachmann, Olivares-Amaya, Atahan-Evrenk, Amador-Bedolla, S{\'a}nchez-Carrera, Gold-Parker, Vogt, Brockway, and Aspuru-Guzik]{hachmann2011}
Johannes Hachmann, Roberto Olivares-Amaya, Sule Atahan-Evrenk, Carlos Amador-Bedolla, Roel~S S{\'a}nchez-Carrera, Aryeh Gold-Parker, Leslie Vogt, Anna~M Brockway, and Al{\'a}n Aspuru-Guzik.
\newblock The harvard clean energy project: large-scale computational screening and design of organic photovoltaics on the world community grid.
\newblock \emph{The Journal of Physical Chemistry Letters}, 2\penalty0 (17):\penalty0 2241--2251, 2011.

\bibitem[Hartenfeller et~al.(2011)Hartenfeller, Eberle, Meier, Nieto-Oberhuber, Altmann, Schneider, Jacoby, and Renner]{hartenfeller2011}
Markus Hartenfeller, Martin Eberle, Peter Meier, Cristina Nieto-Oberhuber, Karl-Heinz Altmann, Gisbert Schneider, Edgar Jacoby, and Steffen Renner.
\newblock A collection of robust organic synthesis reactions for in silico molecule design.
\newblock \emph{Journal of chemical information and modeling}, 51\penalty0 (12):\penalty0 3093--3098, 2011.

\bibitem[Hastings(1970)]{hastings1970}
W~Keith Hastings.
\newblock Monte carlo sampling methods using markov chains and their applications.
\newblock 1970.

\bibitem[Huang et~al.(2022)Huang, Fu, Gao, Zhao, Roohani, Leskovec, Coley, Xiao, Sun, and Zitnik]{huang2022}
Kexin Huang, Tianfan Fu, Wenhao Gao, Yue Zhao, Yusuf Roohani, Jure Leskovec, Connor~W Coley, Cao Xiao, Jimeng Sun, and Marinka Zitnik.
\newblock Artificial intelligence foundation for therapeutic science.
\newblock \emph{Nature chemical biology}, 18\penalty0 (10):\penalty0 1033--1036, 2022.

\bibitem[Janet et~al.(2020)Janet, Ramesh, Duan, and Kulik]{janet2020}
Jon~Paul Janet, Sahasrajit Ramesh, Chenru Duan, and Heather~J Kulik.
\newblock Accurate multiobjective design in a space of millions of transition metal complexes with neural-network-driven efficient global optimization.
\newblock \emph{ACS central science}, 6\penalty0 (4):\penalty0 513--524, 2020.

\bibitem[Jin et~al.(2018)Jin, Barzilay, and Jaakkola]{jin2018}
Wengong Jin, Regina Barzilay, and Tommi Jaakkola.
\newblock Junction tree variational autoencoder for molecular graph generation.
\newblock In \emph{International conference on machine learning}, pp.\  2323--2332. PMLR, 2018.

\bibitem[Jin et~al.(2020)Jin, Barzilay, and Jaakkola]{jin2020}
Wengong Jin, Regina Barzilay, and Tommi Jaakkola.
\newblock Hierarchical generation of molecular graphs using structural motifs.
\newblock In \emph{International conference on machine learning}, pp.\  4839--4848. PMLR, 2020.

\bibitem[Kim et~al.(2021)Kim, Ahn, Lee, and Shin]{kim2021self}
Junsu Kim, Sungsoo Ahn, Hankook Lee, and Jinwoo Shin.
\newblock Self-improved retrosynthetic planning.
\newblock In \emph{International Conference on Machine Learning}, pp.\  5486--5495. PMLR, 2021.

\bibitem[Kishimoto et~al.(2019)Kishimoto, Buesser, Chen, and Botea]{kishimoto2019}
Akihiro Kishimoto, Beat Buesser, Bei Chen, and Adi Botea.
\newblock Depth-first proof-number search with heuristic edge cost and application to chemical synthesis planning.
\newblock \emph{Advances in Neural Information Processing Systems}, 32, 2019.

\bibitem[Koch et~al.(2015)Koch, Gehringer, and Laufer]{koch2015inhibitors}
Pierre Koch, Matthias Gehringer, and Stefan~A Laufer.
\newblock Inhibitors of c-jun n-terminal kinases: an update.
\newblock \emph{Journal of medicinal chemistry}, 58\penalty0 (1):\penalty0 72--95, 2015.

\bibitem[Koscher et~al.(2023)Koscher, Canty, McDonald, Greenman, McGill, Bilodeau, Jin, Wu, Vermeire, Jin, et~al.]{koscher2023}
Brent~A Koscher, Richard~B Canty, Matthew~A McDonald, Kevin~P Greenman, Charles~J McGill, Camille~L Bilodeau, Wengong Jin, Haoyang Wu, Florence~H Vermeire, Brooke Jin, et~al.
\newblock Autonomous, multiproperty-driven molecular discovery: From predictions to measurements and back.
\newblock \emph{Science}, 382\penalty0 (6677):\penalty0 eadi1407, 2023.

\bibitem[Levin et~al.(2023)Levin, Fortunato, Tan, and Coley]{levin2023}
Itai Levin, Michael~E Fortunato, Kian~L Tan, and Connor~W Coley.
\newblock Computer-aided evaluation and exploration of chemical spaces constrained by reaction pathways.
\newblock \emph{AIChE journal}, 69\penalty0 (12):\penalty0 e18234, 2023.

\bibitem[Li et~al.(2018)Li, Vinyals, Dyer, Pascanu, and Battaglia]{li2018}
Yujia Li, Oriol Vinyals, Chris Dyer, Razvan Pascanu, and Peter Battaglia.
\newblock Learning deep generative models of graphs.
\newblock \emph{arXiv preprint arXiv:1803.03324}, 2018.

\bibitem[Liu et~al.(2023)Liu, Xue, Xie, Xia, Tripp, Maziarz, Segler, Qin, Zhang, and Liu]{liu2023retrosynthetic}
Guoqing Liu, Di~Xue, Shufang Xie, Yingce Xia, Austin Tripp, Krzysztof Maziarz, Marwin Segler, Tao Qin, Zongzhang Zhang, and Tie-Yan Liu.
\newblock Retrosynthetic planning with dual value networks.
\newblock In \emph{International Conference on Machine Learning}, pp.\  22266--22276. PMLR, 2023.

\bibitem[Liu et~al.(2018)Liu, Allamanis, Brockschmidt, and Gaunt]{liu2018}
Qi~Liu, Miltiadis Allamanis, Marc Brockschmidt, and Alexander Gaunt.
\newblock Constrained graph variational autoencoders for molecule design.
\newblock \emph{Advances in neural information processing systems}, 31, 2018.

\bibitem[Luo et~al.(2024)Luo, Gao, Wu, Peng, Coley, and Ma]{luo2024projecting}
Shitong Luo, Wenhao Gao, Zuofan Wu, Jian Peng, Connor~W Coley, and Jianzhu Ma.
\newblock Projecting molecules into synthesizable chemical spaces.
\newblock \emph{arXiv preprint arXiv:2406.04628}, 2024.

\bibitem[Lyu et~al.(2019)Lyu, Wang, Balius, Singh, Levit, Moroz, O’Meara, Che, Algaa, Tolmachova, et~al.]{lyu2019}
Jiankun Lyu, Sheng Wang, Trent~E Balius, Isha Singh, Anat Levit, Yurii~S Moroz, Matthew~J O’Meara, Tao Che, Enkhjargal Algaa, Kateryna Tolmachova, et~al.
\newblock Ultra-large library docking for discovering new chemotypes.
\newblock \emph{Nature}, 566\penalty0 (7743):\penalty0 224--229, 2019.

\bibitem[Ma et~al.(2018)Ma, Chen, and Xiao]{ma2018}
Tengfei Ma, Jie Chen, and Cao Xiao.
\newblock Constrained generation of semantically valid graphs via regularizing variational autoencoders.
\newblock \emph{Advances in Neural Information Processing Systems}, 31, 2018.

\bibitem[Madras(2013)]{madras2013history}
Bertha~K Madras.
\newblock History of the discovery of the antipsychotic dopamine d2 receptor: a basis for the dopamine hypothesis of schizophrenia.
\newblock \emph{Journal of the History of the Neurosciences}, 22\penalty0 (1):\penalty0 62--78, 2013.

\bibitem[McCorkindale(2023)]{mccorkindale2023}
William McCorkindale.
\newblock \emph{Accelerating the Design-Make-Test cycle of Drug Discovery with Machine Learning}.
\newblock PhD thesis, 2023.

\bibitem[Merrell(2023)]{merrell2023}
Paul Merrell.
\newblock Example-based procedural modeling using graph grammars.
\newblock \emph{ACM Transactions on Graphics (TOG)}, 42\penalty0 (4):\penalty0 1--16, 2023.

\bibitem[Merrell \& Manocha(2010)Merrell and Manocha]{merrell2010}
Paul Merrell and Dinesh Manocha.
\newblock Model synthesis: A general procedural modeling algorithm.
\newblock \emph{IEEE transactions on visualization and computer graphics}, 17\penalty0 (6):\penalty0 715--728, 2010.

\bibitem[Merrell et~al.(2011)Merrell, Schkufza, Li, Agrawala, and Koltun]{merrell2011}
Paul Merrell, Eric Schkufza, Zeyang Li, Maneesh Agrawala, and Vladlen Koltun.
\newblock Interactive furniture layout using interior design guidelines.
\newblock In \emph{ACM SIGGRAPH 2011 Papers}, SIGGRAPH '11, New York, NY, USA, 2011. Association for Computing Machinery.
\newblock ISBN 9781450309431.
\newblock \doi{10.1145/1964921.1964982}.
\newblock URL \url{https://doi.org/10.1145/1964921.1964982}.

\bibitem[Metropolis et~al.(1953)Metropolis, Rosenbluth, Rosenbluth, Teller, and Teller]{metropolis1953}
Nicholas Metropolis, Arianna~W Rosenbluth, Marshall~N Rosenbluth, Augusta~H Teller, and Edward Teller.
\newblock Equation of state calculations by fast computing machines.
\newblock \emph{The journal of chemical physics}, 21\penalty0 (6):\penalty0 1087--1092, 1953.

\bibitem[M{\"u}ller et~al.(2006)M{\"u}ller, Wonka, Haegler, Ulmer, and Van~Gool]{muller2006}
Pascal M{\"u}ller, Peter Wonka, Simon Haegler, Andreas Ulmer, and Luc Van~Gool.
\newblock Procedural modeling of buildings.
\newblock In \emph{ACM SIGGRAPH 2006 Papers}, pp.\  614--623. 2006.

\bibitem[Polikarpova et~al.(2016)Polikarpova, Kuraj, and Solar-Lezama]{polikarpova2016}
Nadia Polikarpova, Ivan Kuraj, and Armando Solar-Lezama.
\newblock Program synthesis from polymorphic refinement types.
\newblock \emph{ACM SIGPLAN Notices}, 51\penalty0 (6):\penalty0 522--538, 2016.

\bibitem[Samanta et~al.(2020)Samanta, De, Jana, G{\'o}mez, Chattaraj, Ganguly, and Gomez-Rodriguez]{samanta2020}
Bidisha Samanta, Abir De, Gourhari Jana, Vicen{\c{c}} G{\'o}mez, Pratim Chattaraj, Niloy Ganguly, and Manuel Gomez-Rodriguez.
\newblock Nevae: A deep generative model for molecular graphs.
\newblock \emph{Journal of machine learning research}, 21\penalty0 (114):\penalty0 1--33, 2020.

\bibitem[Sanchez-Lengeling \& Aspuru-Guzik(2018)Sanchez-Lengeling and Aspuru-Guzik]{lengeling2018}
Benjamin Sanchez-Lengeling and Alán Aspuru-Guzik.
\newblock Inverse molecular design using machine learning: Generative models for matter engineering.
\newblock \emph{Science}, 361\penalty0 (6400):\penalty0 360--365, 2018.
\newblock \doi{10.1126/science.aat2663}.
\newblock URL \url{https://www.science.org/doi/abs/10.1126/science.aat2663}.

\bibitem[Schkufza et~al.(2013)Schkufza, Sharma, and Aiken]{schkufza2013}
Eric Schkufza, Rahul Sharma, and Alex Aiken.
\newblock Stochastic superoptimization.
\newblock \emph{SIGPLAN Not.}, 48\penalty0 (4):\penalty0 305–316, mar 2013.
\newblock ISSN 0362-1340.
\newblock \doi{10.1145/2499368.2451150}.
\newblock URL \url{https://doi.org/10.1145/2499368.2451150}.

\bibitem[Segler \& Waller(2017)Segler and Waller]{segler2017}
Marwin~HS Segler and Mark~P Waller.
\newblock Neural-symbolic machine learning for retrosynthesis and reaction prediction.
\newblock \emph{Chemistry--A European Journal}, 23\penalty0 (25):\penalty0 5966--5971, 2017.

\bibitem[Shi et~al.(2020)Shi, Huang, Feng, Zhong, Wang, and Sun]{shi2020}
Yunsheng Shi, Zhengjie Huang, Shikun Feng, Hui Zhong, Wenjin Wang, and Yu~Sun.
\newblock Masked label prediction: Unified message passing model for semi-supervised classification.
\newblock \emph{arXiv preprint arXiv:2009.03509}, 2020.

\bibitem[Silver et~al.(2017)Silver, Schrittwieser, Simonyan, Antonoglou, Huang, Guez, Hubert, Baker, Lai, Bolton, et~al.]{silver2017}
David Silver, Julian Schrittwieser, Karen Simonyan, Ioannis Antonoglou, Aja Huang, Arthur Guez, Thomas Hubert, Lucas Baker, Matthew Lai, Adrian Bolton, et~al.
\newblock Mastering the game of go without human knowledge.
\newblock \emph{nature}, 550\penalty0 (7676):\penalty0 354--359, 2017.

\bibitem[Simonovsky \& Komodakis(2018)Simonovsky and Komodakis]{simonovsky2018}
Martin Simonovsky and Nikos Komodakis.
\newblock Graphvae: Towards generation of small graphs using variational autoencoders.
\newblock In \emph{Artificial Neural Networks and Machine Learning--ICANN 2018: 27th International Conference on Artificial Neural Networks, Rhodes, Greece, October 4-7, 2018, Proceedings, Part I 27}, pp.\  412--422. Springer, 2018.

\bibitem[Smith(1997)]{smith1997}
Warren~D Smith.
\newblock Computational complexity of synthetic chemistry--basic facts.
\newblock Technical report, Citeseer, 1997.

\bibitem[Solar-Lezama et~al.(2005)Solar-Lezama, Rabbah, Bod{\'\i}k, and Ebcio{\u{g}}lu]{solar2005}
Armando Solar-Lezama, Rodric Rabbah, Rastislav Bod{\'\i}k, and Kemal Ebcio{\u{g}}lu.
\newblock Programming by sketching for bit-streaming programs.
\newblock In \emph{Proceedings of the 2005 ACM SIGPLAN conference on Programming language design and implementation}, pp.\  281--294, 2005.

\bibitem[Sun et~al.(2017)Sun, Jeliazkova, Chupakhin, Golib-Dzib, Engkvist, Carlsson, Wegner, Ceulemans, Georgiev, Jeliazkov, et~al.]{sun2017excape}
Jiangming Sun, Nina Jeliazkova, Vladimir Chupakhin, Jose-Felipe Golib-Dzib, Ola Engkvist, Lars Carlsson, J{\"o}rg Wegner, Hugo Ceulemans, Ivan Georgiev, Vedrin Jeliazkov, et~al.
\newblock Excape-db: an integrated large scale dataset facilitating big data analysis in chemogenomics.
\newblock \emph{Journal of cheminformatics}, 9:\penalty0 1--9, 2017.

\bibitem[Sun et~al.(2024)Sun, Guo, Yuan, Thost, Owens, Grosz, Selvan, Zhou, Mohiuddin, Pedretti, et~al.]{sun2024}
Michael Sun, Minghao Guo, Weize Yuan, Veronika Thost, Crystal~Elaine Owens, Aristotle~Franklin Grosz, Sharvaa Selvan, Katelyn Zhou, Hassan Mohiuddin, Benjamin~J Pedretti, et~al.
\newblock Representing molecules as random walks over interpretable grammars.
\newblock \emph{arXiv preprint arXiv:2403.08147}, 2024.

\bibitem[Swanson et~al.(2024)Swanson, Liu, Catacutan, Arnold, Zou, and Stokes]{swanson2024}
Kyle Swanson, Gary Liu, Denise~B Catacutan, Autumn Arnold, James Zou, and Jonathan~M Stokes.
\newblock Generative ai for designing and validating easily synthesizable and structurally novel antibiotics.
\newblock \emph{Nature Machine Intelligence}, 6\penalty0 (3):\penalty0 338--353, 2024.

\bibitem[Talton et~al.(2011)Talton, Lou, Lesser, Duke, Mech, and Koltun]{talton2011}
Jerry~O Talton, Yu~Lou, Steve Lesser, Jared Duke, Radom{\'\i}r Mech, and Vladlen Koltun.
\newblock Metropolis procedural modeling.
\newblock \emph{ACM Trans. Graph.}, 30\penalty0 (2):\penalty0 11--1, 2011.

\bibitem[Torren-Peraire et~al.(2024)Torren-Peraire, Hassen, Genheden, Verhoeven, Clevert, Preuss, and Tetko]{torren2024}
Paula Torren-Peraire, Alan~Kai Hassen, Samuel Genheden, Jonas Verhoeven, Djork-Arn{\'e} Clevert, Mike Preuss, and Igor~V Tetko.
\newblock Models matter: the impact of single-step retrosynthesis on synthesis planning.
\newblock \emph{Digital Discovery}, 3\penalty0 (3):\penalty0 558--572, 2024.

\bibitem[Tu et~al.(2022)Tu, Shorewala, Ma, and Thost]{tu2022}
Hongyu Tu, Shantam Shorewala, Tengfei Ma, and Veronika Thost.
\newblock Retrosynthesis prediction revisited.
\newblock In \emph{NeurIPS 2022 AI for Science: Progress and Promises}, 2022.

\bibitem[Vinkers et~al.(2003)Vinkers, de~Jonge, Daeyaert, Heeres, Koymans, van Lenthe, Lewi, Timmerman, Van~Aken, and Janssen]{vinkers2003}
H~Maarten Vinkers, Marc~R de~Jonge, Frederik~FD Daeyaert, Jan Heeres, Lucien~MH Koymans, Joop~H van Lenthe, Paul~J Lewi, Henk Timmerman, Koen Van~Aken, and Paul~AJ Janssen.
\newblock Synopsis: synthesize and optimize system in silico.
\newblock \emph{Journal of medicinal chemistry}, 46\penalty0 (13):\penalty0 2765--2773, 2003.

\bibitem[Volkamer et~al.(2023)Volkamer, Riniker, Nittinger, Lanini, Grisoni, Evertsson, Rodr{\'\i}guez-P{\'e}rez, and Schneider]{volkamer2023}
Andrea Volkamer, Sereina Riniker, Eva Nittinger, Jessica Lanini, Francesca Grisoni, Emma Evertsson, Raquel Rodr{\'\i}guez-P{\'e}rez, and Nadine Schneider.
\newblock Machine learning for small molecule drug discovery in academia and industry.
\newblock \emph{Artificial Intelligence in the Life Sciences}, 3:\penalty0 100056, 2023.

\bibitem[Yao et~al.(2021)Yao, S{\'a}nchez-Lengeling, Bobbitt, Bucior, Kumar, Collins, Burns, Woo, Farha, Snurr, et~al.]{yao2021}
Zhenpeng Yao, Benjam{\'\i}n S{\'a}nchez-Lengeling, N~Scott Bobbitt, Benjamin~J Bucior, Sai Govind~Hari Kumar, Sean~P Collins, Thomas Burns, Tom~K Woo, Omar~K Farha, Randall~Q Snurr, et~al.
\newblock Inverse design of nanoporous crystalline reticular materials with deep generative models.
\newblock \emph{Nature Machine Intelligence}, 3\penalty0 (1):\penalty0 76--86, 2021.

\bibitem[You et~al.(2018)You, Liu, Ying, Pande, and Leskovec]{you2018}
Jiaxuan You, Bowen Liu, Zhitao Ying, Vijay Pande, and Jure Leskovec.
\newblock Graph convolutional policy network for goal-directed molecular graph generation.
\newblock \emph{Advances in neural information processing systems}, 31, 2018.

\bibitem[Yu et~al.(2011)Yu, Yeung, Tang, Terzopoulos, Chan, and Osher]{yu2011}
Lap-Fai Yu, Sai-Kit Yeung, Chi-Keung Tang, Demetri Terzopoulos, Tony~F. Chan, and Stanley~J. Osher.
\newblock Make it home: automatic optimization of furniture arrangement.
\newblock \emph{ACM Trans. Graph.}, 30\penalty0 (4), jul 2011.
\newblock ISSN 0730-0301.
\newblock \doi{10.1145/2010324.1964981}.
\newblock URL \url{https://doi.org/10.1145/2010324.1964981}.

\bibitem[Zhang et~al.(2021)Zhang, Stone, Deshmukh, Ippolito, Ghahremanpour, Tirado-Rives, Spasov, Zhang, Takeo, Kudalkar, et~al.]{zhang2021}
Chun-Hui Zhang, Elizabeth~A Stone, Maya Deshmukh, Joseph~A Ippolito, Mohammad~M Ghahremanpour, Julian Tirado-Rives, Krasimir~A Spasov, Shuo Zhang, Yuka Takeo, Shalley~N Kudalkar, et~al.
\newblock Potent noncovalent inhibitors of the main protease of sars-cov-2 from molecular sculpting of the drug perampanel guided by free energy perturbation calculations.
\newblock \emph{ACS central science}, 7\penalty0 (3):\penalty0 467--475, 2021.

\bibitem[Zhavoronkov et~al.(2019)Zhavoronkov, Ivanenkov, Aliper, Veselov, Aladinskiy, Aladinskaya, Terentiev, Polykovskiy, Kuznetsov, Asadulaev, et~al.]{zhavoronkov2019}
Alex Zhavoronkov, Yan~A Ivanenkov, Alex Aliper, Mark~S Veselov, Vladimir~A Aladinskiy, Anastasiya~V Aladinskaya, Victor~A Terentiev, Daniil~A Polykovskiy, Maksim~D Kuznetsov, Arip Asadulaev, et~al.
\newblock Deep learning enables rapid identification of potent ddr1 kinase inhibitors.
\newblock \emph{Nature biotechnology}, 37\penalty0 (9):\penalty0 1038--1040, 2019.

\bibitem[Zimmerman et~al.(2020)Zimmerman, Anastas, Erythropel, and Leitner]{zimmerman2020}
Julie~B Zimmerman, Paul~T Anastas, Hanno~C Erythropel, and Walter Leitner.
\newblock Designing for a green chemistry future.
\newblock \emph{Science}, 367\penalty0 (6476):\penalty0 397--400, 2020.

\end{thebibliography}
\bibliographystyle{iclr2025_conference}

\newpage
\appendix
\section{Syntactic Templates}\label{app:eda}
\begin{figure}[h!]
       \centering
     \includegraphics[width=0.8\textwidth]{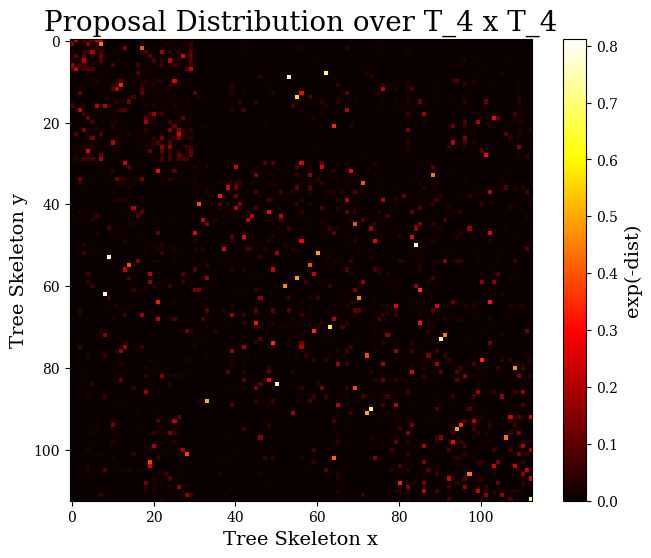}
     \caption{We adopt the tree edit distance as the $\text{dist}$ function. We see that $\hat{\mathcal{T}}_4$ has sufficient transition coverage for bootstrapping our space of syntactic templates.}
     \label{fig:heatmap}    
\end{figure}

Syntactic templates form the essential ingredients for syntax-guided synthesis, as they significantly reduce the number of possible programs. In practice, syntactical templates are provided by users who operate with real-world constraints or experts who can help narrow the search space to \textit{desirable} templates. The exact criteria for selecting templates are problem-dependent. To prove our concept in a more generalizable workflow, we bootstrap our set of syntactic templates $\hat{\mathcal{T}}$ in a data-driven way by obtaining the syntactic templates present in the training set. We then simulate real-world constraints by setting $\hat{\mathcal{T}_k} \leftarrow \{T \in \hat{\mathcal{T}} \mid \text{$T$ has at most $k$ internal nodes}\}$ and optimize within the induced design space $\hat{\partial \gP}_k$ (Section \ref{sec:inner}). We tabulate summary statistics in Table \ref{tab:summary} for the number of unique syntactic templates and the number of topological orders. We see that the empirical distribution is biased towards \textit{simpler} syntactic templates, which reflects real-world constraints and is a key enabler of our amortized approach. We train the parameters $(\Theta,\Phi,\Omega)_k$ of our policies $(\tau, \pi_{\gR}, \pi_{\gB})$, respectively, for $k=3,4,5,6$ on our (pre)training dataset $\gD$. For samples in $\gD$ with more than $k>6$ reactions, we snap it to the closest $T \in \hat{\mathcal{T}_k}$ according to the tree edit distance. We find $k=3$ using full topological decoding (illustration in Figure \ref{fig:fig3}) is best for Synthesizable Analog Generation and $k=4$ with random sampling of the decoding beams is a good compromise between accuracy and efficiency for Synthesizable Molecule Design. We also note that the number of unique templates grows sub-exponentially, and in fact the number of templates for a fixed number of reactions starts diminishing for $k>6$. To make sure this does not cause issues, we ensured there is still sufficient coverage to formulate a Markov Chain on $\hat{\mathcal{T}}_k$, which is crucial for our bilevel algorithms. For example, Figure \ref{fig:heatmap} visualizes the empirical proposal distribution $J(T_1,T_2), \forall T_1, T_2 \in \hat{\mathcal{T}}_4 \times \hat{\mathcal{T}}_4$. Importantly, key hyperparameters like $\beta$ and $n_{\text{edits}}$ enable control over exploration vs exploitation.

\begin{figure}[htbp]%
  \centering
  \subfloat[][]{
\begin{tabular}{lll}
\toprule
\# Rxns & \multicolumn{1}{l}{$|\hat{\mathcal{T}}_{k\setminus k-1}|$ {($|\mathcal{T}_{k\setminus {k-1}}|$)}} & \# Topo. Orders (Max, Mean, Std)$_{\hat{\mathcal{T}}_{k\setminus {k-1}}}$ \\ \midrule
1                    & \multicolumn{1}{l}{2 {(2)}}             & 2, 1.5, 0.5                             \\
2                    & \multicolumn{1}{l}{6 {(6)}}             & 8, 4.17, 2.79                           \\
3                    & \multicolumn{1}{l}{22 {(22)}}            & 80, 19.59, 20.55                        \\
4                    & \multicolumn{1}{l}{83 {(90)}}           & 896, 152.02, 215.53                     \\
5                    & \multicolumn{1}{l}{209 {(394)}}           & 19200, 2506.25, 3705.77                 \\ \bottomrule
                                   
\end{tabular}
  }%
  \qquad
  \subfloat[][]{
  \begin{tabular}{ll}
  \toprule
\# Rxns & $|\hat{\mathcal{T}}_{k\setminus k-1}|$ \\ \midrule
6                    & 298           \\
7                    & 243           \\
8                    & 112           \\
9                    & 63          \\
10                   & 42          \\
11                   & 22          \\
12                   & 11          \\
13                   & 4          \\
14                   & 2         \\ \bottomrule
\end{tabular}
  }
  \qquad
  \subfloat[][]{
  
  \begin{tabular}{lll}
  \toprule
{\# Rxns}  & {\# Topo. Masks ()$_{\hat{\mathcal{T}}_{k\setminus k-1}}$} & {\# Topo. Masks ()$_{{\mathcal{T}}_{k\setminus k-1}}$}  \\ \midrule
{1}                           &{  5, 4, 1} &{ 5, 4, 1}  \\
{2}                           &{  11, 7.67, 2.56}  &{ 11, 7.67, 2.56} \\
{3}                           &{  26, 14.36, 5.86} &{ 26, 14.36, 5.86} \\
{4}                        &{  56, 27.99, 12.47} &{ 56, 26.73, 12.78} \\
{5}                     &{  131, 65.07, 26.36} &{ 131, 49.74, 27.09}\\
{6}                     &{   287, 165.12, 61.43} &{ 287, 92.67, 56.29} \\ \bottomrule
\end{tabular}
  }
  
  \caption{(a) Summary statistics of the number of syntactic templates {(both empirical and theoretically possible)} and possible topological decoding node orders for $k=1,2,\ldots,5$; (b) Summary statistics for only the number of syntactic templates since enumerating all topological sorts becomes intractable; {(c) Summary statistics for the number of topological masks (subset of nodes closed under parent(.))}}  
  \label{tab:summary}%
\end{figure}

{
\section{Extrapolation to Unseen Templates}\label{app:extrapolate}

\subsection{Data Preparation}
In this section, we investigate whether SynthesisNet can extrapolate to unseen templates, and effectively incorporate them for synthesizable analog generation and molecular design. We setup an ablation study as follows:
\begin{enumerate}
    \item Reverse sort the templates by frequency using our dataset $\mathcal{D}_0$. 
    \item Collect every \textit{fourth} template into a hold-out set $\mathcal{T}_{\text{test}}$ for $k=4$.
    \item Construct $\mathcal{D}_0 ' := \{(P,B) \in \mathcal{D}_0 \mid  T_{P,B} \in \mathcal{T}_4\setminus \mathcal{T}_{\text{test}}\}$ where $T_{P,B}\in \mathcal{T}$ is the syntactical template of $(P, B)$.
    \item Run Algo. \ref{alg:pretrain} on $\mathcal{D}_0 '$ to obtain $\mathcal{D}'$.
    \item Train ablation policy networks $\{\pi_{\mathcal{B}}',\pi_{\mathcal{R}}'\}$ using $\mathcal{D}'$.
    \item Evaluate task performances with same $\tau$ as before.
\end{enumerate}

We select hold-out templates in a frequency-stratified manner, ensuring the frequency distribution of $\hat{\mathcal{T}}_{\text{test}}$ is similar to that of $\hat{\mathcal{T}}$. Since smaller templates appear more frequently, the sizes of of templates are also indirectly stratified this way. Since we choose the least frequent from each consecutive group of 4 (Step 2), we note on average templates in $\hat{\mathcal{T}}_{\text{test}}$ tend to be slightly larger than $\hat{\mathcal{T}}$, so results for test templates may be lower in Table \ref{tab:ep-analog}.

\subsection{Results on Synthesizable Analog Generation}
\begin{table}[!htb]
\caption{Apart from swapping out the policy networks, we use the same experimental setup as Table \ref{tab:1}. For fair comparison, we also retrained Ours with $k=4$ templates (whereas Table \ref{tab:1} used $k=3$).}
\label{tab:ep-analog}
\centering
\small

\resizebox{\textwidth}{!}{

\begin{tabular}{@{}llccccccccc@{}}
\toprule
\multicolumn{1}{c}{}        & \multicolumn{1}{c}{}                                &               & \multicolumn{3}{c}{Avg. Sim. $\uparrow$} & \multicolumn{3}{c}{SA $\downarrow$} & \multicolumn{2}{c}{Diversity $\uparrow$} \\
\multicolumn{1}{c}{Dataset} & \multicolumn{1}{c}{Method}                          & RR $\uparrow$ & Top-1        & Top-3       & Top-5       & Top-1      & Top-3      & Top-5     & Top-3               & Top-5              \\ \midrule
Test Set                    & Ours:EP ($\recognizer$)                             & 52\%          & 0.815        & 0.616       & 0.548       & 3.140      & 2.964      & 2.892     & 0.585               & 0.646              \\
                            & Ours ($\recognizer$)                                & 56\%          & 0.827        & 0.633       & 0.555       & 3.100      & 3.019      & 2.918     & 0.543               & 0.628              \\
                            & Ours:EP $\mathcal{T}_{\text{test}}$ ($\recognizer$) & 20\%          & 0.636        & 0.539       & 0.473       & 2.844      & 3.023      & 2.987     & 0.564               & 0.675              \\
                            & Ours $\mathcal{T}_{\text{test}}$ ($\recognizer$)    & 20\%          & 0.626        & 0.552       & 0.493       & 3.360      & 3.135      & 3.070     & 0.542               & 0.634              \\ \midrule
ChEMBL                      & Ours ($\recognizer$)                                & 7.6\%         & 0.531        & 0.443       & 0.396       & 2.544      & 2.510      & 2.460     & 0.675               & 0.727              \\
                            & Ours (MCMC)                                         & 9.2\%         & 0.532        & 0.486       & 0.432       & 2.364      & 2.310      & 2.263     & 0.765               & 0.759              \\
                            & Ours:EP (MCMC)                                      & 8.5\%         & 0.519        & 0.421       & 0.367       & 2.644      & 2.420      & 2.382     & 0.618               & 0.640              \\ \bottomrule
\end{tabular}

}

\end{table}
We evaluate the synthesizable analog task performances using ablation networks. We want to test whether the ablation networks integrate effectively with templates outside its structural support. Thus, we use the same $\tau$ as before for the $\tau$ experiments in Table \ref{tab:ep-analog}, allowing access to the full template set but forcing $\{\pi_{\mathcal{B}}',\pi_{\mathcal{R}}'\}$ to extrapolate when performing inference over $\hat{\mathcal{T}}_\text{test}$. 

The performance takes only a minor dip on the Test Set, compensating for slightly lower Avg. Sim. with higher Diversity and comparable SA. We further zoom in on the subset of $\mathcal{D}_0$ with structure among the held-out template classes. We emphasize this is very difficult for $\{\pi_{\mathcal{B}}',\pi_{\mathcal{R}}'\}$ to do, which has not seen any examples from those structural classes. It actually appears Ours:EP has the slight edge over Ours on the held-out template set, with slightly better SA and greater diversity. We attribute this to a regularization effect induced by removing (slightly, due to Step 2.) more complex program structures. There are still enough complex templates left that this does not harm performance, highlighting the robustness of our model in this setting. We believe the fact the ablation model can maintain comparable performance implies the following:
\begin{itemize}
    \item \textbf{Structural Extrapolation}: It is capable of inference of programs outside the structural support of its training distribution in this case.
    \item \textbf{Template Set Robustness}: Our model is not very sensitive to the default size of the template set, since using only 75\% of it already brings it close to diminishing returns.
\end{itemize}

The dip in performance is more noticeable for the predominantly unsynthesizable dataset ChEMBL, with lower metrics across the board. We suspect the reason is due to ChEMBL containing more complex molecules that require longer synthetic routes. This is also apparent from the lower analog diversity, suggesting training on more template variety helps.

We believe the difficulty of the task (ratio of synthesizable vs unsynthesizable molecules) can inform whether the method is sensitive to the template set it sees during training. Similar ablation studies to this one can highlight when additional resources should be allocated to expanding the training set and when it is sufficient to simply incorporate more templates at test time.

\subsection{Results on Synthesizable Molecular Design}
\begin{figure}[htbp]
\caption{We select the first Oracle from each Table in App. \ref{app:full-results} to compare Ours with Ours (EP). Aside from the ablation networks, we use the same experimental settings as Table \ref{tab:average-results}.}
\label{tab:ep-tdc}
  \centering
  \subfloat[][]{
  
  \resizebox{\textwidth}{!}{
\begin{tabular}{@{}cc|cccccccccc|cccccccccc@{}}
\toprule
          &           & \multicolumn{10}{c|}{GSK3$\beta$}                                                                                                                                                                                                   & \multicolumn{10}{c}{Median 1}                                                                                                                                                                                                      \\ \midrule
          &           & \multicolumn{1}{c|}{}             & \multicolumn{3}{c|}{Top 1}                                           & \multicolumn{3}{c|}{Top 10}                                           & \multicolumn{3}{c|}{Top 100}                     & \multicolumn{1}{c|}{}             & \multicolumn{3}{c|}{Top 1}                                          & \multicolumn{3}{c|}{Top 10}                                           & \multicolumn{3}{c}{Top 100}                      \\
          & category  & \multicolumn{1}{c|}{Oracle Calls} & Score         & SA             & \multicolumn{1}{c|}{AUC}            & Score          & SA             & \multicolumn{1}{c|}{AUC}            & Score          & SA             & AUC            & \multicolumn{1}{c|}{Oracle Calls} & Score        & SA             & \multicolumn{1}{c|}{AUC}            & Score          & SA             & \multicolumn{1}{c|}{AUC}            & Score          & SA             & AUC            \\ \midrule
Ours (EP) & synthesis & \multicolumn{1}{c|}{6921}         & \textbf{0.99} & \textbf{1.975} & \multicolumn{1}{c|}{0.872}          & 0.965          & 2.321          & \multicolumn{1}{c|}{0.818}          & 0.94           & \textbf{2.237} & 0.744          & \multicolumn{1}{c|}{9050}         & \textbf{0.4} & 4.12           & \multicolumn{1}{c|}{0.358}          & \textbf{0.344} & 4.434          & \multicolumn{1}{c|}{\textbf{0.305}} & \textbf{0.301} & 4.394          & \textbf{0.257} \\
Ours      & synthesis & \multicolumn{1}{c|}{4886}         & 0.98          & 2.045          & \multicolumn{1}{c|}{\textbf{0.891}} & \textbf{0.967} & \textbf{2.302} & \multicolumn{1}{c|}{\textbf{0.848}} & \textbf{0.944} & 2.27           & \textbf{0.778} & \multicolumn{1}{c|}{8303}         & \textbf{0.4} & \textbf{3.353} & \multicolumn{1}{c|}{\textbf{0.371}} & 0.342          & \textbf{4.161} & \multicolumn{1}{c|}{\textbf{0.305}} & 0.298          & \textbf{4.256} & 0.252          \\ \bottomrule
\end{tabular}
}
  }%
  \qquad
  \subfloat[][]{
  
    \resizebox{\textwidth}{!}{
\begin{tabular}{@{}cc|cccccccccc|cccccccccc@{}}
\toprule
          &           & \multicolumn{10}{c|}{Osimertinib MPO}                                                                                                                                                                                               & \multicolumn{10}{c}{Perindopril MPO}                                                                                                                                                                                                \\ \midrule
          &           & \multicolumn{1}{c|}{}             & \multicolumn{3}{c|}{Top 1}                                            & \multicolumn{3}{c|}{Top 10}                                          & \multicolumn{3}{c|}{Top 100}                     & \multicolumn{1}{c|}{}             & \multicolumn{3}{c|}{Top 1}                                            & \multicolumn{3}{c|}{Top 10}                                          & \multicolumn{3}{c}{Top 100}                      \\
          & category  & \multicolumn{1}{c|}{Oracle Calls} & Score          & SA             & \multicolumn{1}{c|}{AUC}            & Score          & SA            & \multicolumn{1}{c|}{AUC}            & Score          & SA             & AUC            & \multicolumn{1}{c|}{Oracle Calls} & Score          & SA             & \multicolumn{1}{c|}{AUC}            & Score          & SA             & \multicolumn{1}{c|}{AUC}           & Score          & SA             & AUC            \\ \midrule
Ours (EP) & synthesis & \multicolumn{1}{c|}{10000}        & 0.852          & 2.322          & \multicolumn{1}{c|}{\textbf{0.831}} & \textbf{0.849} & 2.475         & \multicolumn{1}{c|}{\textbf{0.823}} & \textbf{0.839} & 2.484          & \textbf{0.802} & \multicolumn{1}{c|}{10000}        & \textbf{0.626} & \textbf{3.382} & \multicolumn{1}{c|}{\textbf{0.562}} & \textbf{0.598} & \textbf{3.375} & \multicolumn{1}{c|}{\textbf{0.54}} & 0.56           & 3.349          & \textbf{0.509} \\
Ours      & synthesis & \multicolumn{1}{c|}{10000}        & \textbf{0.859} & \textbf{2.263} & \multicolumn{1}{c|}{0.826}          & 0.847          & \textbf{2.21} & \multicolumn{1}{c|}{0.81}           & 0.832          & \textbf{2.249} & 0.769          & \multicolumn{1}{c|}{10000}        & 0.622          & 3.338          & \multicolumn{1}{c|}{0.547}          & 0.591          & 3.378          & \multicolumn{1}{c|}{0.524}         & \textbf{0.558} & \textbf{3.137} & 0.485          \\ \bottomrule
\end{tabular}
}
  }
\end{figure}
We also evaluate the synthesizable molecular design performances using ablation networks. We want to evaluate whether the ablation models can guide the optimization trajectory as a surrogate generator of synthesizable analogs. We see comparable performances across all four Oracles in Table \ref{tab:ep-tdc}, and for each Ours (EP) having the edge on some metrics while Ours having the edge on other metrics. This suggests both are capable enough to serve as the inner subroutine of our bilevel genetic framework, although the models may have different biases on the kind of analogs it generates which affects the optimization trajectories. Since Ours (EP) may generate less structurally diverse analogs, it can converge in fewer Oracle calls, resulting in less Oracle calls and slightly higher AUCs. Meanwhile, Ours produce more diverse analogs, which enables the acquisition of higher confidence analogs. We see Ours to have the edge on SA across Median 1 (Top 1) and the MPO (Top 100) Oracles. This may be because higher confidence regions tend to be where the simpler molecules are, resulting in simpler analogs hence lower SA. Overall, the results show the robustness of our model.

}
\section{Syntax Tree Recognition} \label{app:recognizer}
In this section, we answer key questions like: (1) How does the relationship change with the addition of $\gT$? (2) How strong is the correlation between $\gX$ and $\gT$? (3) How justified are the most confident predictions made by $\recognizer{}{}$? We investigate the relationship between $\mathcal{M}$ and $\gT$. We seek to understand the extent to which the true mapping $\mathcal{M} \rightarrow \gT$  is well-defined. The first part is quantitative analysis, and the second part is a qualitative study.

\subsection{t-SNE and MDS Plots}
\begin{figure}[h!]
\caption{t-SNE on molecules in top (3,4,5,6) skeleton classes}
  \label{fig:tsne}
 \centering
 \begin{subfigure}[b]{0.49\textwidth}
     \centering
     \includegraphics[width=\textwidth]{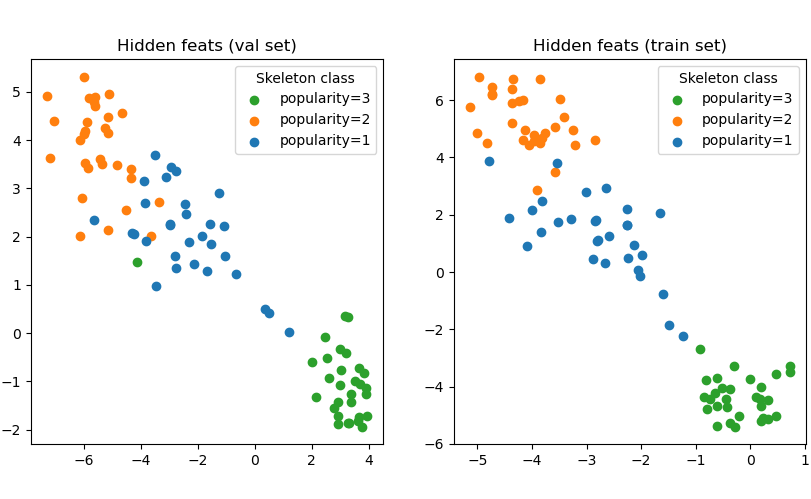}
     \caption{Top 3 (sample 30 each)}
     \label{fig:tsne-3}
 \end{subfigure}
 \hfill
 \begin{subfigure}[b]{0.49\textwidth}
     \centering
     \includegraphics[width=\textwidth]{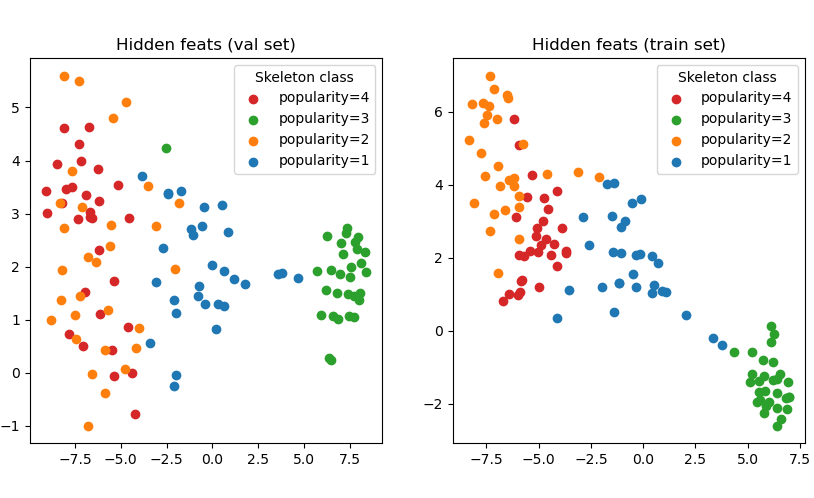}
     \caption{Top 4 (sample 30 each)}
     \label{fig:tsne-4}
 \end{subfigure}
 \hfill  
  \begin{subfigure}[b]{0.49\textwidth}
     \centering
     \includegraphics[width=\textwidth]{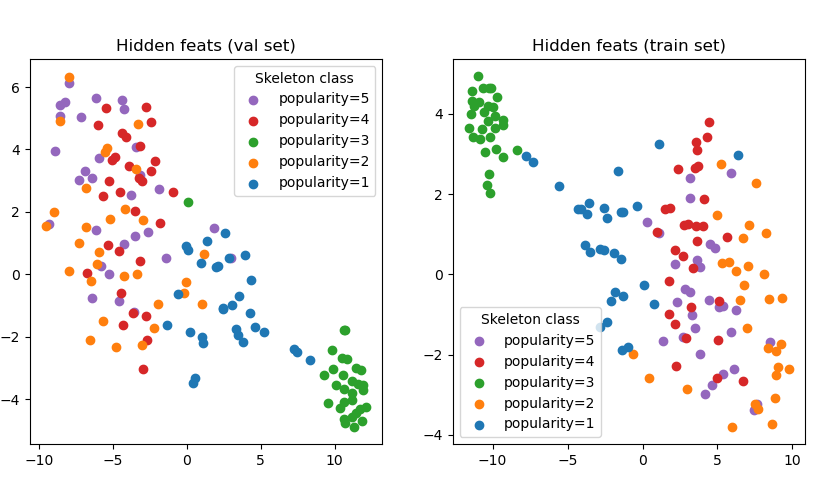}
     \caption{Top 5 (sample 30 each)}
     \label{fig:tsne-5}
 \end{subfigure}
 \hfill   
  \begin{subfigure}[b]{0.49\textwidth}
     \centering
     \includegraphics[width=\textwidth]{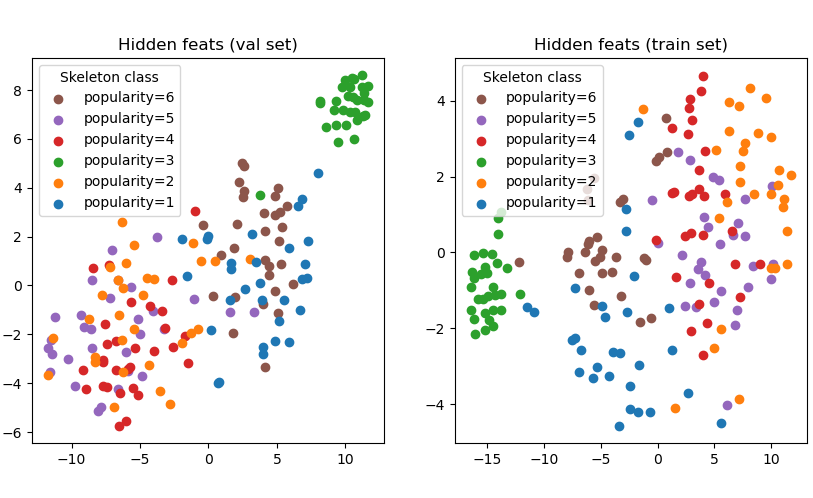}
     \caption{Top 6 (sample 30 each)}
     \label{fig:tsne-6}
 \end{subfigure}
 \hfill   
\end{figure}

We use the t-distributed stochastic neighbor embedding (t-SNE) on the final layer hidden representations of our MLP $\recognizer{}{}$ to visualize how our recognition model discriminates between molecules of different syntax tree classes. From Figure \ref{fig:tsne}, we see the MLP is able to discriminate amongst the top 3 or 4 most popular skeleton classes, visually partitioning the representation space. However, beyond that the representations on the validation set begin to coalesce, i.e., the model begins overfitting.

\begin{figure}[h!]
\caption{MDS on molecules in top (10,20,100) skeleton classes}
  \label{fig:mds}
 \centering
 \hfill
 \begin{subfigure}[b]{0.32\textwidth}
     \centering
     \includegraphics[width=\textwidth]{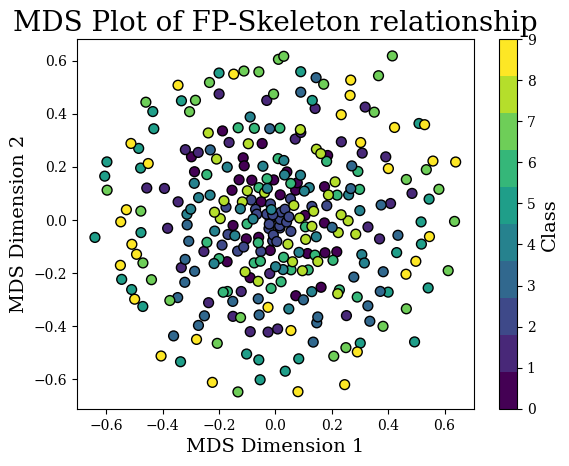}
     \caption{Top 10 (sample 30 each)}
     \label{fig:mds-sks-10}
 \end{subfigure}
 \hfill  
  \begin{subfigure}[b]{0.32\textwidth}
     \centering
     \includegraphics[width=\textwidth]{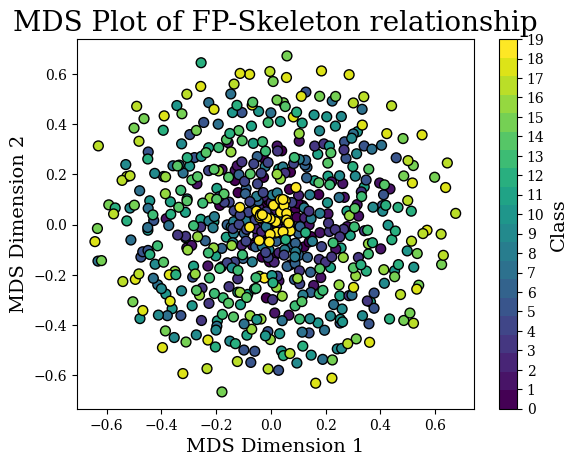}
     \caption{Top 20 (sample 30 each)}
     \label{fig:mds-sks-20}
 \end{subfigure}
 \hfill   
  \begin{subfigure}[b]{0.32\textwidth}
     \centering
     \includegraphics[width=\textwidth]{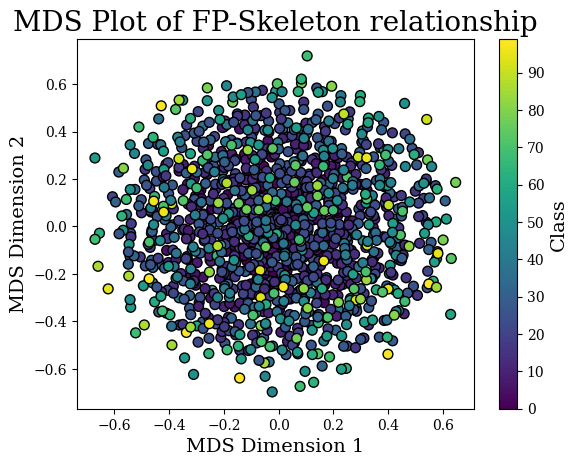}
     \caption{Top 100 (sample 30 each)}
     \label{fig:mds-sks-100}
 \end{subfigure}
 \hfill   
\end{figure}

Since gradient descent is stochastic, we also use multi-dimensional scaling (MDS) using the Morgan Fingerprint Manhattan distance on a subset of our dataset to visualize the relative positioning between molecules of different syntax tree classes (sorted based on popularity). From the plots in Figure \ref{fig:mds}, we observe some interesting trends:
\begin{itemize}
    \item \textbf{Similarly positioned points tend to have similar colors.}
    \item The darker end of the spectrum corresponding to the most popular classes generally cluster together in the middle.         
    \item The classes do not form disjoint partitions in space. As the ranked popularity increases, the points tend to disperse outwards. There are exception classes, e.g., the yellow set of points in Figure \ref{fig:mds-sks-20} that cluster in the center.
\end{itemize}
Based on these findings, it's reasonable to conclude a recognition classifier by itself is overly naive. However, the useful inductive bias that similar molecules are more likely to share the same syntactic template indicates the \textbf{localness} property still holds. Our method is designed with this property in mind: we encourage iterative refinement of the syntactic template when doing analog generation.

\begin{figure}[h!]
 \centering
 \begin{subfigure}[b]{0.23\textwidth}
     \includegraphics[width=\textwidth]{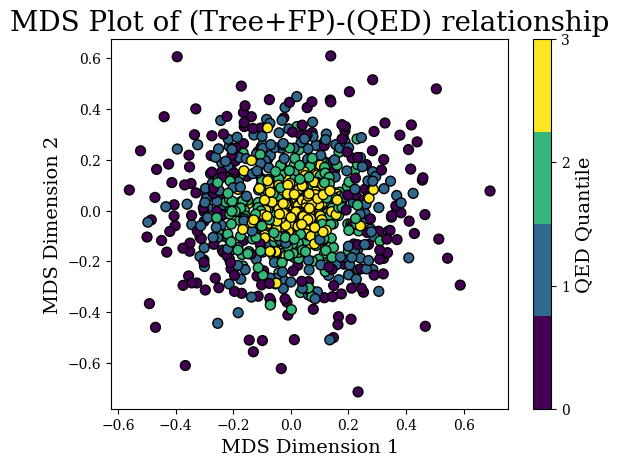}
 \end{subfigure}
\hfill
    \begin{subfigure}[b]{0.23\textwidth}
     \includegraphics[width=\textwidth]{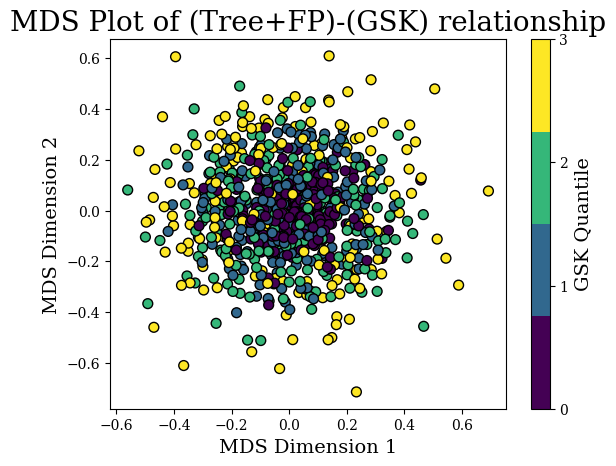}
 \end{subfigure}
 \hfill
     \begin{subfigure}[b]{0.23\textwidth}
     \includegraphics[width=\textwidth]{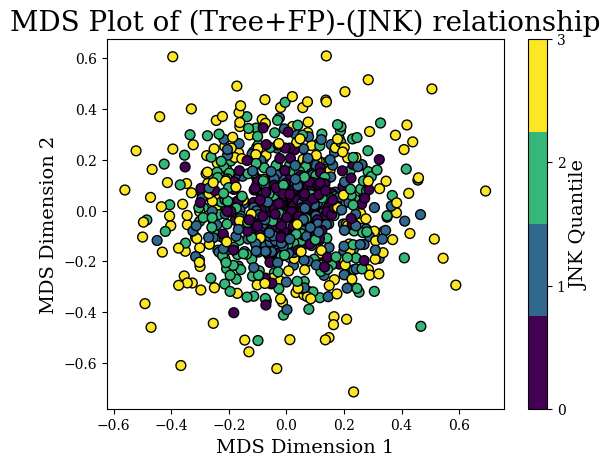}
 \end{subfigure}
 \hfill
  \begin{subfigure}[b]{0.23\textwidth}
     \includegraphics[width=\textwidth]{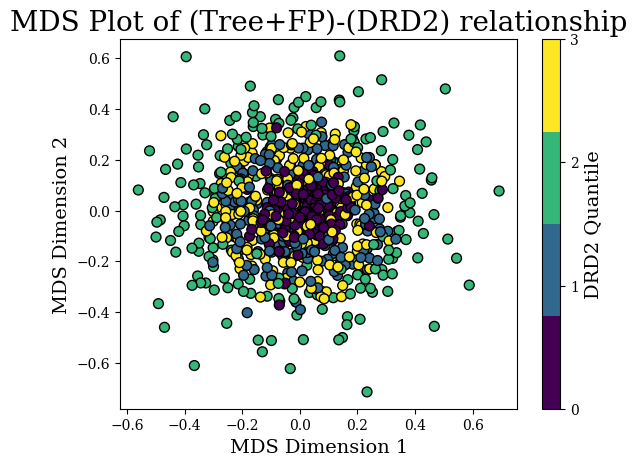}
 \end{subfigure}
 \newline
 \begin{subfigure}[b]{0.23\textwidth}
     \includegraphics[width=\textwidth]{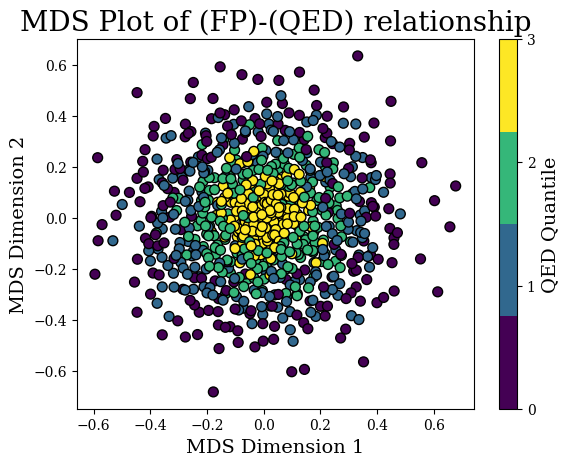}
 \end{subfigure}
 \hfill
    \begin{subfigure}[b]{0.23\textwidth}
     \includegraphics[width=\textwidth]{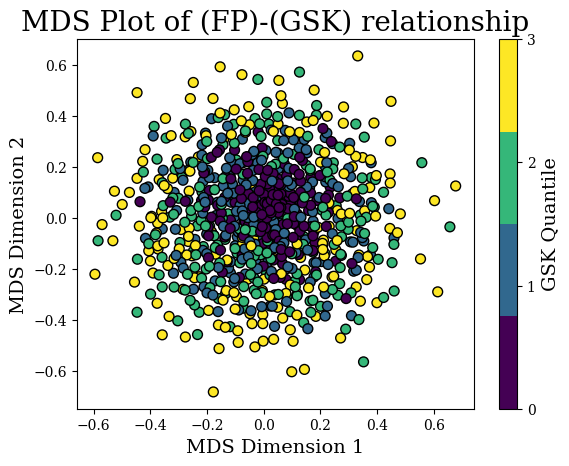}
 \end{subfigure}
 \hfill
     \begin{subfigure}[b]{0.23\textwidth}
     \includegraphics[width=\textwidth]{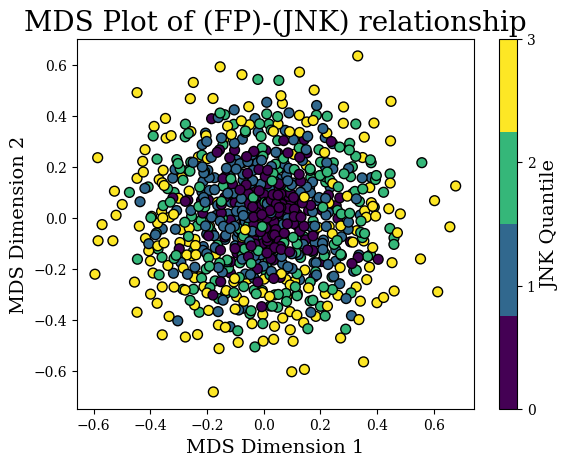}
 \end{subfigure}
 \hfill
  \begin{subfigure}[b]{0.23\textwidth}
     \includegraphics[width=\textwidth]{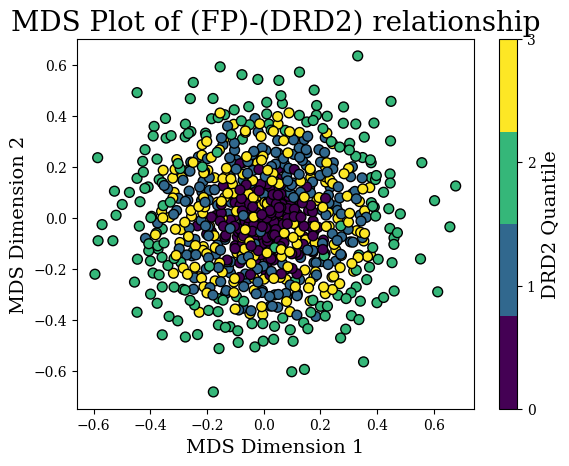}
 \end{subfigure}
  \caption{We visualize the structure-property relationship as a scatterplot of 2D structures vs property values. (Top) Structure is $\gX \times \gT$. We use MDS with the dissimilarity $d_{\gX \times \gT}((\vx_1, T_1),(\vx_2, T_2)) = ||\vx_1 - \vx_2||_1 + \text{Tree-Edit-Distance}(T_1,T_2)$. (Bottom) Structure is only $\gX$.}
  \label{fig:mds}
\end{figure}

We also use MDS to investigate the structure-property relationship to understand the joint effect $\gT$ and $\gX$ has on different properties of interest. As shown in Figure \ref{fig:mds}, we see overall, the functional landscape varies significantly from property to property, but the general trend is that decoupling $\gT$ from $\gX$ does not change the structure-property relationship much. Whereas analog generation requires a more granular understanding of the synergy between $\gX$ and $\gT$, molecular optimization does not. Instead, the evolutionary strategy should be kept fairly consistent between the original design space $(\gX)$ and $(\gX\times \gT)$. However, the top row exhibits lower entropy, with the empirical distribution looking ``less Gaussian". To capture this nuance, the evolutionary algorithm should combine both global and local optimization steps. We meet this observation with a bilevel optimization strategy that combines semantic crossover with syntactic mutation.

\subsection{Expert Case Study}
In this section, we enter the perspective of the recognition model learning the mapping from molecules to syntax tree skeletons. The core difference between this exercise and a common organic chemistry exam question \citep{flynn2014} is the option to abstract out the specific chemistry. Since the syntax only determines the skeletal nature of the molecule, the specific low-level dynamics don't matter. As long as the model can pick up on skeletal similarities between molecules, it will be confident in its prediction. We did the following exercise to understand if cases where the recognition model is most confident on unseen molecules can be attributed to training examples. We took the following steps:\\\\
For each true skeleton class $T$
 \begin{enumerate}
    \item Inference the recognition model on 10 random validation set molecules belonging to $T$.
    \item Pick the top 2 molecules the model was most confident belongs to $T$.
    \item For each molecule $M$.
    \begin{enumerate}
        \item Find the 2 nearest neighbors to $M$ belonging to $T$ in the training set.
    \end{enumerate}
    
\end{enumerate}

\begin{figure}[h!]
 \caption{COc1ncnc(N2C(=O)c3cc([N+](=O)[O-])c(O)cc3N=C2C2NC(=O)OC23CCC3)c1C which recognition model predicts is in its true class with 87.5\% probability}

\label{fig:ex1}
 \centering
  \begin{subfigure}[b]{0.49\textwidth}
     \centering
     \includegraphics[width=\textwidth]{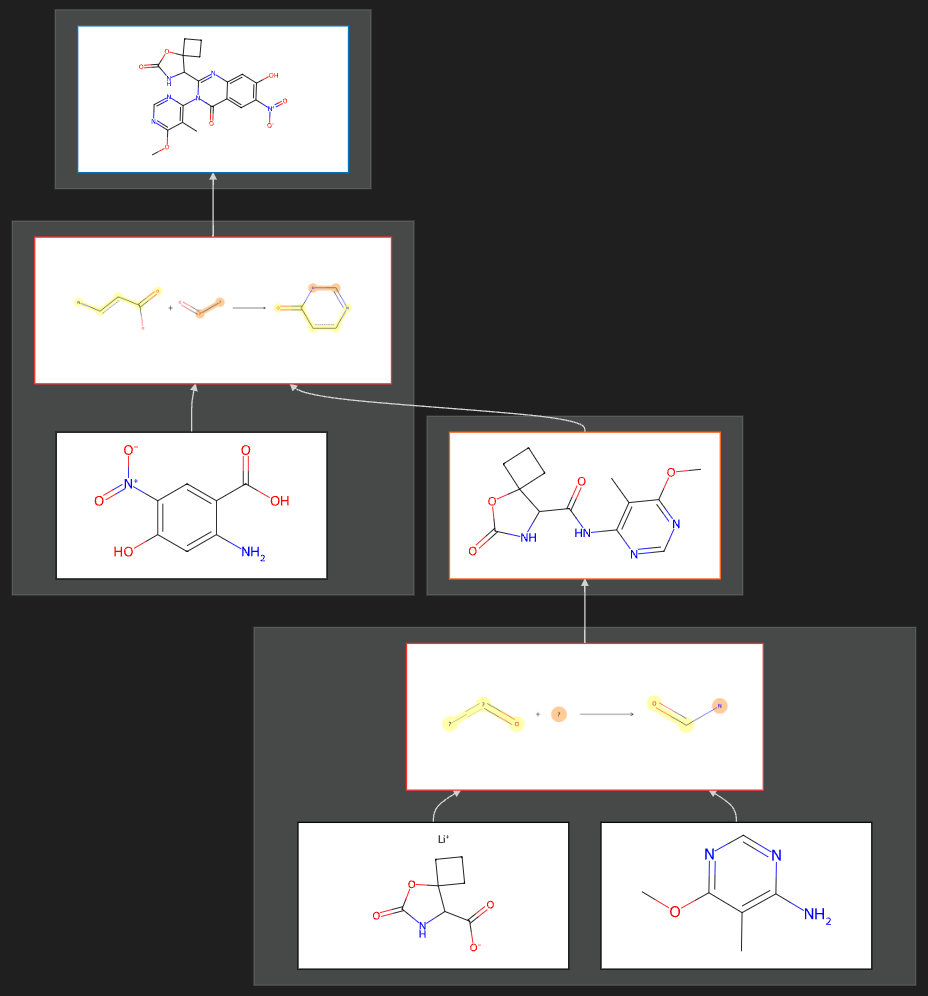}
     \caption{Query molecule}
     \label{fig:ex1-query}
 \end{subfigure}
 \hfill
 \begin{subfigure}[b]{0.49\textwidth}
     \centering
     \includegraphics[width=\textwidth]{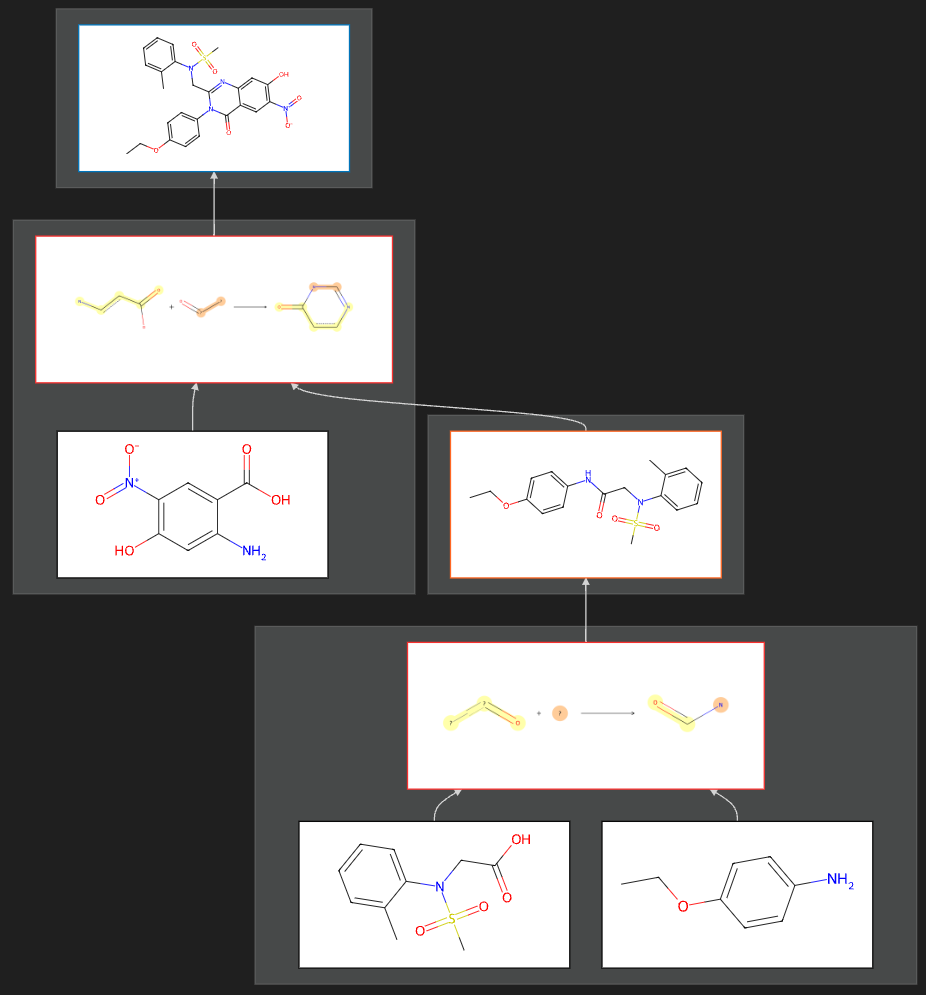}
     \caption{Nearest neighbor in training set}
     \label{fig:ex1-nn}
 \end{subfigure}
 \end{figure}

\begin{figure}[h!]
 \caption{COC(Cc1ccccc1)CN(C1CCCOc2ccccc21)S(=O)(=O)Cc1ccon1 which recognition model predicts is in its true class with 86.2\% probability}
\label{fig:ex2}
 \centering
  \begin{subfigure}[b]{0.32\textwidth}
     \centering
     \includegraphics[width=\textwidth]{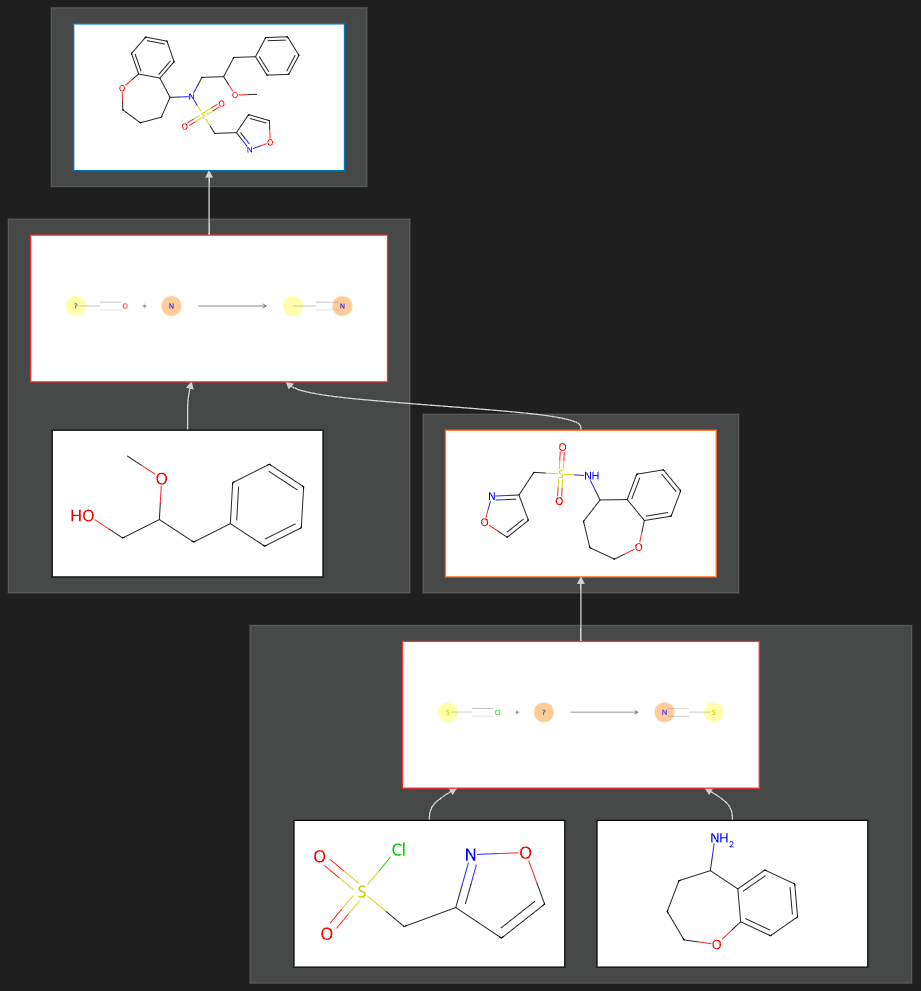}
     \caption{Query molecule}
     \label{fig:ex2-query}
 \end{subfigure}
 \hfill
  \begin{subfigure}[b]{0.32\textwidth}
     \centering
     \includegraphics[width=\textwidth]{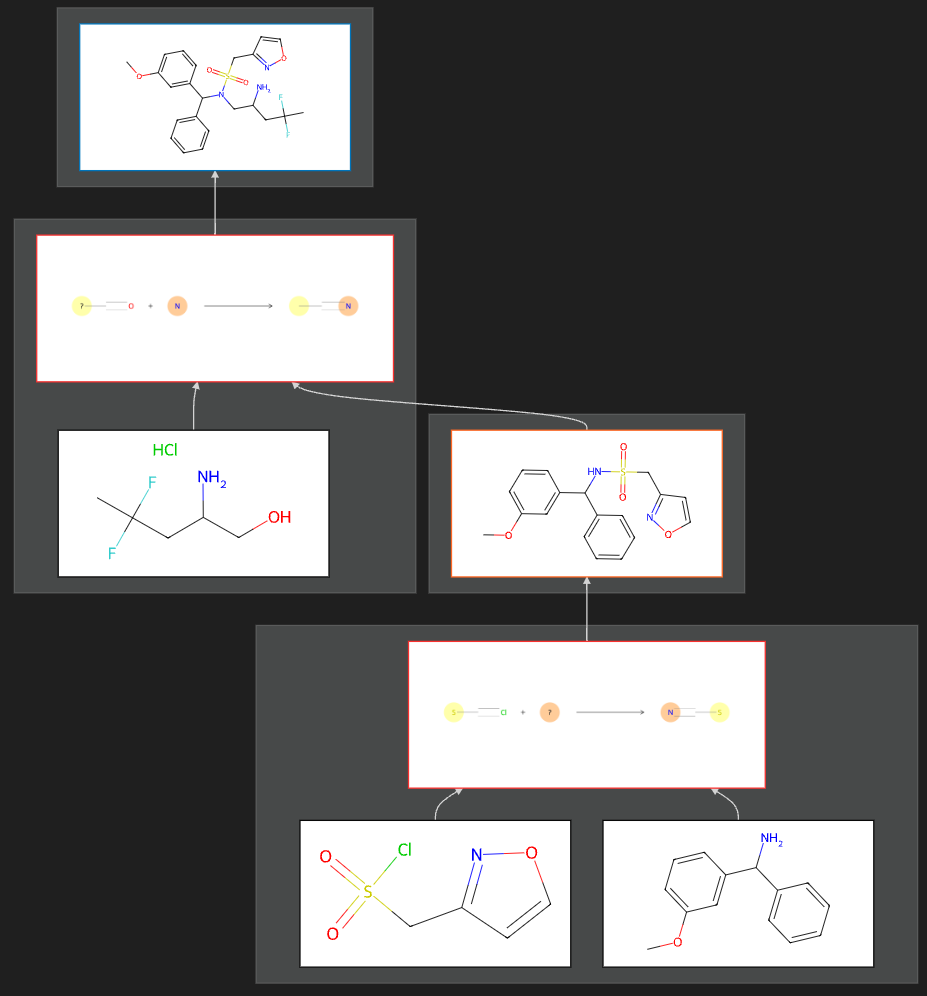}
     \caption{Nearest neighbor in training set}
     \label{fig:ex2-nn-1}
 \end{subfigure}
   \begin{subfigure}[b]{0.32\textwidth}
     \centering
     \includegraphics[width=\textwidth]{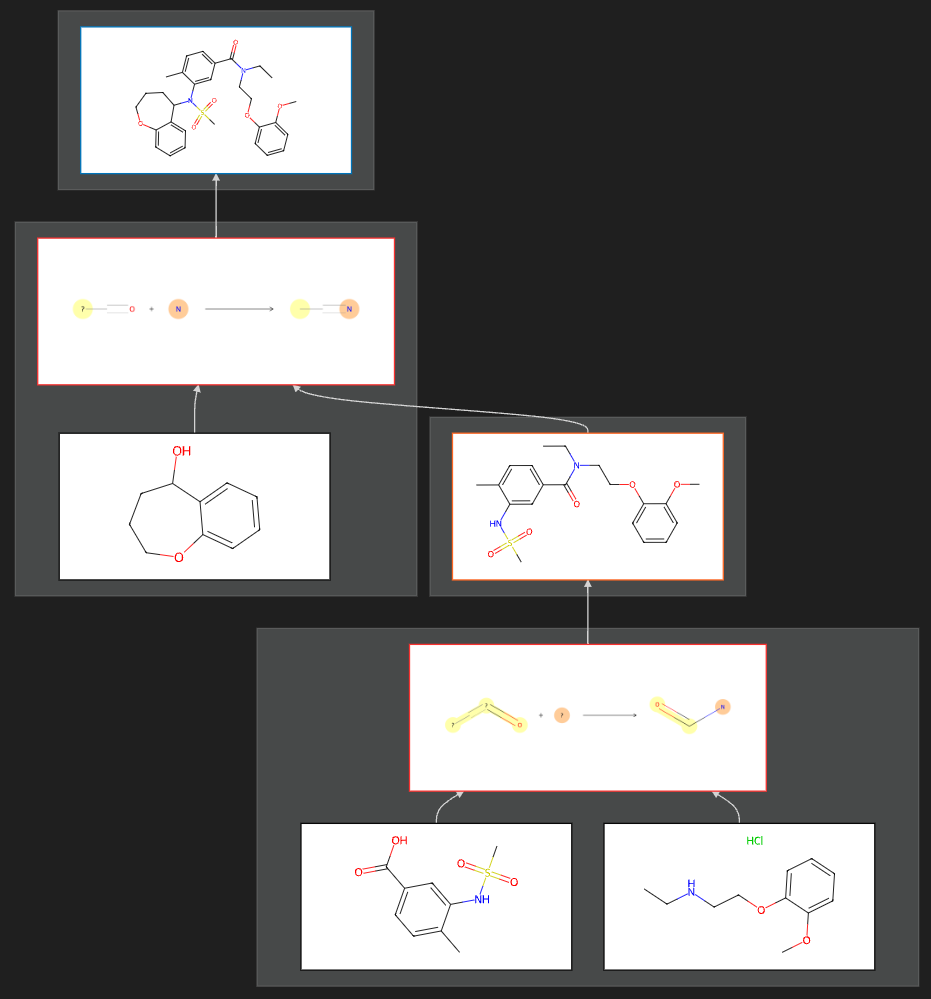}
     \caption{2nd Nearest neighbor}     
     \label{fig:ex2-nn-2}
 \end{subfigure}
 \end{figure}

Shown in Figures Figure \ref{fig:ex1} and Figure \ref{fig:ex2} is the output of these steps for a common skeleton class which requires two reaction steps.

In Figure \ref{fig:ex1}, we see that the query molecule's nearest neighbor is an output from the \textit{same program} but different building blocks. Both feature the same core fused ring system involving a nitrogen. Given that the model has seen Figure \ref{fig:ex1-nn} (and other similar instances), it should associate this core feature with a ring formation reaction step. Taking a step deeper, the respective precursors also share the commonality of having an amide linkage in the middle. Amides are key structural elements that the recognition model can identify. Both precursors underwent the same amide linkage formation step, despite the building blocks being different. Thus, the model's high confidence on the query molecule can be attributed directly to Figure \ref{fig:ex1-nn}.

In Figure \ref{fig:ex2}, there is more ``depth" to the matter. We see a skeletal similarity across all three molecules: a nitrogen in the center with three substituents. Although it's noteworthy that the nitrogen participates in a sulfonamide group in all three cases, using this fact to inform the syntax tree would be a mistake. This is because in Figure \ref{fig:ex2-query} and Figure \ref{fig:ex2-nn-1}, the sulfonamide group is the result of an explicit sulfonamide formation reaction, where a sulfonyl chloride reacts with an amine. However, in Figure \ref{fig:ex2-nn-2} the sulfonamide group is already present in a building block. Thus, we see where the recognition model taking as input the circular fingerprint of this molecule could overfit. Nonetheless, the nitrogen with three substituents necessitates at least one reaction is required. The necessity for a second reaction can be attributed to the ether linkage present in both Figure \ref{fig:ex2-query} and Figure \ref{fig:ex2-nn-2}. The recognizer would be able to justify an additional reaction after it has seen the bicyclic ring structure joined with the sulfonamide group sufficiently many times before. In summary, the model will often be presented multiple complex motifs, but only a subset of them may be responsible for reaction steps. The exact number of reactions needed can only be determined via actually doing the search, but high-level indicators (such as the nitrogen with three substituents) allow the recognition model to abstract out the semantic details and construe a ``first guess" of what the syntax tree is.

\section{{Expanded Related Works}}\label{app:program}
\subsection{Background on Program Synthesis}
Program synthesis is the problem of synthesizing a function $f$ from a set of primitives and operators to meet some correctness specification. For example, if we want to synthesize a program to find the max of two numbers, the correctness specification $\phi_{\text{max}} \coloneqq  f(x,y) \ge x \wedge f(x,y) \ge y \wedge (f(x,y) = x \lor f(x,y) = y) $. As our approach is inspired from ideas in program synthesis, we briefly cover some basic background. A program synthesis problem entails three things: 
\begin{enumerate}[itemsep=0em, parsep=0em]
    \item Background theory $\mathsf{T}$ which is the vocabulary for constructing formulas, a typical example being linear integer arithmetic: which has boolean and integer variables, boolean and integer constants, connectives ($\wedge$, $\lor$, $\neg$, $\rightarrow$), operators ($+$), comparisons ($\le$), conditional (If-Then-Else)
    \item Correctness specification: a logical formula involving the output of $f$ and $\mathsf{T}$
    \item Set of expressions $L$ that $f$ can take on described by a context-free grammar $G_L$.
\end{enumerate}
Program synthesis is often formulated as deducing a constructive proof to the statement: for all inputs, there exists an output such that $\phi$ holds. The constructive proof itself is then the program. At the low-level, program synthesis methods repeatedly calls a SAT solver with the logical formula $\neg \phi$. If UNSAT is returned, this means $f$ is valid. Syntax-guided synthesis \citep{alur2013, schkufza2013} (SyGuS) is a framework for meeting the correctness specification with a syntactic template. Syntactic templates explicitly constrains $G_L$, significantly reducing the number of implementations $f$ can take on. Sketching is an example application where programmers can sketch the skeletal outline of a program for synthesizers to fill in the rest \citep{solar2005}. More directly related to our problem's formulation is inductive synthesis, which seeks to generate $f$ to match input/output examples. The problem of synthesis planning for a molecule $M$ is a special case of the programming-by-example paradigm, where we seek to synthesize a program consistent with a single input/output pair: ($\{B\}$, $M$). Inductive synthesis search algorithms have been developed to search through the combinatorial space of derivations of $G_L$. In particular, stochastic inductive synthesis use techniques like MCMC to tackle complex synthesis problems where enumerative methods do not scale to. MCMC has been used to optimize for the opcodes in a program \citep{schkufza2013} or for the abstract syntax tree directly \citep{alur2013}. In our case, the space of possible program semantics is so large that we decouple the syntax from the semantics, performing stochastic synthesis over only the syntax trees. We also borrow ideas from functional program synthesis, where top-down strategies are preferred over bottom-up ones to better leverage the connection between a high-level specification and a concrete implementation \citep{polikarpova2016}. Similar to how top-down synthesis enables aggressive pruning of the search space via type checking, retrosynthesis algorithms leverages the target molecule $M$ to prune the search space via template compatability checks.
\subsection{{Execution-guided Program Synthesis}}\label{app:program-execution} We would like to note the distinction between our program synthesis formulation and other formulations. Retrosynthesis is essentially already guided by the execution state at every step. Each expansion in the search tree executes a deterministic reaction template to obtain the new intermediate molecule. Planners based on single-step models \citep{chen2020}, for example, assume the Markov Property by training models to directly predict a template given \textit{only} the intermediate \citep{torren2024,tu2022}. In program synthesis, meanwhile, the state space is a set of partial programs with actions corresponding to growing the program. The execution of the program (or verification against the specification) does not happen until a complete program is obtained. In recent years, neural program synthesis methods found using auxiliary information in the form of the \textit{execution state} of a program can help indirectly inform the search \citep{bunel2018,chen2018,ellis2019} since it gives a sense on what the program can compute so far. This insight does not apply to retrosynthesis, since retrosynthesis already executes on the fly. It also does not apply for the methods introduced in Section \ref{sec:2-3} that construct a synthetic tree in a bottom-up manner, for the same reason (the only difference is they use forward reaction templates, with a much smaller set of robust reaction templates) to obtain the execution state each step. However, as described in Section \ref{sec:3-3-2}, our approach combines the computational advantages of restricting to a small set of forward reaction templates with the inductive bias of retrosynthetic analysis. Our policy is to predict \textit{forward} reaction templates in a \textit{top-down} manner. This formulation is common in top-down program synthesis, where an action corresponds to selecting a hole in the program. Similarly, our execution of the program does not happen until the tree is filled in. However, we leverage the insight that the execution state helps in an innovative way, as discussed in \ref{app:aux}.

{
\subsection{{Inspirations from Retrosynthesis and Alternate Formulations}}\label{app:program-retrosynthesis}
We begin by elaborating the distinctions between retrosynthesis methods and methods for synthesizable molecular design. Then, we identify a few recent works from retrosynthesis that can inspire cross-pollination of ideas. Finally, we end with alternate formulations of the problem that are also valuable to consider for future cross-examination.

{\textbf{Intended Use Case}}

{Retrosynthesis aims to find a synthetic route for a given target molecule, without reference to how the target molecule is obtained or further optimizations on the target molecule. The target molecule is a compound that may serve any application or use case that we will not get into here, but importantly it \textit{is} the problem to solve. We refer readers to \cite{gao2020} for further descriptions.}

{Synthesizable molecular design aims to be a standalone molecular optimization workflow that explicitly constrains the design space to be synthetically accessible building blocks and reactions \cite{vinkers2003}. This is often coupled with property oracles that evaluate the designs, which guides the optimization towards parts of the design space with higher fitness \cite{gao2022}.}

{\textbf{MDP Formulation}}

{Retrosynthesis can be formulated as a tree-shaped MDP, where each state is a molecule (initial state being the target, terminal states being building blocks) and each action is a reversed (``retro") reaction. The tree shape of the MDP is due to the fact the retro reaction (action) produces a set of reactants (states) \cite{liu2023retrosynthetic}. Retrosynthetic planners often tackle the MDP by combining a single-step model (predicting retro reactions) and a multi-step planner (e.g. A* search \cite{liu2018}, MCTS \cite{segler2017}, depth-first search \cite{kishimoto2019}). A solution to the MDP is a \textit{tree} of actions, i.e. a synthetic tree, where all sequences of actions in this tree lead to terminal states.}

{Synthesizable molecular design feature a broader set of methods but can be defined as a discrete optimization problem over synthetic trees directly: $\argmax_x f(F(x))$ where $x$ is a synthetic tree, $F$ the root molecule of $x$ and $f$ is the fitness function. As the design space is intractably large, prior approaches discussed in \ref{sec:2-2} formulate the problem as a serial MDP, where each state is a synthetic tree and each action an edit operation (add, merge, etc.) to the synthetic tree. Though simple, we argue such a formulation is ill-advised, for reasons we discussed and demonstrated in the main text. We use an alternate formulation, inspired by program synthesis, that considers each state as a partially sketched program and each action as completing a hole of the program.}

{\textbf{Learning Goal}}

{Retrosynthesis methods aim to learn a policy $\pi(a|s)$, where $s$ is a molecule and $a$ a retro reaction, which can be template-based or template-free (we won't go into that here). Traditional works learn the policy from public datasets of synthetic routes (e.g. USPTO), but recent works have explored novel strategies for learning $\pi$ by combining offline and online training. The offline training is usually done on a reaction dataset to initialize a policy network and/or reaction model. The online training iteratively adapts the policy network by acquiring more data using a planning algorithm, possibly guided by the current policy network. More specifically, \cite{guo2024retrosynthesis} uses MCTS to acquire data, inferring policy and value targets based on the node visit counts. \cite{kim2021self} uses a self-improvement strategy, reminiscent of AlphaGo Zero \cite{silver2017}, that trains independent reference forward and backward reaction models to control the quality and diversity of new reaction pathways acquired by the planner. \cite{liu2023retrosynthetic} follows a similar strategy, but decouples the synthesizability and cost of the value function. They also architect a two-branch policy network that uses a trainable single-step network to optimize the probabilities over reactions from the frozen reference network to better model real-world synthesis considerations. Like \cite{guo2024retrosynthesis}, they use MCTS guided by the current policy network to acquire more data, creating a synergistic feedback loop that results in a holistic, trained policy network which can be plugged into multi-step planners.}

{Synthesizable molecular design methods, meanwhile, are relatively more spread out, with research going into problem formulation and algorithmic frameworks for tackling this more open-ended problem. In SynthesisNet, the policy learning is done entirely offline (as described in Sections \ref{sec:3-3-2} and \ref{sec:4-1-1}) to amortize for the cost of searching during the actual online phases (MCMC and GA), but above techniques from retrosynthesis can also facilitate self-improving the surrogate network. One potential idea for generating more experiences is to take partial program examples from the existing data, use guided planning to complete those examples, then retrain the surrogate network with the augmented dataset. We leave the details for exciting future extensions of our work.}

\textbf{Alternate Formulations}

Towards synthesizable molecular design, alternate formulations from recent developments can also be considered. Decision Transformer \cite{chen2021decision} is a recent work that re-imagines offline RL as a conditional sequence modeling task. Notably, the model conditions on reward-to-go and the history for generating the next action. The Transformer architecture enables long-term modeling of the environment's dynamics, enabling credit assignment and relational modeling of its history. The offline setting tackled by Decision Transformer naturally aligns with our method, where we do (self-)supervised policy training from programs generated offline. Although the program structure carries important hierarchical information about the relations of building blocks and reactions to one another, it's worth considering whether the synthesis tree construction serialization protocol used by prior works \cite{bradshaw2020,gao2021,luo2024projecting,gao2024generative} can be used to formulate a conditional sequence modeling problem. GFlowNet \cite{bengio2021flow,bengio2023gflownet} is also a recent work that formulates a flow network over a tree-based MDP, where the incoming and outgoing flow are proportional the probability of subsequent actions and learned using flow matching objectives. The goal of this model is to learn amortized samplers for reward functions, producing both diverse and high-quality samples constructed in a step-by-step manner following an MDP environment. The learning occurs on offline trajectories with observed rewards. The formulation using reward functions can be a direction of future work for our framework, which currently only considers rewards in the online phases. However, if we had considered rewards to be provided upfront along with the data generation procedure, we can adopt GFlowNet to amortize the expensive work done by MCMC, and directly sample programs. We leave this study for future works. We believe cross-examining alternate formulations and recent methodologies to be essential for finding future inspirations for extending the innovation horizon of methods used to tackle synthesizable molecular design.

}

\section{Derivation of Grammar} \label{app:grammar}
We now define the grammar $G_{\gP}$ describing the set of implementations our program can take on. A context-free grammar is a tuple $G_{\gP}\coloneqq (\mathcal{N}, \Sigma, \gP, \gX)$ that contains a set $\mathcal{N}$ of non-terminal symbols, a set $\Sigma$ of terminal symbols, a starting node $\gX$, and a set of production rules which define how to expand non-terminal symbols. Recall we are given a set of reaction templates $\gR$ and building blocks $\gB$. Templates are either uni-molecular ($\coloneqq \gR_1$) or bi-molecular ($\coloneqq \gR_2$), such that $\gR = \gR_1 \cup \gR_2$. In the original grammar, these take on the following:
\begin{enumerate}
    \item \textbf{Starting symbol}: $T$
    \item \textbf{Non-terminal symbols}: $R_1$, $R_2$, $B$
    \item \textbf{Terminal symbols}:
    \begin{itemize}
        \item $\{R \in \gR_1\}$: Uni-molecular templates
        \item $\{R \in \gR_2\}$: Bi-molecular templates
        \item $\{BB \in \gB\}$: Building blocks
    \end{itemize}
    \item \textbf{Production rules}:
    \begin{enumerate}
        \item $T \rightarrow R_1$
        \item $T \rightarrow R_2$
        \item $R_1 \rightarrow R(B)$ ($\forall R \in \gR_1$)
        \item $R_1 \rightarrow R(R_1)$ ($\forall R \in \gR_1$)
        \item $R_1 \rightarrow R(R_2)$ ($\forall R \in \gR_1$)
        \item $\forall (X_1, X_2) \in \{``R_1", ``R_2", ``B"\} \times \{``R_1", ``R_2", ``B"\}$
        \begin{itemize}
            \item $R_2 \rightarrow R(X_1, X_2)$ ($\forall R \in \gR_2$)
        \end{itemize}        
        \item $B \rightarrow BB$ ($\forall BB \in \gB$)
    \end{enumerate}
\end{enumerate}    

\noindent Example expressions derived from this grammar are ``R3(R3(B1,B2),R2(B3))" and ``R4(R1(B2,B1))" for the programs in Figure \ref{fig:fig1}. 

\noindent Identifying a retrosynthetic pathway can be formulated as the problem of searching through the derivations of this grammar conditioned on a target molecule. This unconstrained approach is extremely costly, since the number of possible derivations can explode.

In our syntax-guided grammar, we are interested in a finite set of syntax trees. The syntax tree of a program depicts how the resulting expression is derived by the grammar. These are either provided by an expert who has to meet experimental constraints, or specified via heuristics (e.g., maximum of $x$ reactions, limiting the tree depth to $y$). For example, the syntax-guided grammar for the set of trees with at most $2$ reactions is specified as follows:
\begin{enumerate}
    \item \textbf{Starting symbol}: $T$
    \item \textbf{Non-terminal symbols}: $R_1$, $R_2$, $B$
    \item \textbf{Terminal symbols}:
    \begin{itemize}
        \item $\{R \in \gR_1\}$: Uni-molecular templates
        \item $\{R \in \gR_2\}$: Bi-molecular templates
        \item $\{BB \in \gB\}$: Building blocks
    \end{itemize}
    \item \textbf{Production rules}:
    \begin{enumerate}
        \item $T \rightarrow R_2(B,B)$
        \item $T \rightarrow R_1(B)$
        \item $T \rightarrow R_1(R_2(B,B))$
        \item $T \rightarrow R_1(R_1(B))$
        \item $T \rightarrow R_2(B,R_1(B))$
        \item $T \rightarrow R_2(B,R_2(B,B))$
        \item $T \rightarrow R_2(R_1(B),B)$
        \item $T \rightarrow R_2(R_2(B,B),B)$
        \item $R_1 \rightarrow R$ ($\forall R \in \gR_1$)
        \item $R_2 \rightarrow R$ ($\forall R \in \gR_2$)
        \item $B \rightarrow BB$ ($\forall BB \in \gB$)
    \end{enumerate}
\end{enumerate}

This significantly reduces the number of possible derivations, but two challenges remain:
\begin{itemize}
    \item How can when pick the initial production rule when the number of syntax trees grow large?
    \textit{We use an iterative refinement strategy, governed by a Markov Chain Process over the space of syntax trees. The simulation is initialized at the structure predicted from our recognition model Appendix \ref{app:recognizer}.}
    \item How can we use the inductive bias of retrosynthetic analysis when applying rules 9, 10, 11? 
    \textit{We formulate a finite horizon MDP over the space of partial programs, where the actions are restricted to decoding only frontier nodes. This topological order to decoding is consistent with the top-down problem solving done in retrosynthetic analysis. Furthermore, our pretraining and decoding algorithm enumerates all sequences consistent with topological order.}
\end{itemize}
These two questions are addressed by the design choices in Section \ref{sec:3-3}.

\section{Policy Network}\label{app:policy-network}
\subsection{Featurization}\label{app:e-1}
Our dataset $\gD$ comprises partial programs ${T \in \partial \gP}$ producing molecules $M$.Then, we compute node features $\mH$ and labels $\mY$ as:
\begin{equation*}
    \vh_n \coloneqq [\text{FP}_{2048}(M), \text{BB}_{2048}(n), \text{RXN}(n)], \quad
    \vy_n \coloneqq \begin{cases}
        \text{RXN}(n), & \text{if $i$ is a reaction node}, \\
        \text{BB}_{256}(n), & \text{otherwise},
    \end{cases}
\end{equation*}
where $\text{FP}_d(\cdot)$ computes the $d$-bit radius 2 Morgan fingerprints, $\text{BB}_d(n)=\text{FP}_{d}(n_{\text{SMILES}})$ if $n$ is attributed with a building block from $\gB$ or $\textbf{0}_{d}$ otherwise and $\text{RXN}(n)=\text{one\_hot}_{91}(n_{\text{\text{RXN}\_ID}})$ if $n$ is attributed with a reaction from $\gR$ or $\textbf{0}_{91}$ otherwise.

If $\mathcal{N}(T)$ and $\mathcal{E}(T)$ denote the node and edge set of $T \in \partial \gP$, then we define, for convenience: 
\begin{align}\label{defs}
    \text{RXN}({T}) &:= \{r \in \mathcal{N}(T) \mid \exists c, p \in \mathcal{N}(T)  \text{ s.t. } (r,c) \in \mathcal{E}(T) \cap (p,r) \in \mathcal{E}(T)\} \\
    \text{BB}({T}) &= \{b \in \mathcal{N}(T) \mid \nexists c \text{ s.t. } (b,c) \in \mathcal{E}(T)\}. \}
\end{align}

\subsection{Loss Function}
Let the superscript $(i)$ indicate the $i$-th sample in the dataset. The loss function is: 
\begin{align*}
    \mathcal{L}_{\gD}(\Phi) &\coloneqq \frac{1}{|\gD|}\sum_{i=1}^{|\gD|}\sum_{n \in \text{RXN}(T^{(i)})}\text{CE}(\pi_\gR(T^{(i)}, \mH^{(i)})_n, \vy^{(i)}_n), \\
    \mathcal{L}_{\gD}(\Omega) &\coloneqq \frac{1}{|\gD|}\sum_{i=1}^{|\gD|}\sum_{n \in \text{BB}(T^{(i)})}\text{MSE}(\pi_\gB(T^{(i)}, \mH^{(i)})_n, \vy^{(i)}_n).
\end{align*}
CE and MSE denote the standard cross entropy loss and mean squared error loss, respectively. For our evaluation metric, we consider accuracy, where the output of $\pi_{\gB}$ is interpreted as the nearest building block with respect to cosine distance.


\subsection{Auxiliary Training Task}\label{app:aux}
In Section \ref{sec:3-3-2}, we defined the representation $T$ to be the parse tree of a partial program. However, we omitted an extra step that was used to preprocess $T$ for training. The motivation for this extra step is discussed deeply in \ref{app:program-execution}. We add an additional step when preprocessing $\gD$: For each $T$ in $\gD$, for each node $r$ corresponding to a reaction, we add a new node $o_r$ corresponding to the intermediate outcome of the reaction. If $\text{RXN}(T)$ is the reaction nodes of $T$, we can construct $T'$ from $T$ as follows: \begin{align}
    \mathcal{N}(T') &\leftarrow \mathcal{N}(T) \cup \text{RXN}(T) \\
    \mathcal{E}(T') &\leftarrow \mathcal{E}(T) \cup \{(\text{parent}(r),o_r), (o_r, r) \forall r \in \text{RXN}(T)\} 
\end{align}

\begin{figure}
\begin{subfigure}[b]{0.49\textwidth}
     \centering
     \includegraphics[width=\textwidth]{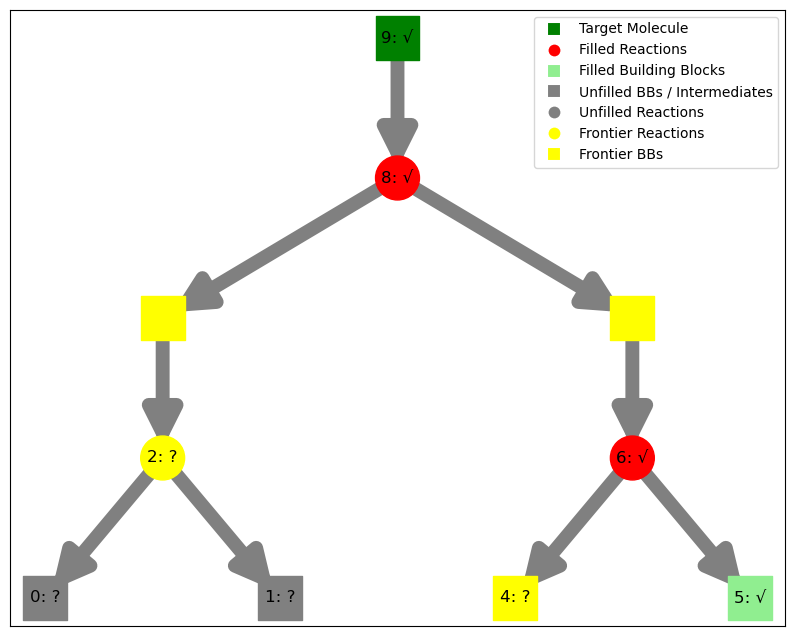}
     \label{fig:auxiliary-example-1}
 \end{subfigure}
 \hfill   
   \begin{subfigure}[b]{0.49\textwidth}
     \centering
     \includegraphics[width=\textwidth]{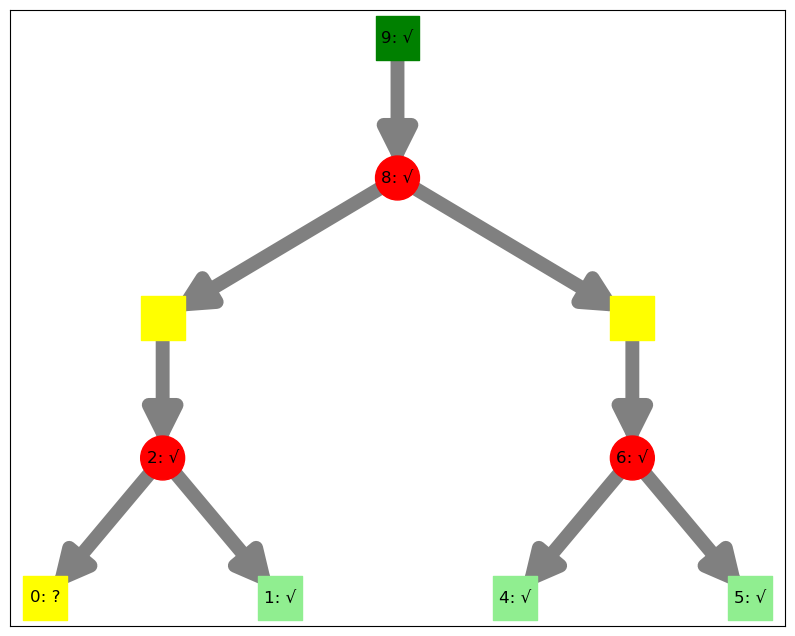}
     \label{fig:auxiliary-example-2}
 \end{subfigure}
 \caption{Examples of $T'$ where prediction targets are the frontier reactions (yellow circles), frontier building blocks (numbered yellow squares) and auxiliary intermediates (un-numbered yellow squares).}
 \label{fig:auxiliary-example}
\end{figure}

Lastly, we attribute each $o_r$ with the intermediate obtained from the original synthetic tree, i.e. executing the output of the program rooted at $r$. We featurize $\{y_{o} := \text{FP}_{256}({o}_{\text{SMILES}})\}$ and add them as additional prediction targets to $\gD$. Examples of $T'$ are given in Figure \ref{fig:auxiliary-example}.

\subsection{Ablation Study: Auxiliary Task}\label{app:ablate}
To understand whether the two key design choices for $\partial \gP'$ are justified, we did two ablations:
\begin{enumerate}
    \item We use the original description of $\partial \gP$ in Section \ref{sec:3-3-2}, i.e. without the auxiliary task.
    \item We use $\partial \gP'$, but without attributing the intermediate nodes (so the set of targets is the same as Ablation 1.)
\end{enumerate}

\begin{figure}[h!]
 \hfill   
   \begin{subfigure}[b]{0.24\textwidth}
     \centering
     \raisebox{0.1\height}{
     \includegraphics[width=\textwidth]{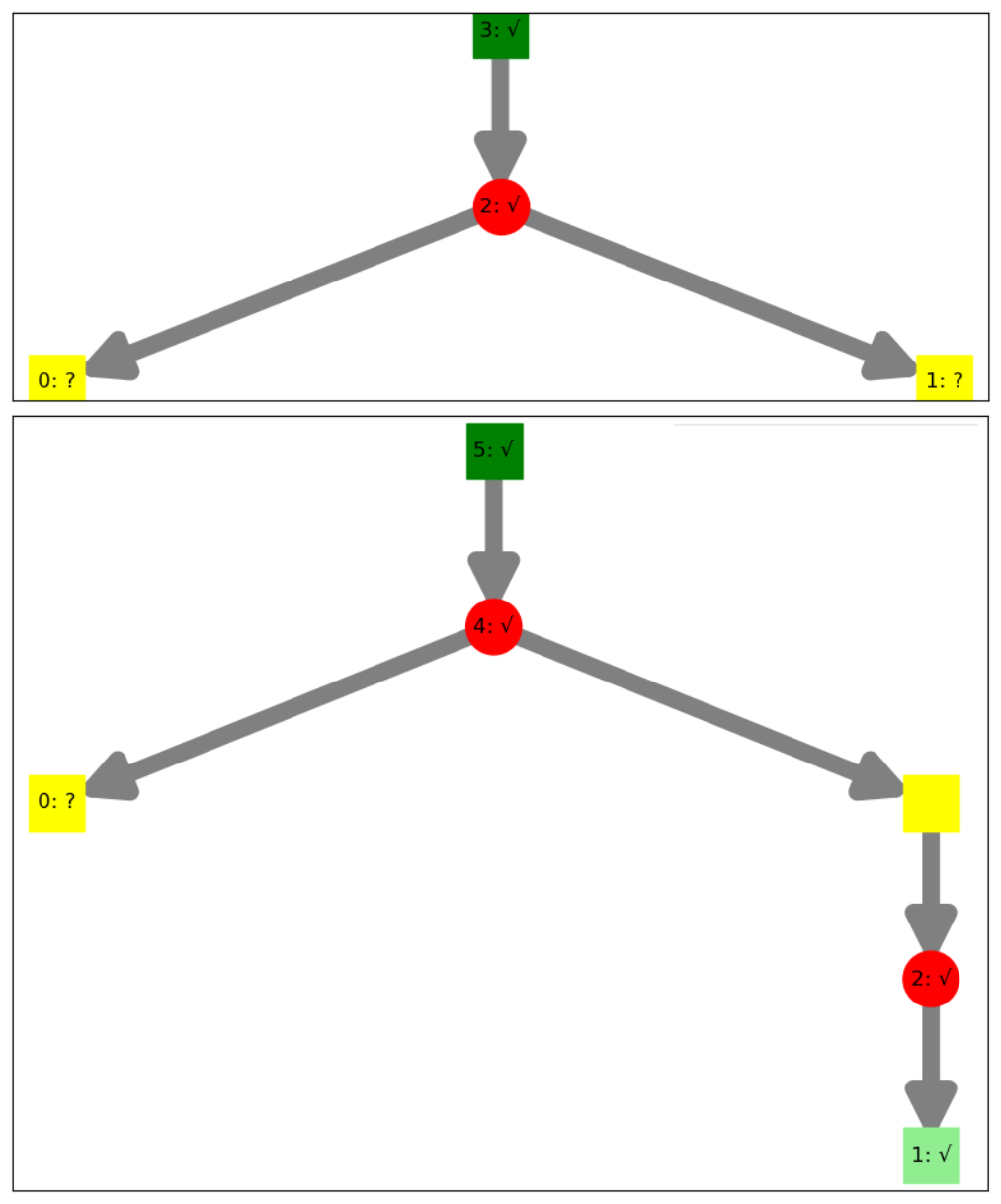}\     
     }
     \caption{Examples from $\partial \gP'$}
     \label{fig:nn}
 \end{subfigure}
 \hfill   
   \begin{subfigure}[b]{0.24\textwidth}
     \centering
     \includegraphics[width=\textwidth]{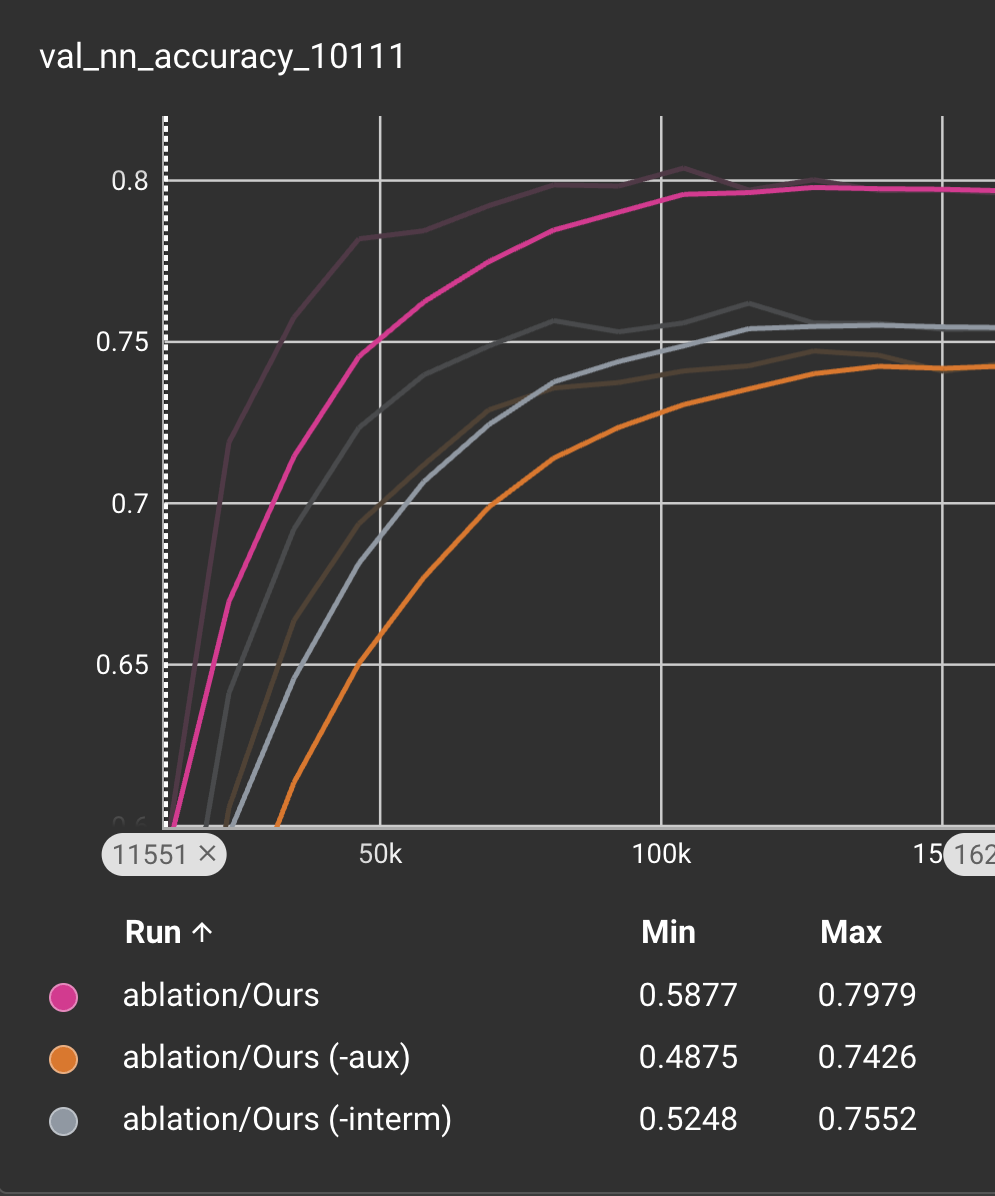}
     \caption{NN accuracy loss over top example Figure \ref{fig:nn}}
     \label{fig:10111}
 \end{subfigure}
 \hfill   
    \begin{subfigure}[b]{0.24\textwidth}
     \centering
     \includegraphics[width=\textwidth]{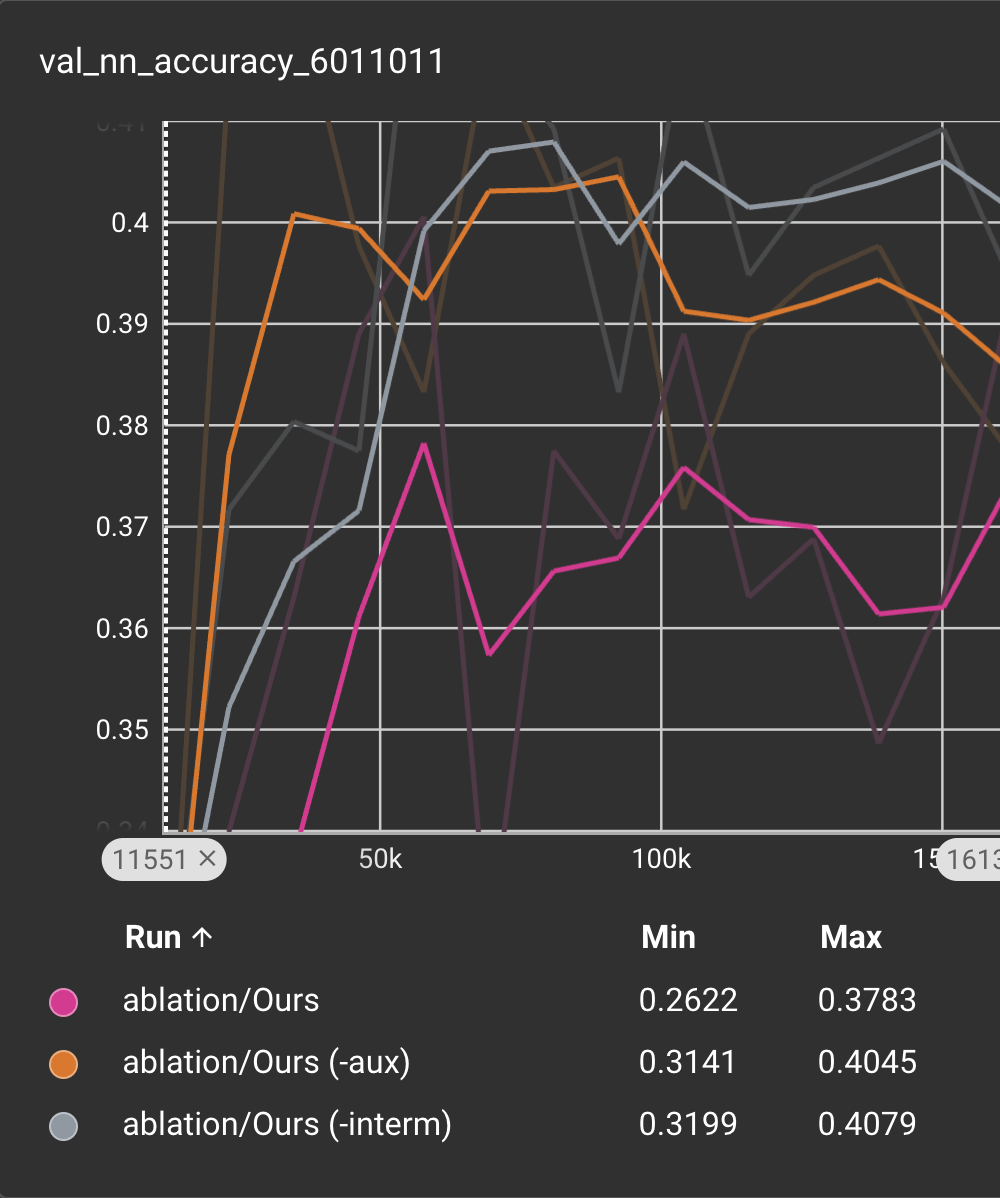}
      \caption{NN accuracy over bottom example Figure \ref{fig:nn}}
     \label{fig:6011011}
 \end{subfigure}
 \hfill   
     \begin{subfigure}[b]{0.24\textwidth}
     \centering
     \includegraphics[width=\textwidth]{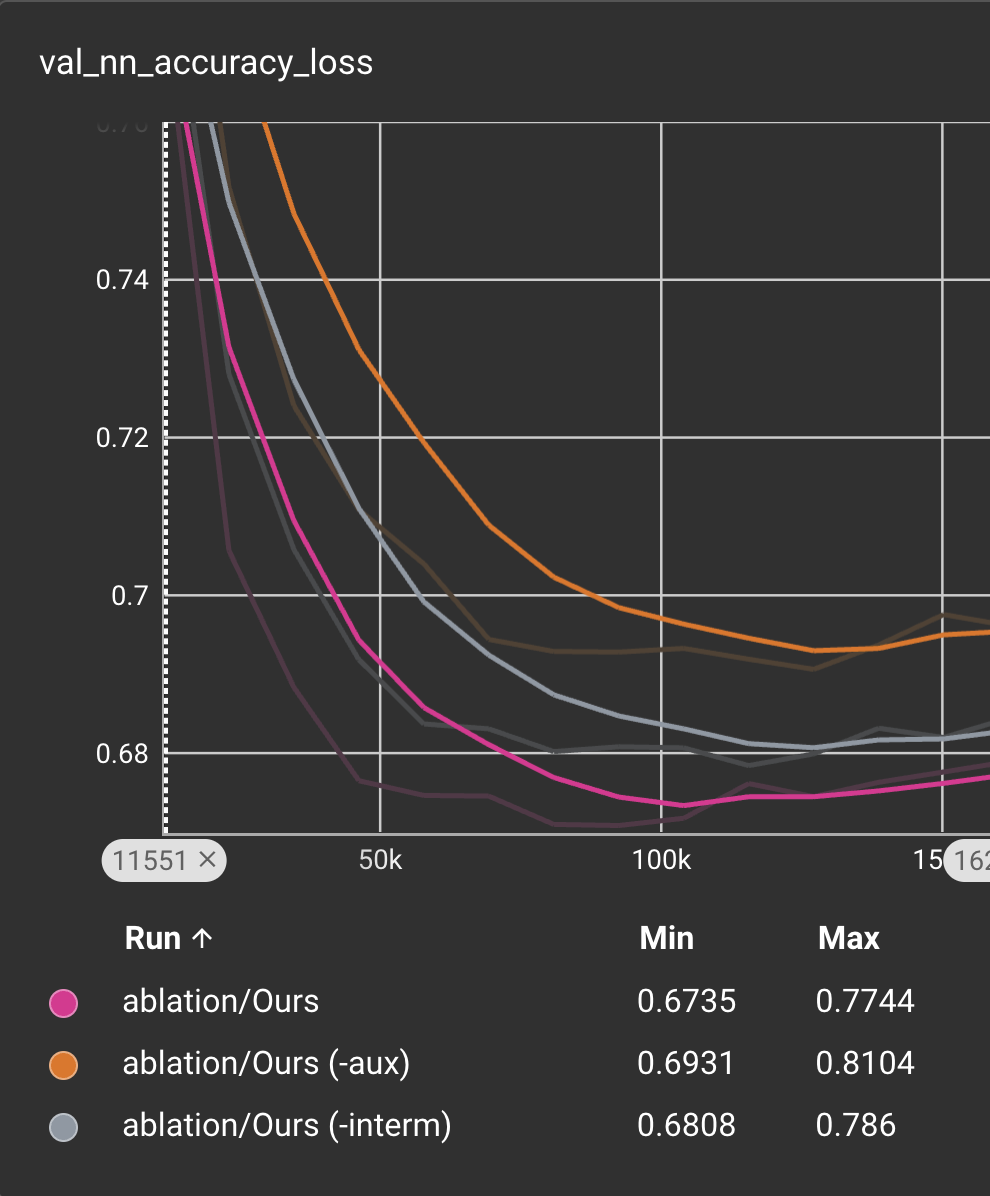}
     \caption{NN accuracy over the validation set}
     \label{fig:ablation}
 \end{subfigure}
 \hfill  

     \caption{We compare the proposed ablations on the NN accuracy metric over the whole dataset as well as on two specific syntactic classes.}
   \centering 

\end{figure}

As shown in Figure \ref{fig:ablation}, using $\partial \gP'$ (Ours) achieves higher NN accuracy. This shows the benefit of learning the auxiliary training task. Meanwhile, ablating the auxiliary task (-aux) and ablating the intermediate node (-interm) does not have meaningful difference, indicating our architecture is robust to graph edits which are semantically equivalent. To understand the comparative advantage vs disadvantage of the auxiliary training task, consider the two examples in Figure \ref{fig:nn}. The first example is equivalent to learning a single-step backward reaction prediction on \textit{forward} templates\footnote{For some templates, the forward template is one-to-one. For others, applying the backward template results in an ill-defined precursor, due to the many-to-one characteristic of these templates.}. Our model clearly benefits from the auxiliary training task, which provides additional examples for learning the backward reaction steps. However, our model fares worse on predicting the first reactant of the top reaction. This may be due to competing resources. Despite the task being the same (and the set of forward templates are fixed), the model has to allocate sufficient capacity for the auxiliary task, whose output domain is much higher dimensional than $\gB$. Ensuring positive transfer from learning the auxiliary task is an interesting extension for future work.

{
\subsection{{Ablation Study: Linear Training}}\label{app:ablate-simple}
}

{

\begin{table}[!htb]
\caption{We follow the same setup as Table \ref{tab:1}, but retrain Ours using a dataset constructed with parameter $k=4$ (whereas we used $k=3$ in Table \ref{tab:1}) to match the linear sampling strategy models, which we refer to as Ours:MC. We evaluate models trained using sampling constants 1 and 10, as described in \ref{sec:4-1-3}.}
\label{tab:mc-analog}
\centering
\small

\resizebox{\textwidth}{!}{

\begin{tabular}{@{}llccccccccc@{}}
\toprule
         &                           &                           & \multicolumn{3}{c}{Avg. Sim. $\uparrow$}                                          & \multicolumn{3}{c}{SA $\downarrow$}                                               & \multicolumn{2}{c}{Diversity $\uparrow$}              \\
Dataset  & Method                    & RR $\uparrow$             & Top-1                     & Top-3                     & Top-5                     & Top-1                     & Top-3                     & Top-5                     & Top-3                     & Top-5                     \\ \midrule
Test Set & Ours ($\recognizer$)      & \multicolumn{1}{l}{56\%}  & \multicolumn{1}{l}{0.827} & \multicolumn{1}{l}{0.633} & \multicolumn{1}{l}{0.555} & \multicolumn{1}{l}{3.100} & \multicolumn{1}{l}{3.019} & \multicolumn{1}{l}{2.918} & \multicolumn{1}{l}{0.543} & \multicolumn{1}{l}{0.628} \\
         & Ours:MC1 ($\recognizer$)  & 36\%                      & 0.732                     & 0.564                     & 0.513                     & 3.048                     & 2.913                     & 2.844                     & 0.609                     & 0.665                     \\
         & Ours:MC10 ($\recognizer$) & 65\%                      & 0.869                     & 0.658                     & 0.609                     & 3.163                     & 3.000                     & 2.928                     & 0.558                     & 0.610                     \\ \midrule
ChEMBL   & Ours ($\recognizer$)      & 7.6\%                     & 0.531                     & 0.443                     & 0.396                     & 2.544                     & 2.510                     & 2.460                     & 0.675                     & 0.727                     \\
         & Ours (MCMC)               & 9.2\%                     & 0.532                     & 0.486                     & 0.432                     & 2.364                     & 2.310                     & 2.263                     & 0.765                     & 0.759                     \\
         & Ours:MC1 (MCMC)           & \multicolumn{1}{l}{2.0\%} & \multicolumn{1}{l}{0.406} & \multicolumn{1}{l}{0.337} & \multicolumn{1}{l}{0.289} & \multicolumn{1}{l}{2.604} & \multicolumn{1}{l}{2.563} & \multicolumn{1}{l}{2.439} & \multicolumn{1}{l}{0.756} & \multicolumn{1}{l}{0.767} \\
         & Ours:MC10 (MCMC)          & \multicolumn{1}{l}{8.5\%} & \multicolumn{1}{l}{0.519} & \multicolumn{1}{l}{0.421} & \multicolumn{1}{l}{0.367} & \multicolumn{1}{l}{2.644} & \multicolumn{1}{l}{2.420} & \multicolumn{1}{l}{2.331} & \multicolumn{1}{l}{0.618} & \multicolumn{1}{l}{0.640} \\ \bottomrule
\end{tabular}

}

\end{table}

\subsubsection{Results on Synthesizable Analog Generation} 

We study whether the efficiency of a sampling-based training strategy comes at a cost of performance. We make two observations from the results in Table \ref{tab:mc-analog}.

\textbf{Constant factor matters.} The constant multiplier $C$ from $\mathcal{D}_0$ to $\mathcal{D}$ not only determines how much data the model sees each pass for the sake of efficiency. It is also be a parameter controlling the tradeoff between over-representing larger vs smaller templates. If it is larger than the largest number of masks (Table \ref{tab:summary}), the Linear strategy is essentially deactivated, since all masks will be used. At a lower value, only small programs with at most the number of masks as $C$ are fully represented. Medium-to-larger programs in $\mathcal{D}_0$ are under-represented, at the rate of the fraction of total masks that $C$ constitutes for its template class. 

\textbf{Performance boost in-distribution.} We find that for $C=10$, the performance is \textit{better} across reconstruction, similarity and diversity with comparable SA to the standard training. Meanwhile for $C=1$, the performance declines sharply. It is likely that standard training is overfitting to masks from larger programs, resulting in poorer generalization. Meanwhile, $C=10$ downsamples those programs, and its sampling can be viewed as data-level regularization against overfitting.

\textbf{Slight performance drop out-of-distribution.} We find both $C=1$ and $C=10$ underperform compared to standard training. For $C=10$, reconstruction and Top-1 similarity are actually comparable, but its similarity, SA and diversity are noticeably worse than standard training. Since ChEMBL feature predominantly unsynthesizable molecules, it is likely that the distribution of molecular fingerprints better reflect those outputs of the more complex programs in $\mathcal{D}_0$, which are downweighted by higher values of $C$.

\subsubsection{Results on Synthesizable Molecular Design}
\begin{figure}[htbp]
\caption{We select the first Oracle from each Table in App. \ref{app:full-results} to compare Ours with Ours (EP). Aside from the ablation networks, we use the same experimental settings as Table \ref{tab:average-results}.}
\label{tab:ep-tdc}
  \centering
  \subfloat[][]{
  
  \resizebox{\textwidth}{!}{

\begin{tabular}{@{}ll|cccccccccc|cccccccccc@{}}
\toprule
            &           & \multicolumn{10}{c|}{GSK3$\beta$}                                                                                                                                                                                                  & \multicolumn{10}{c}{Median 1}                                                                                                                                                                                                      \\ \midrule
            &           & \multicolumn{1}{l|}{}             & \multicolumn{3}{l|}{Top 1}                                           & \multicolumn{3}{l|}{Top 10}                                          & \multicolumn{3}{l|}{Top 100}                     & \multicolumn{1}{c|}{}             & \multicolumn{3}{c|}{Top 1}                                          & \multicolumn{3}{c|}{Top 10}                                           & \multicolumn{3}{c}{Top 100}                      \\
            & category  & \multicolumn{1}{l|}{Oracle Calls} & Score         & SA             & \multicolumn{1}{c|}{AUC}            & Score         & SA             & \multicolumn{1}{c|}{AUC}            & Score          & SA             & AUC            & \multicolumn{1}{c|}{Oracle Calls} & Score        & SA             & \multicolumn{1}{c|}{AUC}            & Score          & SA             & \multicolumn{1}{c|}{AUC}            & Score          & SA             & AUC            \\ \midrule
Ours (MC)   & synthesis & \multicolumn{1}{c|}{5056}         & \textbf{0.98} & \textbf{2.045} & \multicolumn{1}{c|}{\textbf{0.923}} & \textbf{0.97} & \textbf{2.294} & \multicolumn{1}{c|}{\textbf{0.893}} & 0.942          & \textbf{2.294} & \textbf{0.814} & \multicolumn{1}{c|}{7949}         & \textbf{0.4} & 4.12           & \multicolumn{1}{c|}{0.356}          & \textbf{0.342} & \textbf{3.902} & \multicolumn{1}{c|}{0.304}          & 0.295          & \textbf{4.013} & \textbf{0.256} \\
Ours (MC10) & synthesis & \multicolumn{1}{c|}{}             &               &                & \multicolumn{1}{c|}{}               &               &                & \multicolumn{1}{c|}{}               &                &                &                & \multicolumn{1}{c|}{8045}         & \textbf{0.4} & \textbf{3.353} & \multicolumn{1}{c|}{0.357}          & 0.344          & 4.593          & \multicolumn{1}{c|}{0.297}          & \textbf{0.302} & 4.44           & 0.247          \\
Ours        & synthesis & \multicolumn{1}{c|}{4886}         & \textbf{0.98} & \textbf{2.045} & \multicolumn{1}{c|}{0.891}          & 0.967         & 2.302          & \multicolumn{1}{c|}{0.848}          & \textbf{0.944} & 2.27           & 0.778          & \multicolumn{1}{c|}{8303}         & \textbf{0.4} & \textbf{3.353} & \multicolumn{1}{c|}{\textbf{0.371}} & \textbf{0.342} & 4.161          & \multicolumn{1}{c|}{\textbf{0.305}} & 0.298          & 4.256          & 0.252          \\ \bottomrule
\end{tabular}

}
  }%
  \qquad
  \subfloat[][]{
  
    \resizebox{\textwidth}{!}{

\begin{tabular}{@{}ll|cccccccccc|cccccccccc@{}}
\toprule
            &           & \multicolumn{10}{c|}{Osimertinib MPO}                                                                                                                                                                                              & \multicolumn{10}{c}{Perindopril MPO}                                                                                                                                                                                                 \\ \midrule
            &           & \multicolumn{1}{l|}{}             & \multicolumn{3}{c|}{Top 1}                                           & \multicolumn{3}{c|}{Top 10}                                          & \multicolumn{3}{c|}{Top 100}                     & \multicolumn{1}{l|}{}             & \multicolumn{3}{c|}{Top 1}                                            & \multicolumn{3}{c|}{Top 10}                                           & \multicolumn{3}{c}{Top 100}                      \\
            & category  & \multicolumn{1}{c|}{Oracle Calls} & Score         & SA             & \multicolumn{1}{c|}{AUC}            & Score         & SA             & \multicolumn{1}{c|}{AUC}            & Score          & SA             & AUC            & \multicolumn{1}{c|}{Oracle Calls} & Score          & SA             & \multicolumn{1}{c|}{AUC}            & Score          & SA             & \multicolumn{1}{c|}{AUC}            & Score          & SA             & AUC            \\ \midrule
Ours (MC)   & synthesis & \multicolumn{1}{c|}{9402}         & 0.865         & 2.282          & \multicolumn{1}{c|}{0.830}          & 0.853         & \textbf{2.187} & \multicolumn{1}{c|}{0.813}          & 0.841          & 2.189          & 0.771          & \multicolumn{1}{c|}{10000}        & 0.572          & \textbf{3.101} & \multicolumn{1}{c|}{0.541}          & 0.567          & \textbf{3.072} & \multicolumn{1}{c|}{0.521}          & 0.555          & 3.077          & 0.486          \\
Ours (MC10) & synthesis & \multicolumn{1}{c|}{5056}         & \textbf{0.98} & \textbf{2.045} & \multicolumn{1}{c|}{\textbf{0.923}} & \textbf{0.97} & 2.294          & \multicolumn{1}{c|}{\textbf{0.893}} & \textbf{0.942} & \textbf{2.294} & \textbf{0.814} & \multicolumn{1}{c|}{10000}        & 0.596          & 3.263          & \multicolumn{1}{c|}{0.542}          & 0.574          & 3.147          & \multicolumn{1}{c|}{0.523}          & 0.556          & 3.058          & \textbf{0.489} \\
Ours        & synthesis & \multicolumn{1}{c|}{10000}        & 0.859         & 2.263          & \multicolumn{1}{c|}{0.826}          & 0.847         & 2.21           & \multicolumn{1}{c|}{0.81}           & 0.832          & 2.249          & 0.769          & \multicolumn{1}{c|}{10000}        & \textbf{0.622} & 3.338          & \multicolumn{1}{c|}{\textbf{0.547}} & \textbf{0.591} & 3.378          & \multicolumn{1}{c|}{\textbf{0.524}} & \textbf{0.558} & \textbf{3.137} & 0.485          \\ \bottomrule
\end{tabular}

}
  }
\end{figure}
\textbf{Depends on the task.} We also include preliminary results on synthesizable molecular design. We find the results hold up for the MC models. Encouragingly, for the difficult task of Osimertinib MPO, we see for $C=10$, the results are substantially better. At the same time, Ours remain better for Perindopril MPO, which suggests each training strategy may suit different tasks. We also see for easier tasks like GSK and Median 1, the $C=1$ model and Ours are essentially interchangeable. This suggests even a low reconstruction accuracy suffices for these Oracles.

\textbf{Implications.} The stratified sampling is designed to enhance supervised policy learning on larger programs while preventing the combinatorial explosion in Algo. \ref{alg:pretrain} and to some extent, overfitting. However, this may benefit easier tasks (e.g. distributions of synthesizable molecules) while disadvantaging harder tasks. Thus, $C$ should be tuned for optimal trade-off between efficiency and downstream synthesizable analog generation performance.

}
\subsection{Model Architecture}
\noindent We opt for two Graph Neural Networks (for $\Phi, \Omega$), each with 5 modules. Each module uses a TransformerConv layer  \citep{shi2020} (we use 8 attention heads), a ReLU activation, and a Dropout layer. We adopt sinusoidal positional embeddings via numbering nodes using the postorder traversal (to preserve the pairwise node relationships for the same skeleton). Then, we pretrain $\Phi, \Omega$ with $\gD$.

{
\subsection{{Training \& Convergence Analysis}}\label{app:convergence}
\begin{figure}[h!]
 \hfill   
   \begin{subfigure}[b]{0.15\textwidth}
     \centering
     \raisebox{0.6\height}{
     \includegraphics[width=\textwidth]{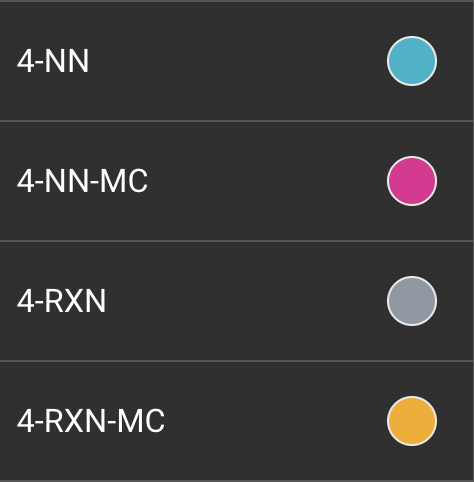}\     
     }
     \caption{Legend}
     \label{fig:nn}
 \end{subfigure}
 \hfill   
   \begin{subfigure}[b]{0.43\textwidth}
     \centering
     \includegraphics[width=\textwidth]{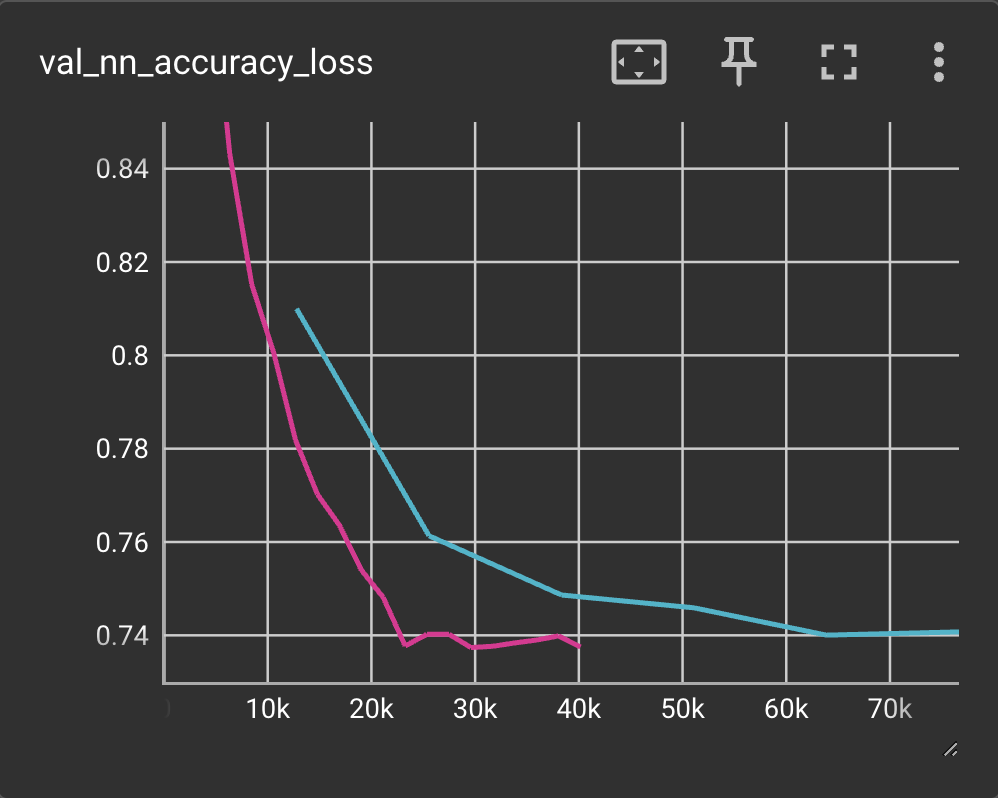}
     \caption{NN training step vs NN accuracy}
     \label{fig:nn-convergence}
 \end{subfigure}
 \hfill   
    \begin{subfigure}[b]{0.365\textwidth}
     \centering
     \includegraphics[width=\textwidth]{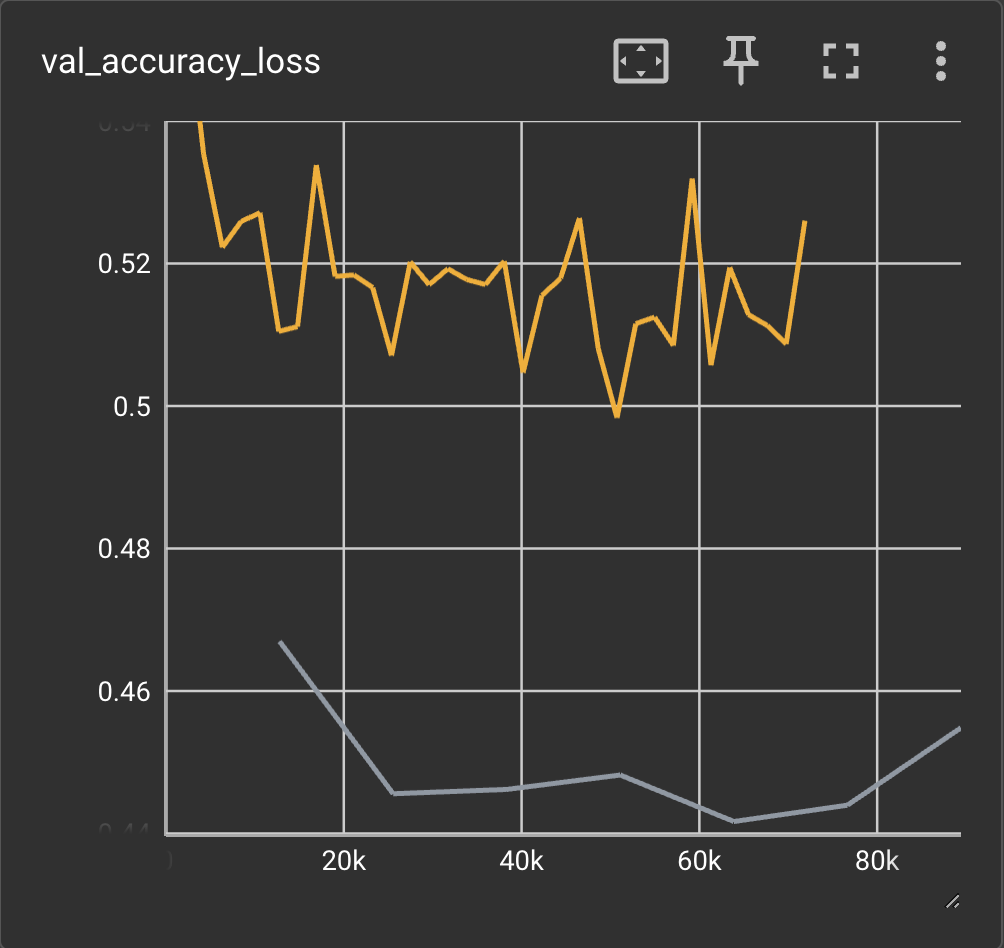}
      \caption{RXN training step vs accuracy}
     \label{fig:rxn-convergence}
 \end{subfigure}
 \hfill   

     \caption{We plot the number of training steps needed to converge our models under the standard and linear (-MC) training strategies for $C=1$.}
     \label{fig:convergence}
   \centering 

\end{figure}
We elaborate on \ref{sec:4-1-3} further with a quantitative comparison of training costs with SynNet \cite{gao2022}. The key difference is $\mathcal{D}_{\text{synnet}}$ batches by \textit{the size of a synthetic tree}, whereas we batch by the synthetic trees (in program form).\\\\
\textbf{SynNet Training Complexity:} SynNet serializes the construction of a synthetic tree, so a training epoch does $O(\sum_{\mathcal{D}_0} |T|)$ passes, where $|T|$ is the number of nodes (or number of edges, as they are different by one). The cost of a MLP forward and backward pass for SynNet is $O(L\cdot H^2+F\cdot H)$. The total complexity per epoch is $O((\sum |T|) (L\cdot H^2+F\cdot H))$. \\\\
\textbf{SynthesisNet Linear Training Complexity:} As discussed in \ref{sec:4-1-3}, a training epoch does $O(\mathcal{D}_0)$ passes, where the constant factor can be adjusted. The cost of a GNN forward and backward pass on a tree $T$ is $O(L\cdot |T| \cdot (F \cdot H+H^2))$, where $L$, $F$, $H$ are number of layers, feature and hidden dimension, respectively. The total complexity per epoch is $O(L\cdot (\sum |T|) \cdot (F \cdot H+H^2))$ since we train on the forest of trees over $\mathcal{D}_0$. This is equivalent to $O((\sum |T|) (L\cdot H^2+F\cdot H))$.

We can conclude the per-epoch complexity following the linear training strategy is equivalent to that of SynNet's. However, what matters in practice is the convergence rate, so we also include a quantitative comparison between convergence plots. We find that even SynthesisNet's standard training strategy is practically equivalent to SynNet in the number of passes needed to converge the model. As a disclaimer, we only include SynthesisNet's numbers using the default setting of $k=4$, where $|\mathcal{D}_0|\approx 135k$ and $|\mathcal{D}|\approx 818k$. We also show results of linear training with constant factor of $1$ (i.e. $|\mathcal{D}|=|\mathcal{D}_0|$).

\textbf{Convergence Comparison:} Both SynNet and SynthesisNet uses a batch size of 64. We see from Figure \ref{fig:convergence} that SynthesisNet requires $\approx 170k$ training steps (batches) to converge (combining both BB and RXN networks) following standard training. Meanwhile, we refer readers to Figure 13 of \cite{gao2021}, where the convergence plots show at least (with the most generous interpretation) 1M steps needed to converge \textit{each} of the action, reactant1, reaction and reactant2 networks. Pooling the training of all networks, we give a \textit{very} generous estimate of $1M$ batches for SynNet to converge. What's left is to figure the average scaling size factor from a batch of trees (ours) to a batch of tree nodes (SynNet), which computes as $818k/135k \approx 6$, which implies $\approx 6\cdot 170k \approx 1M$ steps. We can conclude the SynthesisNet with \textit{standard} training is comparable if not more efficient in the number of training steps to converge the model. Should efficiency be of further concern, we suggest using the Linear Training strategy, and adjusting the sample constant factor accordingly.

\textbf{Training and Inference Time:} Converging the RXN and BB networks took us $\approx$ 1 and 5 hours on a single NVIDIA RTX A6000. A single inference call to the surrogate takes a few seconds.

}

\subsection{Attention Visualization}\label{app:attn}
We elucidate how our policy network leverages the full horizon of the MDP to dynamically adjust the propagation of information throughout the decoding process. Since our decoding algorithm decodes once for every topological order of the nodes, the actual attention dynamics can vary significantly. Thus, we show a prototypical decoding order where:
\begin{enumerate}
    \item All reactions are decoded before building blocks.
    \item If decoding a reaction, the reaction node which $\pi_\gR$ predicts with the highest probability is decoded.
    \item If decoding a building block, the node where the embedding from $\pi_\gB$ has minimal distance to a building block is decoded.
\end{enumerate}

\begin{figure}[htbp]
\centering
\hfill
\begin{subfigure}{0.45\textwidth}
  \raisebox{0.15\height}{
  \hspace{-1em}
 \includegraphics[width=\textwidth]{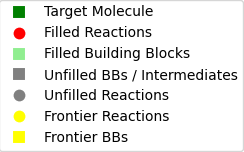}
 }
 \caption{Legend}
 \label{fig:legend}
\end{subfigure}
\begin{subfigure}{0.45\textwidth}
 \centering
 \includegraphics[width=\textwidth]{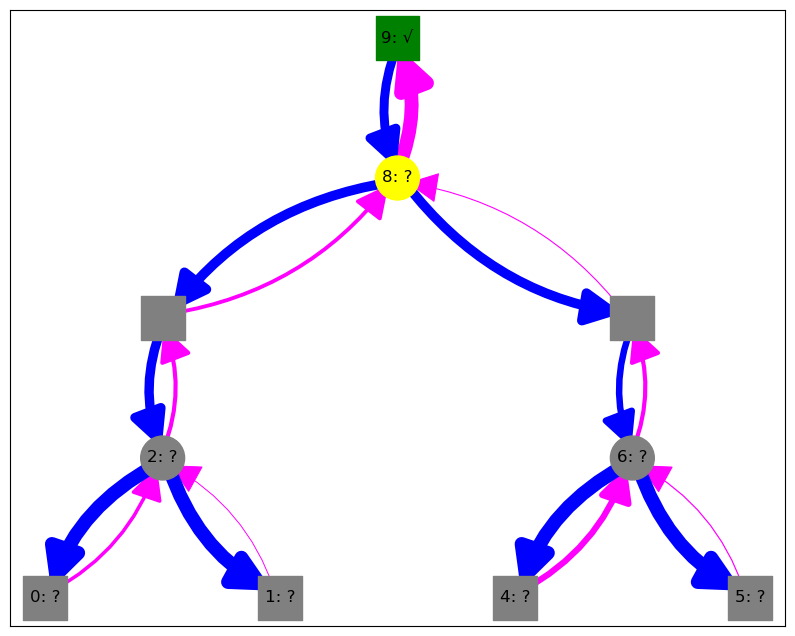}
 \caption{Step 1, decoding candidates: 8}
 \label{fig:attention-1-1}
\end{subfigure}
\hfill   
\begin{subfigure}{0.45\textwidth}
 \centering
 \includegraphics[width=\textwidth]{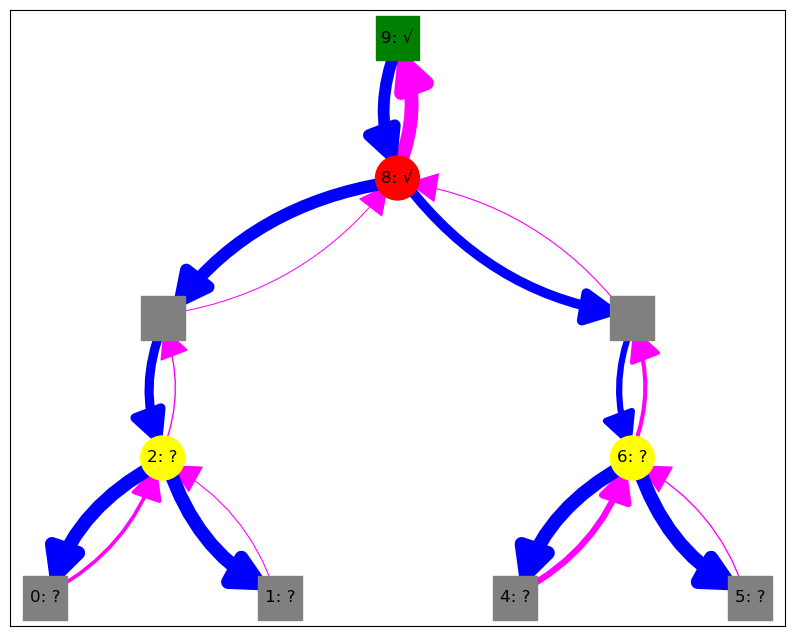}
 \caption{Step 2, decoding candidates: 2, 6}
 \label{fig:attention-1-2}
\end{subfigure}
\hfill   
\begin{subfigure}{0.45\textwidth}
 \centering
 \includegraphics[width=\textwidth]{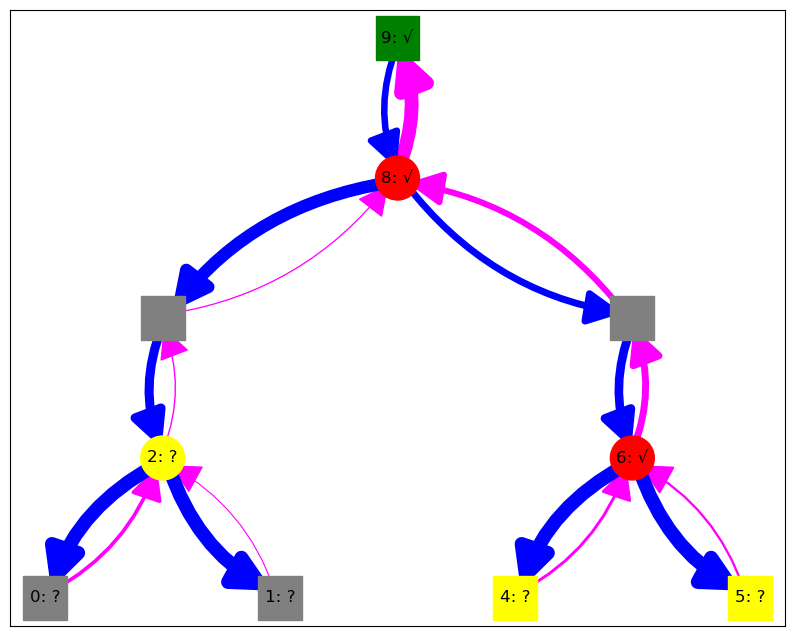}
 \caption{Step 3, decoding candidates: 2}
 \label{fig:attention-1-3}
\end{subfigure}
\hfill   
\begin{subfigure}{0.45\textwidth}
 \centering
 \includegraphics[width=\textwidth]{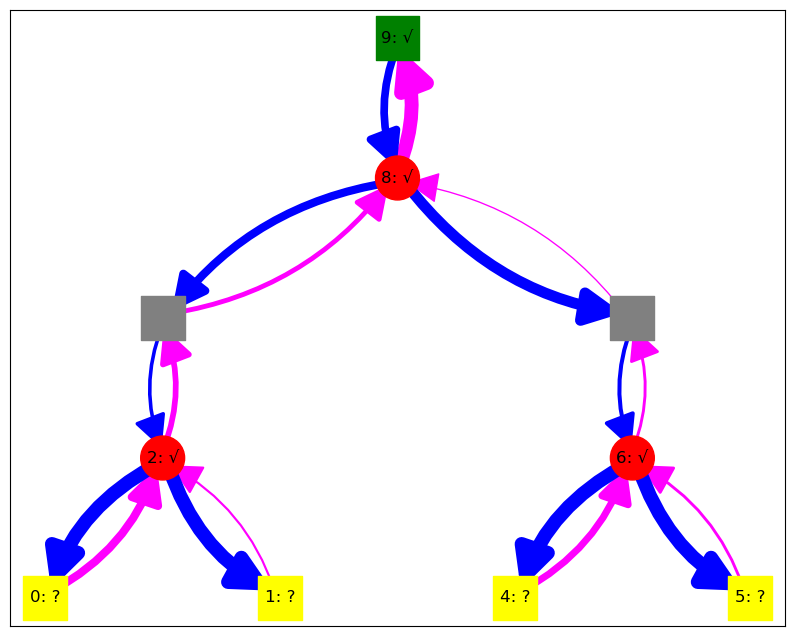}
 \caption{Step 4, decoding candidates: 0, 1, 4, 5}
 \label{fig:attention-1-4}
\end{subfigure}
\hfill   
\begin{subfigure}{0.45\textwidth}
 \centering
 \includegraphics[width=\textwidth]{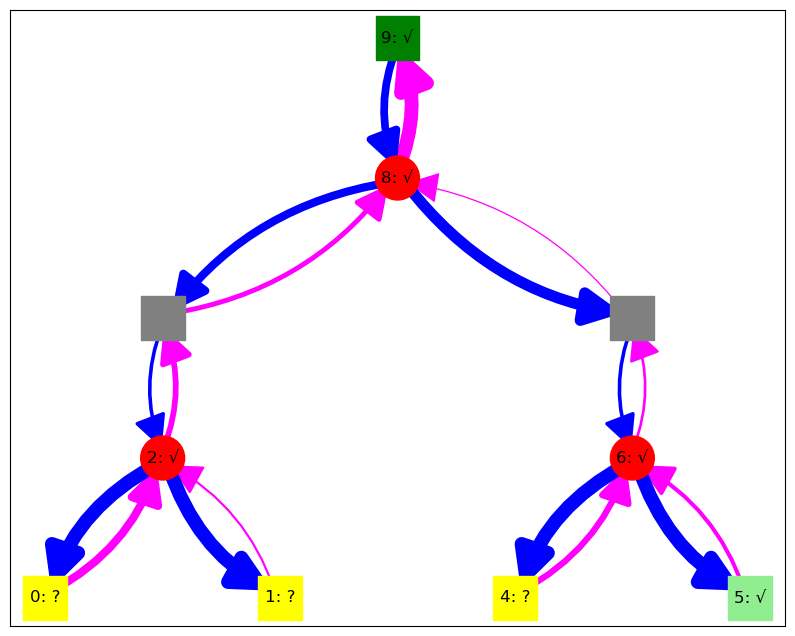}
 \caption{Step 5, decoding candidates: 0, 1, 4}
 \label{fig:attention-1-5}
\end{subfigure}
\hfill   
\begin{subfigure}{0.45\textwidth}
 \centering
 \includegraphics[width=\textwidth]{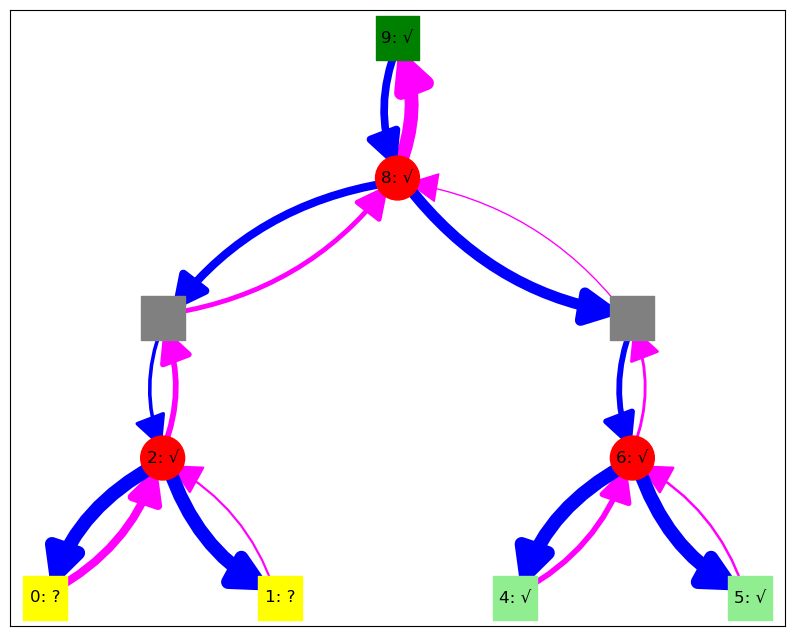}
 \caption{Step 6, decoding candidates: 0, 1}
 \label{fig:attention-1-6}
\end{subfigure}
\hfill   
\begin{subfigure}{0.45\textwidth}
 \centering
 \includegraphics[width=\textwidth]{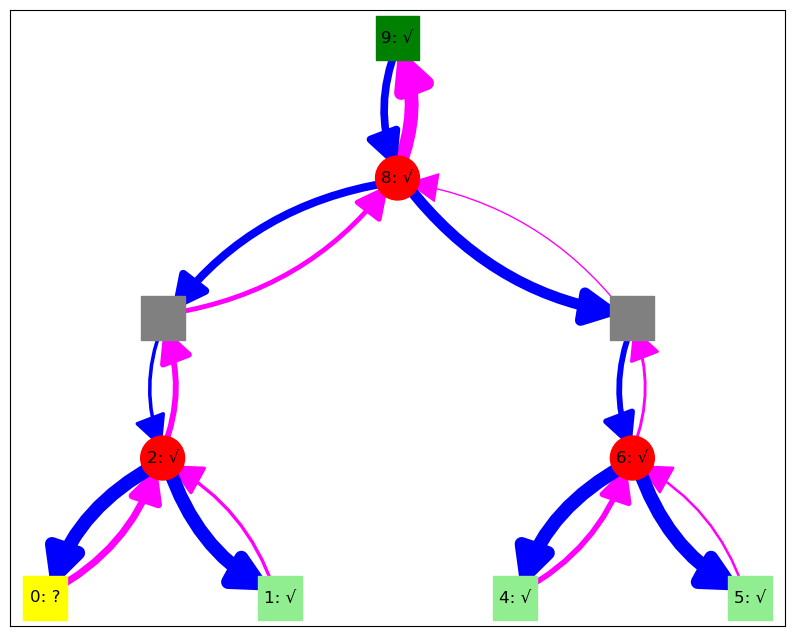}
 \caption{Step 7, decoding candidates: 0}
 \label{fig:attention-1-7}
\end{subfigure}
\hfill   
\caption{Case Study of Attention Flow}
\label{fig:attention-1}
\end{figure}

In \citep{shi2020}, each TransformerConv layer $l$ produces an attention weight for each edge, $[\alpha^{(l)}_{i,j}]$ where $\sum_{j} \alpha^{(l)}_{i,j} = 1$. We average over all layers to obtain the mean attention weight for each directed edge, i.e., we set the thickness of each edge $(i,j)$ in each subfigure of {Figure \ref{fig:attention-1}} to be proportional $\sum_l \alpha^{(l)}_{i,j}$.

We make some generation observations:
\begin{itemize}
    \item The information flow along child-parent edges indicate usage of the full horizon. This is the main feature of our approach compared to traditional search methods like retrosynthesis.
    \item Our positional embeddings enables asymmetric modeling. SMIRKS templates specify the order of reactants and is usually not arbitrary. We observe that more often than not, the parent attends to its left child more than the right child. This may be a consequence of template definition conventions, where the first reactant is the major precursor. The subtree under the node more likely to be the major precursor is more important for predicting the reaction.
\end{itemize}

Now, we do a detailed walkthrough the 7-step decoding process to understand the evolution of the information flow. Each subfigure corresponds to the state of the MDP after a number of decoding steps, with the candidates of decoding colored in yellow. The attention scores are computed during the inference of $\Phi$ or $\Omega$ and averaged.

\begin{enumerate}
    \item In Figure \ref{fig:attention-1-1}, we see that $8$ attends significantly to the target, unsurprisingly. $8$ also attends to both its children, and attends more to its left child, which is a prior consistent with our general observation.
    \item In Figure \ref{fig:attention-1-2}, we see that after a specific reaction is instantiated at $8$, the attention dynamics somewhat change. The edge from $8$ to its left child thickens, while the edge from the left child to $8$ thins. This is likely because now that the identity of $8$ is known, it no longer needs to attend to its left child. The reciprocal relationship now intensifies, as the first reactant of $8$ now attends to $8$.
    \item In Figure \ref{fig:attention-1-3}, after the reaction at $6$ is decoded, we see the information propagate back up the tree and to the other subtree to inform $2$. We see the edge along the path from $6-8$ thickens, indicating the representation of $8$ is informed with new information, and in turn propagates it to $2$.
    \item In Figure \ref{fig:attention-1-4}, after the reaction at $2$ is decoded, we see the same phenomenon happen, where the information flow again propagates back up and to the other subtree. However, we see this comes with a tradeoff, as $6$ attends to its parent less, and instead reverts to its original attention strength to its children. We hypothesize the identity of $2$ has a strong effect on the posterior of $6$. This is an example where branching out to try more possible orders of decoding would facilitate a more complete algorithm.    
    \item In Figure \ref{fig:attention-1-5}, we see how determining $5$ causes $6$ to attend more to $5$ than it does to its parent. Knowledge of $5$ allows the explaining away of $4$.
    \item In Figure \ref{fig:attention-1-6}, we note instances of a general phenomenon: the second reactant is decoded followed by the first. Empirically, the distribution of the second reactant has lower entropy than the first. $4$ was inferred after $5$ as the knowledge of its parent reaction and sibling reactant likely constrains its posterior significantly.
    \item In Figure \ref{fig:attention-1-7}, we see a similar phenomenon where the representation of $2$ attends slightly more to $1$ after it is decoded.
\end{enumerate}
In summary, the syntax structure of the full horizon is crucial during the decoding process. The attention scores allow us to visualize the dynamic propagation of information as nodes are decoded. Our observations highlights the flexibility of this approach compared to an infinite horizon formulation.

\begin{table}[h]
\caption{Hyperparameters of our GA.}
\label{tab:ga_param}
\centering
\begin{tabular}{llc}
\toprule
& Parameter & Value \\
\midrule
\multirow{6}{*}{General} & Max.\ generations & 200 \\  
& Population size & 128 \\ 
& Offspring size & 512 \\ 
& Seed initialization & random \\
& Fingerprint size & 2048 \\
& Early stopping warmup & 30 \\ 
& Early stopping patience & 10 \\ 
& Early stopping $\Delta$ & 0.01 \\ 
\midrule 
\multirow{4}{*}{Semantic evolution}
& Parent selection prob.\ of $i$  & $\propto$ $(\mathrm{rank}(i) + 10)$ \\
& Num.~crossover bits $n_{\text{cross}}$ & $\mathcal{N}(1024, 205)$ \\ 
& Num.~mutate bits $n_{\text{flip}}$ & 12 \\ 
& Prob.~mutate bits $p_{\text{flip}}$ & 0.5 \\
\midrule 
Syntactic mutation
& Num.~tree edits $n_{\text{edit}}$ & $\mathcal{U}\{1, 2, 3\}$  \\
\midrule 
\multirow{2}{*}{Surrogate}
& Max.\ topological orders & 5 \\ 
& Sampling strategy & greedy \\ 
\bottomrule
\end{tabular}
\end{table}

\section{Genetic Algorithm}\label{app:ga}

Our genetic algorithm (GA) is designed to mimic SynNet's \citep{gao2021}, and settings are given in Table \ref{tab:ga_param}. We fix the same number of offspring fitness evaluations per generation to ensure a fair comparison, strategically allocating the evaluations between offspring generated using semantic evolution and those generated using syntactic mutation.



\subsection{Semantic Evolution}\label{app:ga-sem}

Given two parents $\vx_1$ and $\vx_2$, semantic evolution samples a child $\vx_*$ as follows. We combine $n_{\text{cross}}$ random bits from $\vx_1$ and the other $2048 - n_{\text{cross}}$ bits from $\vx_2$ and then, with probability $p_{\text{flip}}$, flipping $n_{\text{flip}}$ random bits of the crossover result.

\subsection{Syntactic Mutation} \label{app:ga-syn}

Given a child $\vx_*$ from Appendix \ref{app:ga-sem}, syntactic mutation performs $n_{\text{edit}}$ edits on $T = \recognizer{}(\vx_*)$ to obtain a syntactic analog $T$. With equal probability, each edit either adds or removes a random leaf. To do so, we enumerate all possible additions and removals, and ignore the ones that produce an empty tree or a tree with more than 4 reaction nodes. The edit is uniformly sampled from all such choices, or no operation is performed if no viable choices exist. Using the surrogate, the siblings $(\vx_*, T)$ and $(\vx_*, T)$ are then turned into two fingerprints, and one of with the higher expected improvement under a Gaussian process (GP) is selected. Our GP uses a radial basis function kernel with length scale 1 and is fitted using the population and offspring from the preceding generation.


\subsection{Surrogate Checkpoint}

The surrogate checkpoint was trained as described in Appendix \ref{app:policy-network}. To lower the runtime of the GA, we only reconstruct using a random subset of the input skeleton's possible topological orders. For each topological order, we follow a greedy decoding scheme where reactions are decoded before building blocks, as described in Appendix \ref{app:attn}. 

\section{Full Results on TDC Oracles}\label{app:full-results}

\begin{table}[!ht]
\centering
\caption{Guacamol structural target-directed benchmarks: Median 1 \& 2 (average similarity to multiple molecules) and Celecoxib Rediscovery (hit expansion around Celecoxib).}
\renewcommand*{\arraystretch}{1.3}
\resizebox{\textwidth}{!}{%
\begin{tabular}{ll|rlllllllll|rlllllllll|ccclcclccl}
\hline
 & & \multicolumn{10}{c|}{Median 1} & \multicolumn{10}{c|}{Median 2} & \multicolumn{10}{c}{Celecoxib} \\ \hline
 & & \multicolumn{1}{c|}{} & \multicolumn{3}{c|}{Top 1} & \multicolumn{3}{c|}{Top 10} & \multicolumn{3}{c|}{Top 100} & \multicolumn{1}{c|}{} & \multicolumn{3}{c|}{Top 1} & \multicolumn{3}{c|}{Top 10} & \multicolumn{3}{c|}{Top 100} & \multicolumn{1}{c|}{} & \multicolumn{3}{c|}{Top 1} & \multicolumn{3}{c|}{Top 10} & \multicolumn{3}{c}{Top 100} \\
 & category & \multicolumn{1}{c|}{Oracle Calls} & \multicolumn{1}{c}{Score} & \multicolumn{1}{c}{SA} & \multicolumn{1}{c|}{AUC} & \multicolumn{1}{c}{Score} & \multicolumn{1}{c}{SA} & \multicolumn{1}{c|}{AUC} & \multicolumn{1}{c}{Score} & \multicolumn{1}{c}{SA} & \multicolumn{1}{c|}{AUC} & \multicolumn{1}{c|}{Oracle Calls} & \multicolumn{1}{c}{Score} & \multicolumn{1}{c}{SA} & \multicolumn{1}{c|}{AUC} & \multicolumn{1}{c}{Score} & \multicolumn{1}{c}{SA} & \multicolumn{1}{c|}{AUC} & \multicolumn{1}{c}{Score} & \multicolumn{1}{c}{SA} & \multicolumn{1}{c|}{AUC} & \multicolumn{1}{c|}{Oracle Calls} & Score & SA & \multicolumn{1}{c|}{AUC} & Score & SA & \multicolumn{1}{c|}{AUC} & Score & SA & \multicolumn{1}{c}{AUC} \\ \hline
synnet & synthesis & \multicolumn{1}{r|}{5245} & 0.245 (15|3) & 4.066 (17|4) & \multicolumn{1}{l|}{0.237 (15|3)} & 0.228 (13|3) & 3.807 (11|3) & \multicolumn{1}{l|}{0.219 (12|2)} & 0.198 (14|3) & 3.687 (6|3) & 0.188 (12|2) & \multicolumn{1}{r|}{4350.6} & 0.259 (11|3) & 2.364 (5|3) & \multicolumn{1}{l|}{0.253 (8|2)} & 0.244 (10|3) & {\ul 2.473 2|1} & \multicolumn{1}{l|}{0.237 (8|2)} & 0.214 (11|3) & {\ul \textbf{2.459 (1|1)}} & 0.206 (8|2) & \multicolumn{1}{c|}{7377.4} & 0.526 (9|3) & 2.234 (14|4) & \multicolumn{1}{l|}{0.487 (9|3)} & 0.479 (9|3) & 2.262 (7|3) & \multicolumn{1}{l|}{0.443 (8|3)} & 0.411 (9|3) & 2.466 (12|4) & 0.376 (8|3) \\
\hline
\end{tabular}

}
\label{tab:full-1}
\end{table}

\begin{table}[!ht]
\caption{Bioactivity Oracles for GSK3B, JNK3, and DRD2}
\centering
\renewcommand*{\arraystretch}{1.3}
\resizebox{\textwidth}{!}{%
\begin{tabular}{ll|rlllllllll|ccclcclccl|ccclcclccl}
\hline
 & & \multicolumn{10}{c|}{GSK3$\beta$} & \multicolumn{10}{c|}{JNK3} & \multicolumn{10}{c}{DRD2} \\ \hline
 & & \multicolumn{1}{c|}{} & \multicolumn{3}{c|}{Top 1} & \multicolumn{3}{c|}{Top 10} & \multicolumn{3}{c|}{Top 100} & \multicolumn{1}{c|}{} & \multicolumn{3}{c|}{Top 1} & \multicolumn{3}{c|}{Top 10} & \multicolumn{3}{c|}{Top 100} & \multicolumn{1}{c|}{} & \multicolumn{3}{c|}{Top 1} & \multicolumn{3}{c|}{Top 10} & \multicolumn{3}{c}{Top 100} \\
 & category & \multicolumn{1}{c|}{Oracle Calls} & \multicolumn{1}{c}{Score} & \multicolumn{1}{c}{SA} & \multicolumn{1}{c|}{AUC} & \multicolumn{1}{c}{Score} & \multicolumn{1}{c}{SA} & \multicolumn{1}{c|}{AUC} & \multicolumn{1}{c}{Score} & \multicolumn{1}{c}{SA} & \multicolumn{1}{c|}{AUC} & \multicolumn{1}{c|}{Oracle Calls} & Score & SA & \multicolumn{1}{c|}{AUC} & Score & SA & \multicolumn{1}{c|}{AUC} & Score & SA & \multicolumn{1}{c|}{AUC} & \multicolumn{1}{c|}{Oracle Calls} & Score & SA & \multicolumn{1}{c|}{AUC} & Score & SA & \multicolumn{1}{c|}{AUC} & Score & SA & \multicolumn{1}{c}{AUC} \\ \hline
synnet & synthesis & \multicolumn{1}{r|}{9059.4} & 0.932 (8|3) & 2.576 (5|2) & \multicolumn{1}{l|}{0.855 (6|3)} & 0.901 (8|3) & 2.747 (6|3) & \multicolumn{1}{l|}{0.79 (6|3)} & 0.797 (9|3) & 2.897 (6|4) & {\ul 0.656 (7|2)} & \multicolumn{1}{c|}{6240} & 0.798 (7|3) & 2.116 (3|2) & \multicolumn{1}{l|}{{\ul 0.723 (3|1)}} & 0.715 (8|3) & {\ul \textbf{2.172 (1|1)}} & \multicolumn{1}{l|}{0.631 (5|2)} & 0.563 (10|3) & 2.244 (2|2) & 0.466 (9|2) & \multicolumn{1}{c|}{6472.6} & {\ul \textbf{1.0 (1|1)}} & 2.58 (7|3) & \multicolumn{1}{l|}{0.985 (4|3)} & 0.998 (9|3) & 2.806 (8|4) & \multicolumn{1}{l|}{{\ul \textbf{0.969 (1|1)}}} & 0.986 (12|3) & 2.845 (7|4) & 0.898 (4|2) \\
\hline
\end{tabular}
}
\label{tab:full-2}
\end{table}

\begin{table}[!ht]
\centering
\caption{Guacamol multi-objective Oracles for properties of known drugs: Osimertinib, Fexofenadine, Ranolazine.}
\renewcommand*{\arraystretch}{1.3}
\resizebox{\textwidth}{!}{%
\begin{tabular}{ll|clllllllll|clllllllll|rlllllllll}
\hline
 & & \multicolumn{10}{c|}{Osimertinib MPO} & \multicolumn{10}{c|}{Fexofenadine MPO} & \multicolumn{10}{c}{Ranolazine MPO} \\ \hline
 & & \multicolumn{1}{c|}{} & \multicolumn{3}{c|}{Top 1} & \multicolumn{3}{c|}{Top 10} & \multicolumn{3}{c|}{Top 100} & \multicolumn{1}{c|}{} & \multicolumn{3}{c|}{Top 1} & \multicolumn{3}{c|}{Top 10} & \multicolumn{3}{c|}{Top 100} & \multicolumn{1}{c|}{} & \multicolumn{3}{c|}{Top 1} & \multicolumn{3}{c|}{Top 10} & \multicolumn{3}{c}{Top 100} \\
 & category & \multicolumn{1}{c|}{Oracle Calls} & \multicolumn{1}{c}{Score} & \multicolumn{1}{c}{SA} & \multicolumn{1}{c|}{AUC} & \multicolumn{1}{c}{Score} & \multicolumn{1}{c}{SA} & \multicolumn{1}{c|}{AUC} & \multicolumn{1}{c}{Score} & \multicolumn{1}{c}{SA} & \multicolumn{1}{c|}{AUC} & \multicolumn{1}{c|}{Oracle Calls} & \multicolumn{1}{c}{Score} & \multicolumn{1}{c}{SA} & \multicolumn{1}{c|}{AUC} & \multicolumn{1}{c}{Score} & \multicolumn{1}{c}{SA} & \multicolumn{1}{c|}{AUC} & \multicolumn{1}{c}{Score} & \multicolumn{1}{c}{SA} & \multicolumn{1}{c|}{AUC} & \multicolumn{1}{c|}{Oracle Calls} & \multicolumn{1}{c}{Score} & \multicolumn{1}{c}{SA} & \multicolumn{1}{c|}{AUC} & \multicolumn{1}{c}{Score} & \multicolumn{1}{c}{SA} & \multicolumn{1}{c|}{AUC} & \multicolumn{1}{c}{Score} & \multicolumn{1}{c}{SA} & \multicolumn{1}{c}{AUC} \\
\hline
\end{tabular}
}
\label{tab:full-3}
\end{table}

\begin{table}[!ht]
\caption{Guacamol multi-objective Oracles for properties of known drugs: Perindopril, Amlodipine, Sitagliptin, and Zaleplon. Disclaimer: We used version 1.0.0. PMO \cite{gao2022}, where we adopted the numbers from, used 0.3.6. We'd like to mention there were \protect\href{https://github.com/mims-harvard/TDC/pull/171/}{changes} since 0.3.6 that may affect the Sitagliptin and Zaleplon evaluations.}
\centering
\renewcommand*{\arraystretch}{1.3}
\resizebox{\textwidth}{!}{%
\begin{tabular}{ll|rlllllllll|rlllllllll|rlllllllll|rlllllllll}
\hline
 & & \multicolumn{10}{c|}{Perindopril MPO} & \multicolumn{10}{c|}{Amlodipine MPO} & \multicolumn{10}{c|}{Sitagliptin MPO} & \multicolumn{10}{c}{Zaleplon MPO} \\ \hline
 & & \multicolumn{1}{c|}{} & \multicolumn{3}{c|}{Top 1} & \multicolumn{3}{c|}{Top 10} & \multicolumn{3}{c|}{Top 100} & \multicolumn{1}{c|}{} & \multicolumn{3}{c|}{Top 1} & \multicolumn{3}{c|}{Top 10} & \multicolumn{3}{c|}{Top 100} & \multicolumn{1}{c|}{} & \multicolumn{3}{c|}{Top 1} & \multicolumn{3}{c|}{Top 10} & \multicolumn{3}{c|}{Top 100} & \multicolumn{1}{l|}{} & \multicolumn{3}{c|}{Top 1} & \multicolumn{3}{c|}{Top 10} & \multicolumn{3}{c}{Top 100} \\
 & category & \multicolumn{1}{c|}{Oracle Calls} & \multicolumn{1}{c}{Score} & \multicolumn{1}{c}{SA} & \multicolumn{1}{c|}{AUC} & \multicolumn{1}{c}{Score} & \multicolumn{1}{c}{SA} & \multicolumn{1}{c|}{AUC} & \multicolumn{1}{c}{Score} & \multicolumn{1}{c}{SA} & \multicolumn{1}{c|}{AUC} & \multicolumn{1}{c|}{Oracle Calls} & \multicolumn{1}{c}{Score} & \multicolumn{1}{c}{SA} & \multicolumn{1}{c|}{AUC} & \multicolumn{1}{c}{Score} & \multicolumn{1}{c}{SA} & \multicolumn{1}{c|}{AUC} & \multicolumn{1}{c}{Score} & \multicolumn{1}{c}{SA} & \multicolumn{1}{c|}{AUC} & \multicolumn{1}{c|}{Oracle Calls} & \multicolumn{1}{c}{Score} & \multicolumn{1}{c}{SA} & \multicolumn{1}{c|}{AUC} & \multicolumn{1}{c}{Score} & \multicolumn{1}{c}{SA} & \multicolumn{1}{c|}{AUC} & \multicolumn{1}{c}{Score} & \multicolumn{1}{c}{SA} & \multicolumn{1}{c|}{AUC} & \multicolumn{1}{c|}{Oracle Calls} & \multicolumn{1}{c}{Score} & \multicolumn{1}{c}{SA} & \multicolumn{1}{c|}{AUC} & \multicolumn{1}{c}{Score} & \multicolumn{1}{c}{SA} & \multicolumn{1}{c|}{AUC} & \multicolumn{1}{c}{Score} & \multicolumn{1}{c}{SA} & \multicolumn{1}{c}{AUC} \\
\hline
\end{tabular}
}
\label{tab:full-4}
\end{table}

\begin{figure}[h!]

 \centering
 \begin{minipage}{0.3\linewidth}
  \begin{subfigure}[b]{0.99\textwidth}
     \centering
     \includegraphics[width=\textwidth]{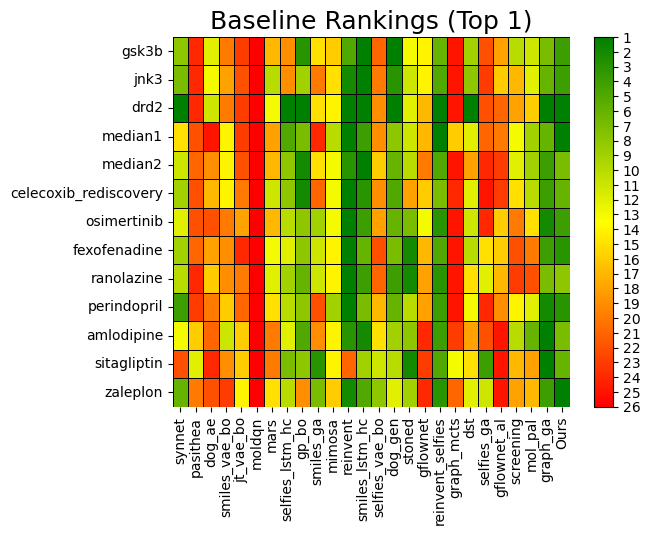}
 \end{subfigure}
 \begin{subfigure}[b]{0.99\textwidth}
     \centering
     \includegraphics[width=\textwidth]{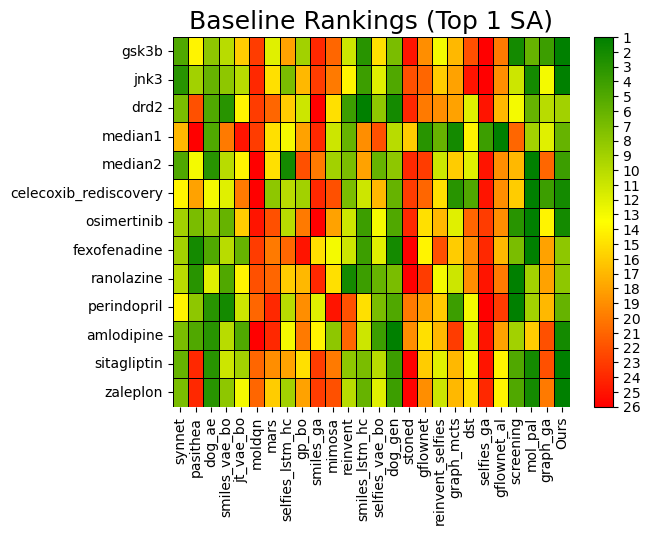}
 \end{subfigure}
  \begin{subfigure}[b]{0.99\textwidth}
     \centering
     \includegraphics[width=\textwidth]{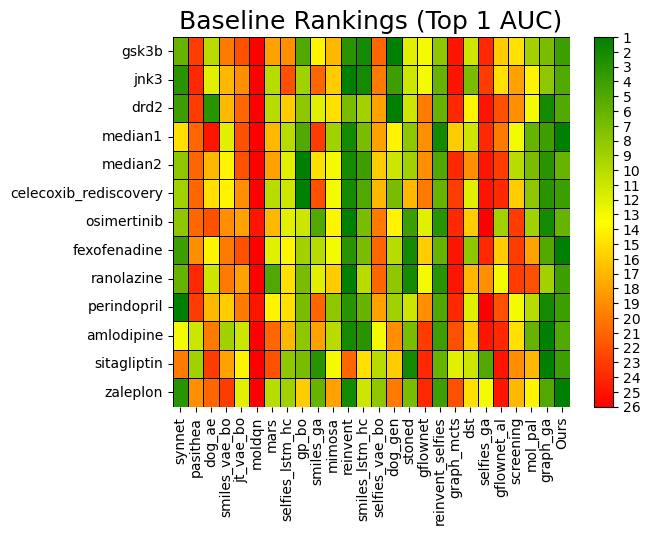}
 \end{subfigure} 
 \end{minipage}
\hspace{0.01\linewidth}
 \begin{minipage}{0.3\linewidth}
  \begin{subfigure}[b]{0.99\textwidth}
     \centering
     \includegraphics[width=\textwidth]{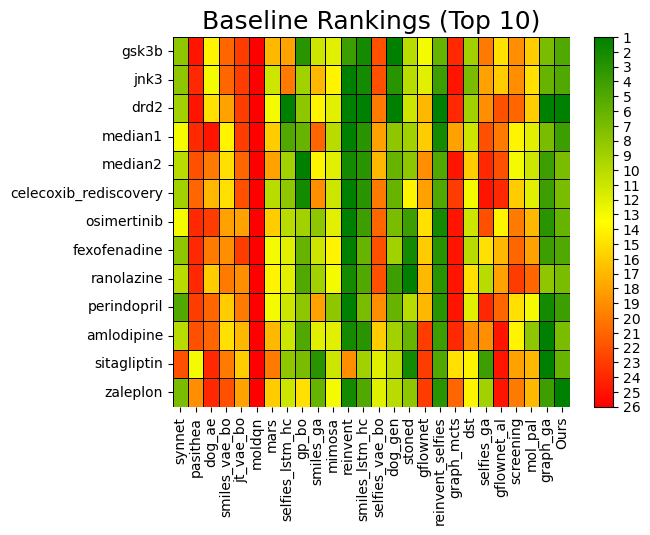}
 \end{subfigure}
 \begin{subfigure}[b]{0.99\textwidth}
     \centering
     \includegraphics[width=\textwidth]{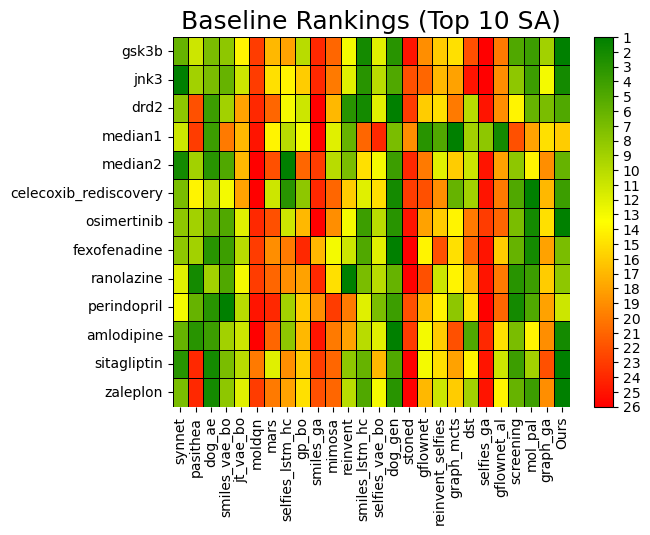}
 \end{subfigure}
  \begin{subfigure}[b]{0.99\textwidth}
     \centering
     \includegraphics[width=\textwidth]{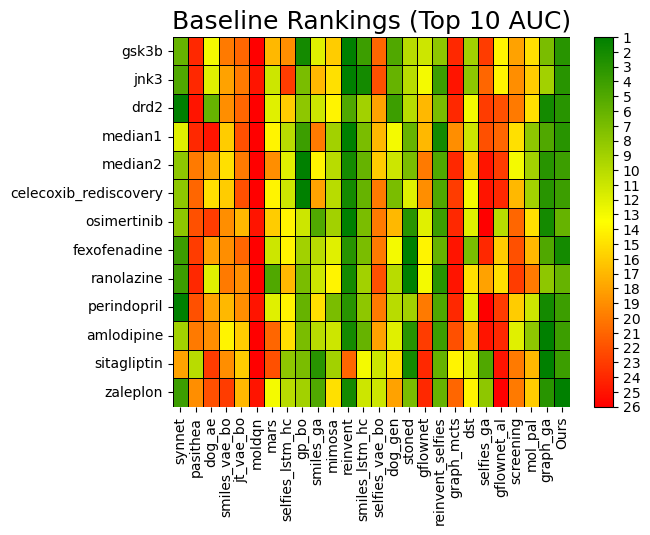}
 \end{subfigure}
 
 \end{minipage}
\hspace{0.01\linewidth}
  \begin{minipage}{0.3\linewidth}
  \begin{subfigure}[b]{0.99\textwidth}
     \centering
     \includegraphics[width=\textwidth]{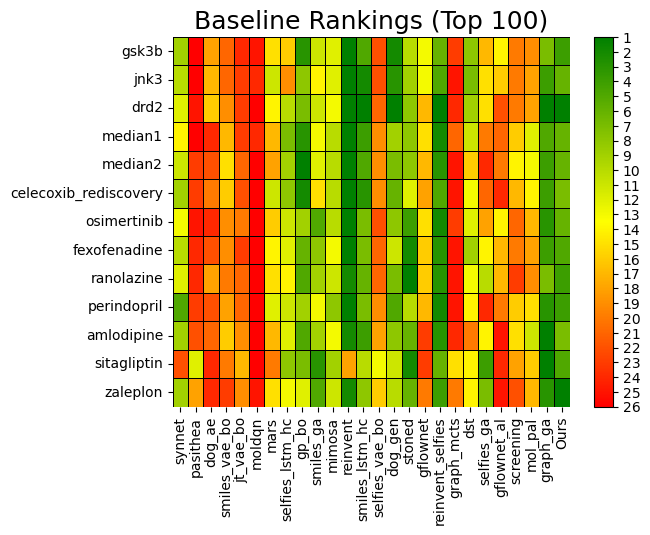}
 \end{subfigure}
 \begin{subfigure}[b]{0.99\textwidth}
     \centering
     \includegraphics[width=\textwidth]{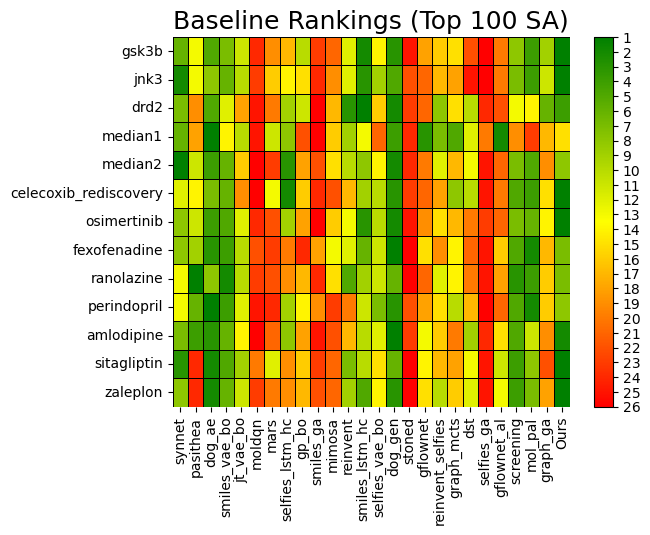}
 \end{subfigure}
  \begin{subfigure}[b]{0.99\textwidth}
     \centering
     \includegraphics[width=\textwidth]{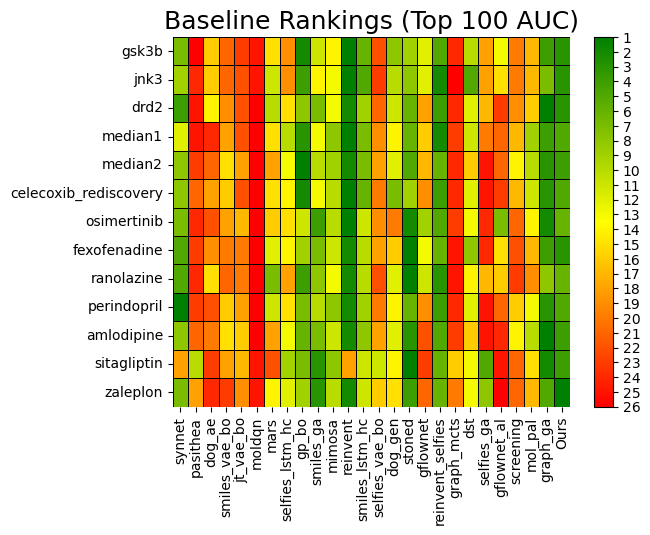}
 \end{subfigure}
 
 \end{minipage}
\caption{Ranking of our method against baselines on Top $k$ Average Scores (top), SA Scores (middle) and AUC (bottom), for $k=1,10,100$ (left, middle, right).}
\label{fig:rankings}

 \end{figure}

Tables \ref{tab:full-1}, \ref{tab:full-2}, \ref{tab:full-3} and \ref{tab:full-4} are comprehensive results against baselines taxonomized in \citep{gao2022}. We evaluate the average score of the Top K molecules, their average synthetic accessibility \citep{ertl2009} and top K AUC (AUC of no. oracle calls vs score plot), for K=1,10,100. Like \citep{gao2022}, we limit to 10000 Oracle calls, truncating and padding to 10000 if convergence occurs before 10000 calls. For each cell, numbers are followed by rankings. $X (R_1|R_2)$ means score $X$ is ranked $R_1$-best amongst all methods for that column and $R_2$-best amongst in-category methods. We visualize the rankings in Figure \ref{fig:rankings} to facilitate easier interpretation of the results.

\section{Docking Case Study with AutoDock Vina}\label{app:docking-study}
\begin{figure}[h!]
 \centering
 \begin{subfigure}[b]{0.99\textwidth}
     \centering
     \includegraphics[width=\textwidth]{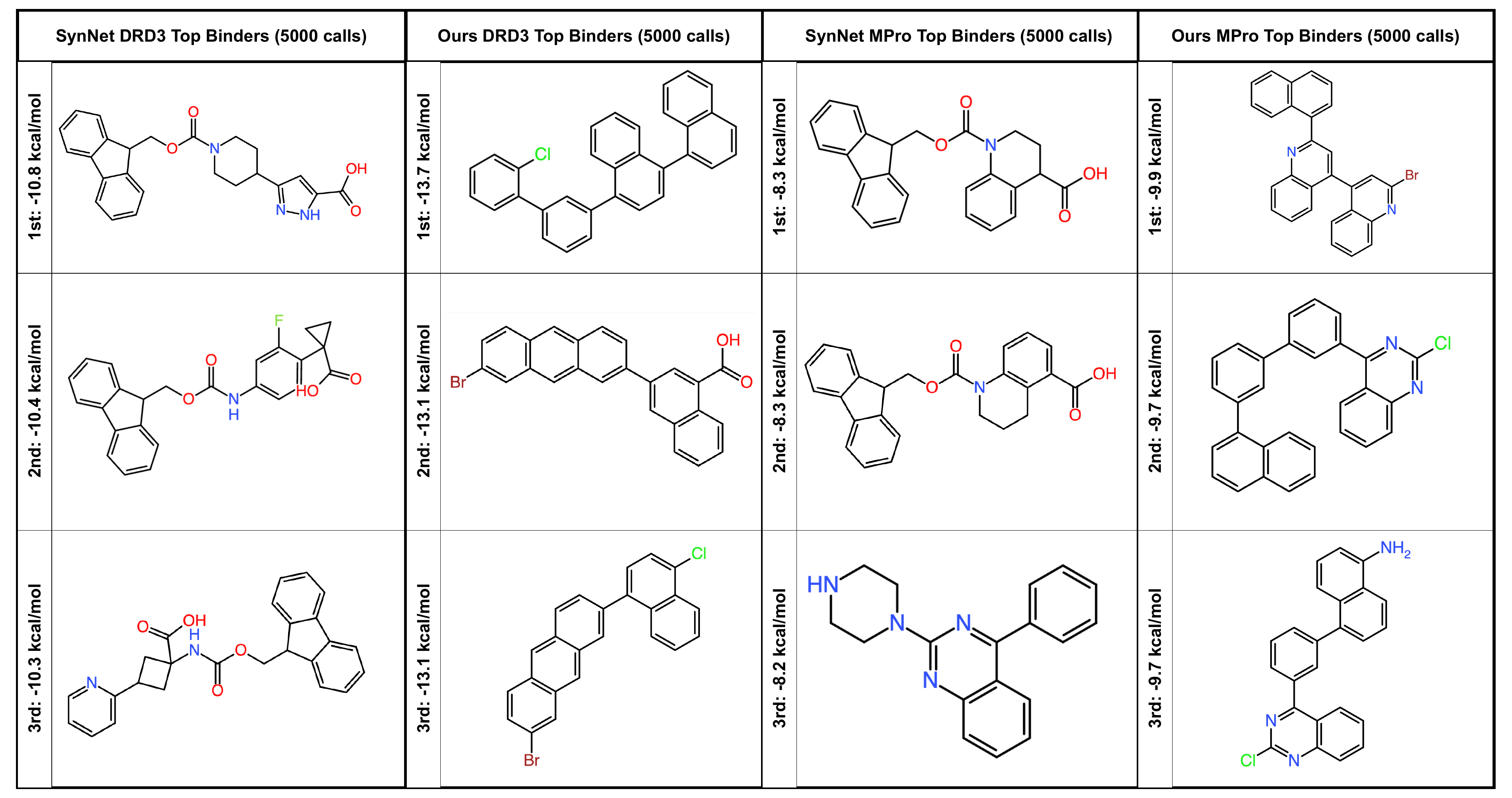}
 \caption{Top 3 molecules with lowest binding energy against DRD3 and M\textsuperscript{pro} from Ours vs SynNet}
 \label{fig:binding-ours}
 \end{subfigure}
 \vspace{10pt}
  \begin{subfigure}[b]{0.99\textwidth}
     \centering
     \includegraphics[width=\textwidth]{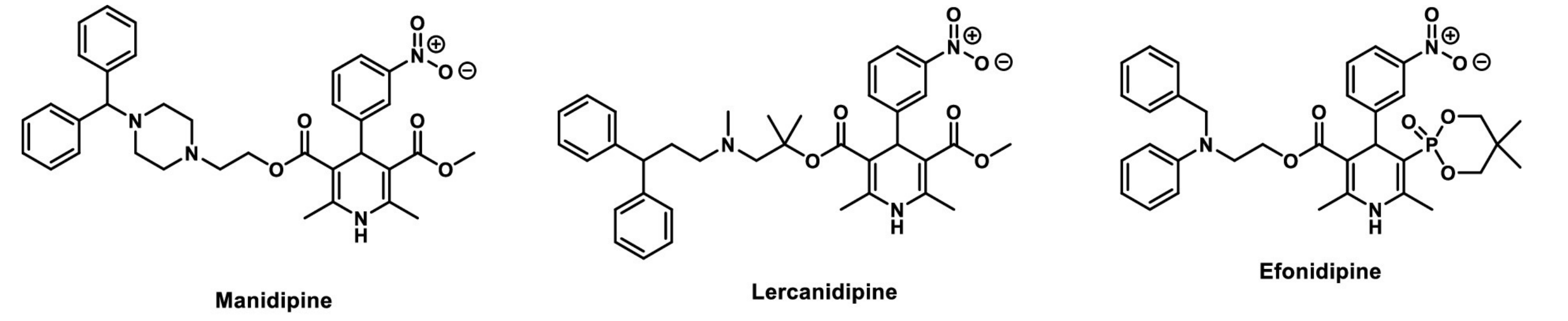} 
 \caption{Top binders against M\textsuperscript{pro} from literature, based on consensus docking scores \citep{ghahremanpour2020}}
 \label{fig:binding-literature}
 \end{subfigure}
\end{figure}

We structurally analyze the top molecules discovered by our method, visualized in Figure \ref{fig:binding-ours}.

For our optimized binders against DRD3, the chlorine substituent and polycyclic aromatic structure suggest good potential for binding through $\pi-\pi$ interactions and halogen bonding. The bromine and carboxyl groups can enhance binding affinity through halogen bonding and hydrogen bonding, respectively. The polycyclic structure further supports $\pi-\pi$ stacking interactions. In general, they have a comparable binding capability to the baseline molecules, but with simpler structures, so the ease of synthesis for the predicted molecules are higher than the baseline molecules.

For our optimized binders against Mpro, the three predicted molecules contain multiple aromatic rings in conjugation with halide groups. The conformation structures of the multiple aligned aromatic rings play a significant role in docking and achieve ideal molecular pose and binding affinity to Mpro, compared to the baseline molecules shown in Figure \ref{fig:binding-literature}. The predicted structures indicate stronger $\pi-\pi$ interaction and halogen bonding compared with the baselines. In terms of ease of synthesis, Bromination reactions are typically straightforward, but multiple fused aromatic rings can take several steps to achieve. In general, the second and third can be easier to synthesize than the top binder due to less aromatic rings performed. However, the literature molecules appeared to be even harder to synthesize due to their high complexity structures. So the predicted molecules obtained a general higher ease of synthesis than the baseline molecules. Compared with the other baseline molecules, e.g. Manidipine, Lercanidipine, Efonidipine (Dihydropyridines), known for their calcium channel blocking activity, but not specifically protease inhibitors, Azelastine, Cinnoxicam, Idarubicin vary widely in their primary activities, not specifically designed for protease inhibition. Talampicillin and Lapatinib are also primarily designed for other mechanisms of action. Boceprevir, Nelfinavir, Indinavir, on the other hand, are known protease inhibitors with structures optimized for binding to protease active sites, so can serve as strong benchmarks. Overall, the binding effectiveness of the predicted molecules are quite comparable to the baseline molecules.

\end{document}